\documentclass[onecollarge]{svjour2} 

\smartqed  

\usepackage{amsfonts }
\usepackage{bbm}
\usepackage[dvipdfmx]{graphicx}
\usepackage{amsmath,amssymb}
\usepackage{mathrsfs}
\usepackage{dcolumn}
\usepackage{bm}
\usepackage{tabularx}
\usepackage{multirow}
\usepackage{color}
\usepackage{ascmac}
\usepackage{diagbox}
\usepackage{makecell}

\usepackage{times}
\usepackage{comment}
\usepackage{ulem}
\usepackage{footnote}

\newcommand{\tb}{\textbf}

\newcommand{\ra}{\rangle}
\newcommand{\la}{\langle}

\newcommand{\eps}{\epsilon}

\newcommand{\eqsp}[2]{\begin{equation}\label{#2}\begin{split}#1\end{split}\end{equation}}
\newcommand{\eq}[2]{\begin{equation}\label{#2}#1\end{equation}}
\newcommand{\INT}[3]{\int^{#2}_{#1}\mathrm{d}#3}
\newcommand{\diff}[2]{\frac{\mathrm{d}#1}{\mathrm{d}#2}}
\newcommand{\pdiff}[2]{\frac{\mathrm{\partial}#1}{\mathrm{\partial}#2}}
\newcommand{\mr}[1]{\mathrm{#1}}

\newcommand{\mbf}[1]{\mathbf{#1}}
\newcommand{\mbb}[1]{\mathbb{#1}}
\newcommand{\Le}{\left}
\newcommand{\Ri}{\right}
\newcommand{\fr}[2]{\frac{#1}{#2}}
\newcommand{\mc}{\mathcal}
\newcommand{\dd}{{\mathrm d}}

\def\XXint#1#2#3{{\setbox0=\hbox{$#1{#2#3}{\int}$}
\vcenter{\hbox{$#2#3$}}\kern-.5\wd0}}

\journalname{Journal of Statistical Physics}

\begin{document}

\title{Mean-field theory of the Santa Fe model revisited: \\a systematic derivation from an exact BBGKY hierarchy \\ for the zero-intelligence limit-order book model}

\author{Taiki Wakatsuki \and Kiyoshi Kanazawa}

\institute{
	\email{wakatuki.taiki.36a@st.kyoto-u.ac.jp}\\
	\email{kiyoshi@scphys.kyoto-u.ac.jp}\\
	Department of Physics, Graduate School of Science, Kyoto University, Kyoto 606-8502, Japan
}

\date{Received: date / Accepted: date}

\maketitle

\begin{abstract}
    The Santa Fe model is an established econophysics model for describing stochastic dynamics of the limit order book from the viewpoint of the zero-intelligence approach. While its foundation was studied by combining a dimensional analysis and a mean-field theory by E. Smith {\it et al.} in Quantitative Finance 2003, their arguments are rather heuristic and lack solid mathematical foundation; indeed, their mean-field equations were derived with heuristic arguments and their solutions were not explicitly obtained. In this work, we revisit the mean-field theory of the Santa Fe model from the viewpoint of kinetic theory---a traditional mathematical program in statistical physics. We study the exact master equation for the Santa Fe model and systematically derive the Bogoliubov-Born-Green-Kirkwood-Yvon (BBGKY) hierarchical equation. By applying the mean-field approximation, we derive the mean-field equation for the order-book density profile, parallel to the Boltzmann equation in conventional statistical physics. Furthermore, we obtain explicit and closed expression of the mean-field solutions. Our solutions have several implications: (1)~Our scaling formulas are available for both $\mu\to 0$ and $\mu\to \infty$ asymptotics, where $\mu$ is the market-order submission intensity. Particularly, the mean-field theory works very well for small $\mu$, while its validity is partially limited for large $\mu$. (2)~The ``method of image'' solution, heuristically derived by Bouchaud-M\'ezard-Potters in Quantitative Finance 2002, is obtained for large $\mu$, serving as a mathematical foundation for their heuristic arguments. (3)~Finally, we point out an error in E. Smith {\it et al.} 2003 in the scaling law for the diffusion constant due to a misspecification in their dimensional analysis.
\end{abstract}
\keywords{Market microstructure \and Econophysics \and Santa Fe model \and Limit order book \and Kinetic theory \and BBGKY hieararchy}

\section{Introduction}
    Understanding the dynamics of financial market microstructure is an important topic in quantitative finance~\cite{StanleyB,SlaninaB}, particularly from the viewpoint of the order flow dynamics~\cite{BouchaudB}. Modern financial markets (such as stock markets and foreign exchange markets) are based on the continuous-time double auction system, whereby three types of orders are available: (1)~limit-order submission, (2)~market-order submission, and (3)~cancellation order. Here, (1)~the limit order represents an order showing their price/volume at which the trader is willing to buy/sell. Limit orders are collected and are summarised as the limit order book (LOB). The highest bid and lowest ask prices are referred to as the best bid and ask prices. When a trader wants to buy/sell immediately, (2)~the trader will issue a market order, by which the trader buys/sells at the corresponding best price. Of course, if the trader who issued a limit order in the past does not want to buy/sell anymore, (3)~the trader is allowed to cancel their limit order by issuing a cancellation order. This is the outline of the order-flow mechanism in financial market microstructure. 

    The Santa Fe model is one of the established stochastic models describing such order flows within the continuous-time Markov process~\cite{BouchaudB,ZIOB-PRL2003,ZIOB-QFin2003}. In this model, the order flow dynamics is simply modelled without any strategic behaviour of traders, where all order arrivals are assumed to follow the Poisson process. Therefore, this model belongs to the family of the zero-intelligence models---an antithesis model to the rational-agent assumption in standard economics. Interestingly, this model exhibits very plausible behaviour consistent with empirical results and is regarded as a ``null model'' (or a starting point) before taking various strategic behaviour into account.

    While the Santa Fe model is a simple model, its theoretical analysis is very challenging as it is genuinely a many-body model. Mathematically, it belongs to an infinite-dimensional Markov process. In the previous literature, approximate finite-dimensional modelling was proposed by introducing a cutoff in the LOB~\cite{Abergel2016,Abergel2013}, to qualitatively prove its Brownian nature in the long-time limit. For more quantitative analysis, a dimensional analysis supplemented by a mean-field theory was proposed by E. Smith {\it et al.} in Quantitative Finance 2003~\cite{ZIOB-QFin2003} to derive scaling laws for market metrics, such as the spread, price impact, and diffusion constant. Though their mean-field theory is not perfect at predicting all scaling laws of the financial metrics, it has been an important cornerstone for theoretical understanding of the Santa Fe model. 

    However, the mean-field theory by E. Smith {\it et al.} 2003~\cite{ZIOB-QFin2003} is very complicated and heuristic without solid ground: First, they applied a simple dimensional analysis to roughly predict the scaling laws of financial metrics. They next derived two mean-field equations heuristically, one for the LOB density profile and the other for the gap size probability density function (PDF), whose analytical solutions were not explicitly obtained. While lacking explicit mean-field solutions, they argued that the scaling laws derived in the simple dimensional analysis are consistent with the mean-field equations. Since their overall logic is neither simple nor clear regarding their application limit, it would have merit to reorganise mathematical foundation of their mean-field theory by using a more exact and systematic formulation. 

    In this work, we revisit the mean-field theory for the most basic version of the Santa Fe LOB model with homogeneous submission intensity distribution by using the kinetic approach~\cite{Dorfman2021}---a traditional framework developed in statistical physics and applied to interdisciplinary fields~\cite{Bertin2006,Schadschneider2010,Carmona2018,Pareschi2013} including financial econophysics~\cite{SocialPhysics,KanazawaPRL2018,KanazawaPRE2018,KanazawaJSP2023}. We start from the stochastic dynamics of multiple queues, and derive the exact master equation for the Santa Fe model, that describes the time-evolution of the LOB state in a closed form. We then apply an exact coarse-graining to the master equation to obtain the Bogoliubov-Born-Green-Kirkwood-Yvon (BBGKY) hierarchal equation~\cite{Resibois1977}. By applying the mean-field approximation, we derive a nonlinear closed equation for the order-book density profile by assuming the small tick-size limit, that has a parallel structure to the Boltzmann equation in conventional kinetic theory. By explicit solving the ``Boltzmann'' equation, we present the following findings: 
    \begin{enumerate}
        \item[(1)] We obtain closed and explicit expressions for the financial metrics, such as the spread, instantaneous price impact, and diffusion constant for both the small $\mu$ limit and the large $\mu$ limit, where $\mu$ is the intensity of market order submissions.
        \item[(2)] Regarding its application limit, the mean-field theory works very well for the small $\mu$ asymptotics. For the large $\mu$ asymptotics, on the other hand, its precise correctness is slightly limited: while the mean-field scalings are correct for most financial metrics, the scalings do not work for the permanent price impact and diffusion constant. This observation is consistent with the previous report.
        \item[(3)] For the $\mu\to \infty$ asymptotics, most of the mean-field scalings are identical to those previously reported in E. Smith {\it et al.} 2003~\cite{ZIOB-QFin2003}. However, their mean-field scaling is incorrect for the diffusion constant due to misspecification of a characteristic constant in their dimensional analysis.         
        \item[(4)] Our mean-field theory argues that, for large $\mu$, the average order book profile formula is exactly equal to the ``method of image'' solution, which was proposed by Bouchaud-M\'ezard-Potters in Quantitative Finance 2002~\cite{ZIOB-Bouchaud2002} by a heuristic argument. Thus, our kinetic framework serves as a mathematical foundation for their method of image solution within the mean-field approximation.  
        \item[(5)] While E. Smith {\it et al.}~\cite{ZIOB-QFin2003} studied two simultaneous mean-field equations, one of them (i.e., the gap-size mean-field equation) is unnecessary to derive the mean-field scalings; the mean-field equation for the LOB density profile has sufficient information when deriving power-law scalings of the financial metrics. Note that the gap-size mean-field equation can also be derived via the BBGKY hierarchy.
    \end{enumerate}
    Thus, we have established the mathematical foundation of the Santa Fe LOB model within the mean-field theory, by exploiting the standard kinetic program in statistical physics. 

    This paper is organised as follows: Section~\ref{sec:model} describes our mathematical notation and review the Santa Fe model. Particularly, we explain our explicit motivation after literature review. In Sec.~\ref{sec:statespace-pathleveldynamics}, we introduce the state space (or the phase space) to formulate the Santa Fe model within the infinite-dimensional Markov dynamics. Also, we formulate its path-level dynamics. In Sec.~\ref{sec:kinetictheory-ME+BBGKY}, we derive the master equation for the Santa Fe model and derive its exact BBGKY hierarchical equation, according to the standard program of kinetic theory. Section~\ref{sec:BoltzmannEq} studies its mean-field theory: the Boltzmann equation for the LOB profile and its approximate solution. We also show that our mean-field solution coincides with the method of image solution, phenomenologically introduced in Ref.~\cite{ZIOB-Bouchaud2002}. Section~\ref{sec:summary-scaling} summarises our scaling results with comparative discussion about Ref.~\cite{ZIOB-QFin2003} and numerical simulations. We finally conclude this work with several remarks in Sec.~\ref{sec:conclusion}.

\section{Setup description of the Santa Fe model and its reveiw}\label{sec:model}
    This section reviews the Santa Fe model for the limit order book dynamics. 

    \subsection{Mathematical notation}
        Let us explain our mathematical notation in this work. $\mathbb{Z}:=\{0,\pm 1, \pm 2,\dots\}$ is the set of all integers, and $\mathbb{N}:=\{1,2,\dots\}$ is the set of all natural numbers. $\mbb{Z}_{1/2}:=\{i/2\mid i \in \mathbb{Z}\}$ is the set of integers and half-integers. $\mathbb{R}$ is the set of all real numbers.
        
        We next explain our notation for probability theory. In this paper, any stochastic variables are distinguished from non-stochastic real numbers by adding hat symbols (i.e., $\hat{A}$ is a stochastic variable and $A$ is a non-stochastic real number). $P_t(A)$ represents the probability distribution function (PDF) at time $t$, where the probability that $\hat{A}\in \left[A,A+dA\right)$ is given by $P_t(A)dA$. The ensemble average of the function $f(\hat{A}_1,\cdots,\hat{A}_N)$ is given by 
        \begin{equation}
            \langle f(\hat{A}_1,\cdots,\hat{A}_N)\rangle:=\int_{-\infty}^{\infty}  \,\prod^N_{i=1}d\hat{A}_iP_t(\hat{A}_1,\cdots,\hat{A}_N)f(\hat{A}_1,\cdots,\hat{A}_N).
        \end{equation}

        In this paper, we use indicator functions for mathematical organisation of case-by-case analysis for stochastic dynamics. We therefore define an indicator function for a proposition $\mathcal{P} (x)$ that depends on a variable $x$. For example, let $\mathcal{P}(x)$ be a proposition that $x$ belongs to the set $\Omega$. We denote the indicator function by 
        \eq{
            \mathbb{I} _{\Omega}(x):=
            \begin{cases}
                1&(x\in \Omega)\\
                0&(x\notin \Omega)\\
            \end{cases}.
        }{}
        Also, let $\mathcal{P}(x)$ be a proposition that $x$ is $y$. The corresponding indicator function is given by the Kronecker delta: 
        \eq{
            \delta_{x,y}:=
            \begin{cases}
                1&(x=y)\\
                0&(x\neq y)\\
            \end{cases}.
        }{}
        See Appendix \ref{sec:App_indicator} for more technical details about the indicator functions.

    \subsection{Setup description of the Santa Fe model}\label{subsec:Mic_dyn}
        The Santa Fe model is an established model for describing the order flow dynamics. The Santa Fe model is a zero-intelligence model, where rational behaviour is not at all assumed for the trader dynamics and all order flow dynamics are assumed to follow simple Poisson processes. 

        The Santa Fe model is an infinite-dimensional Markov process with queues placed at discrete prices $\mbf{p}:=\{i\Delta\}^{\infty}_{i=-\infty}$ with a price interval $\Delta$. For a price $p:=i\Delta\in\mbb{R}$ with an integer $i\in\mbb{N}$, the integer part $i$ is called the {\it price level}. In this report, the price and corresponding price level can be negative for simplicity\footnote{For example, our price can be practically regarded as a log price, which can be negative.}. For each price level $i$, the queue can have bid or ask limit orders. For this setup, three events occur according to a time-homogeneous Poisson process: limit-order submission, cancellation of existing limit orders, and market-order submission occur as a Poisson process. Furthermore, we introduce a cutoff length $L\Delta>0$ from the best price for each event to simplify the dynamics. The cutoff length is set to be very large, and we finally take the $L\to \infty$ limit. 
          
        \subsubsection{Limit-order submission}
            \begin{figure}[t]
                \centering
                    \includegraphics[width=0.9\columnwidth]{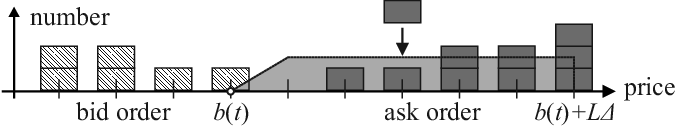}
                    \caption{
                        Limit-order submission dynamics in the Santa Fe model. A new limit order is uniformly submitted from the opposite best price with the intensity $\lambda \Delta$. Note that we introduce a cut-off length $L<\infty$, beyond which new limit-order submission does not occur, and take the $L\to \infty$ limit later. 
                    }
                    \label{fig:schematic-LOBsubmission}
            \end{figure}
            
            Let us consider a limit-order submission (see Fig.~\ref{fig:schematic-LOBsubmission} for a schematic). Any limit-order submission is based on the opposite best price. For the ask side at time $t+\dd t$, for example, an ask limit order can be placed at the price $i\Delta \in \mbf{p}$ within the range
            \eqsp{
                b(t)<i\Delta\leq b(t)+L\Delta
            }{}
            with a Poisson intensity $\lambda \Delta$, where $b(t)$ is the best bid price (i.e., the highest price of bid prices) at time $t$ and $L$ is the cutoff (see Fig.~\ref{fig:schematic-LOBsubmission}). Here, the intensity is the probability per unit time of an event occurrence (i.e., an ask order is submitted at $i\Delta \in \mbf{p}$ with the probability $\dd t\lambda \Delta$). In a parallel manner, a bid limit order can be placed at the price $i\Delta \in \mbf{p}$ within the range
            \eqsp{
                a(t)-L\Delta\leq i\Delta< a(t)
            }{}
            with a Poisson intensity $\lambda\Delta$, where $a(t)$ is the best ask price (i.e., the lowest price of ask prices) at time $t$. In this work, we assume that all limit orders have the unit volume for simplicity. 

            In other words, a new limit order cannot be placed beyond the cutoff, while bid and ask queues are defined beyond. The limit orders will remain until the order is cancelled or executed, even beyond the cutoff. The cutoff is introduced for a technical reason to avoid divergent total intensity during our exact calculations. However, its value is not important; indeed, we finally take the limit $L\to \infty$. 
            
            Note that Refs.~\cite{Abergel2016,Abergel2013} introduced another cutoff to suppress the dimension of the Santa Fe model, approximating it as a finite-dimensional Markov process. While it seems similar to ours at a glance, it is different in the sense that our Santa Fe model has infinite dimension.

        \subsubsection{Cancellation of an existing limit order}
            \begin{figure}[t]
                \centering
                    \includegraphics[width=0.9\columnwidth]{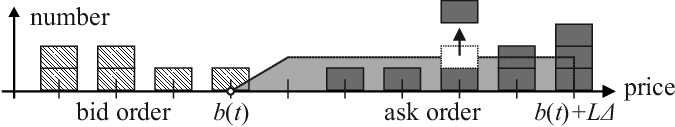}
                    \caption{
                        Order cancellation dynamics in the Santa Fe model. All the orders have the same cancellation intensity $v$. Again, we introduce a cut-off length $L<\infty$, beyond which no cancellation occurs, and take the $L\to \infty$ limit later. 
                    }
                    \label{fig:schematic-cancel}
            \end{figure}
            Any limit order may be cancelled within the cutoff (see Fig.~\ref{fig:schematic-cancel}). In other words, any ask limit order within the range
            \eqsp{
                b(t)<i\Delta\leq b(t)+L\Delta
            }{}
            is cancelled according to the Poisson process with intensity $v$. As a parallel formulation, any bid limit order within the range
            \eqsp{
                a(t)-L\Delta\leq i\Delta< a(t)
            }{}
            is cancelled with intensity $v$. In this work, we assume that cancellation volume is always the unity. 

        \subsubsection{Market-order submission}
            \begin{figure}[t]
                \centering
                \includegraphics[width=0.9\columnwidth]{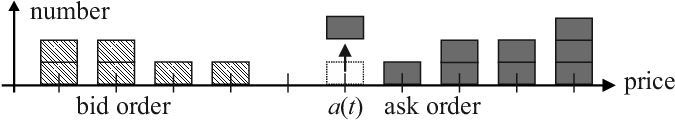}
                \caption{
                    Market-order dynamics in the Santa Fe model. A best-price order is executed by a market order and is removed.
                }
                \label{fig:Schematic-MarketOrder}
            \end{figure}
            
            The final piece is the market-order submission (see Fig.~\ref{fig:Schematic-MarketOrder}). The best bid or ask limit order at $b(t)$ or $a(t)$ is executed by a market-order submission with the unit volume, following the Poisson process with the intensity $\mu$. Since any limit order is cancelled with the intensity $v$, the best bid or ask limit order disappears with the intensity $v+\mu$ in total.

    \subsection{Focus of this work: the continuous asymptotic limit of the Santa Fe model}
        \label{sub_sec:para_model}
        In this study, we focus on the {\it continuous asymptotic limit} of the Santa Fe model for our analytical purpose. In other words, we consider the following asymptotic limit: 
        \begin{itembox}[l]{The continuous asymptotic limit of the Santa Fe model}
            \vspace{-5mm}
            \eqsp{
                n_{\rm st}:=\frac{\lambda\Delta}{v}\to 0,\quad 
                \epsilon:=\frac{v+\mu}{\lambda}\to 0.
            }{eq:parameter_domain}
            \vspace{-5mm}
        \end{itembox}
        The first condition $\Delta \to 0$ represents the {\it small tick-size limit}. For a small tick size $\Delta \to 0$, most of the queues should have one or zero orders because of sparse LOB profile. Indeed, the average number of limit orders $n_{\rm st}$ at the price levels far from the best price levels is estimated by the balance between new limit-order submissions and cancellations, such that 
            \eq{
                v n_{\rm st}\approx \lambda \Delta
                \quad \Longleftrightarrow \quad 
                n_{\rm st} \approx\frac{\lambda\Delta}{v},
            }{}
        where the contribution of market-order executions is ignored. Therefore, the small tick-size limit is correctly formulated by $n_{\rm st}\ll 1$, which finally enables us to characterise the Santa Fe model dynamics by the average LOB profile $\rho^a_t(r)$ as a continuous function with respect to the continuous relative price $r$. 

        The second condition $\epsilon\to 0$ represents the {\it high-liquidity limit}, under which the LOB has many limit orders and kinetic theory is potentially available. Standard statistical physics programs assume the thermodynamic limit $N_{\rm particle}\to \infty$, where $N_{\rm particle}$ is the total number of particles. Since we reorganise the mean-field theory of the Santa Fe model via kinetic theory in this work, we must consider the ``thermodynamic limit'' of the Santa Fe model. The high-liquidity limit $\epsilon\ll 1$ naturally corresponds to the ``thermodynamic limit" of the Santa Fe model. Indeed, let us estimate the total number of ask limit orders $N_{\rm tot}([p,p+l))$ within the range $[p,p+l)$ as 
        \eq{
                N_{\rm tot}([p,p+l)) \approx n_{\rm st}\frac{l}{\Delta} = \frac{\lambda l}{v} > \fr{\lambda l}{\mu+v} = \frac{l}{\eps},
            }{}
        where $p$ is the price far from the best ask price and $l$ is the price range of order one, such that $p\gg a(t)$ and $l=O(1)$. For the high-liquidity limit $\eps\to 0$, there are many limit orders on average, such that $N_{\rm tot}([p,p+l))\gg 1$ for any price range $l=O(1)$. Thus, the high-liquidity limit $\eps \to 0$ plays a similar role of the ``thermodynamic limit'' in conventional statistical physics. 

    \subsection{Review of the original mean-field theory in E. Smith {\it et al.} 2003}\label{sec:review_Smith}
        Let us review the original mean-field theory in E. Smith {\it et al.} Quantitative Finance 2003~\cite{ZIOB-QFin2003}, to finally highlight the difference from ours, particularly for readers unfamiliar with Ref.~\cite{ZIOB-QFin2003}. Their original theory is based on a combination of the dimensional analysis and two mean-field equations. Let us first start with the review of their dimensional analysis, followed by their mean-field equations. Note that we stick to a simpler version of the Santa Fe model where the volume of any limit order and market order is unity. Also, since the main parts are written to be self-contained, readers interested only in our main results can skip this subsection.

        \subsubsection{Review of the original dimensional analysis}
            In Ref.~\cite{ZIOB-QFin2003}, they selected the typical characteristic constants for price and time as 
                \eqsp{
                    \frac{\mu}{\lambda}\sim\text{[dimension of price]}, \quad 
                    \frac{1}{v}\sim\text{[dimension of time]}.
                }{eq:Smith-characteristic-constants}
            Based on this assumption, they estimated the scalings of the spread $s$ and the diffusion constant $D$ as 
                \eqsp{
                    s \propto \frac{\mu}{\lambda}\sim\text{[dimension of price]}, \quad
                    D\propto v\frac{\mu^2}{\lambda^2}\sim\text{[price}^2\text{ per time]}.
                }{eq:Smith-scaling}

        \subsubsection{Review of the original mean-field theory}
            The starting point of their theory is a master equation of the Santa Fe model that was derived by a heuristic argument. Specifically, they assume that the time evolution equation of the PDF $P^a_t(n,r)$ for the number of ask limit orders $n$ at the price distance $r$ from the midprice is given by 
                \eqsp{
                    \frac{\partial P^a_t(n,r)}{\partial t}
                    =&\tilde{\lambda}(r)\Delta\Le[P^a_t(n-u,r)-P^a_t(n,r)\Ri]
                    +v\Le[(n+u)P^a_t(n+u,r)-nP^a_t(n,r)\Ri]\\
                    &+\tilde{\mu}(r)\Le[P^a_t(n+u,r)-P^a_t(n,r)\Ri]+\sum_{\dd m}Q(\dd m)\Le[P^a_t(n,r-\dd m)-P^a_t(n,r)\Ri],
                }{}
            where $\tilde{\lambda}(r)$ is the effective limit-order submission intensity dependent on the relative distance $r$, $\tilde{\mu}(r)$ is the effective execution intensity dependent on the relative distance $r$, and $Q(\dd m)$ is the midprice-change intensity with the jump size $\dd m$. These effective intensities $\tilde{\lambda}(r)$, $\tilde{\mu}(r)$, and $Q(\dd m)$ were phenomenologically introduced satisfying several trivial relations, such as $\lim_{r\to \infty}\tilde{\lambda}(r)=\lambda$, $\lim_{r\to 0}\tilde{\mu}(r)=\mu$, and $\lim_{r\to \infty}\tilde{\mu}_s(r)=0$. In the mean-field approximation, they heuristically identified the specific forms of $\tilde{\lambda}(r)$, $\tilde{\mu}(r)$, and $Q(\dd m)$ as nonlinear functions of $P^a_t(n,r)$. Thus, this mean-field equation is a nonlinear master equation with respect to the one-body PDF $P^a_t(n,r)$.

            They further applied the small tick-size approximation to obtain a simpler nonlinear master equation for the average LOB density profile $\psi(r)$ with the relative price $r$ from the midprice,  
            \eqsp{
                \lambda \Le(1+e^{-\INT{0}{r}{x}\psi(x)}\Ri)=\mu\psi(r)e^{-\INT{0}{r}{x}\psi(x)}+v\psi(r)-D\frac{d^2\psi(r)}{d{r}^2}.
            }{eq:Smith_density}
            While this equation is very intricate, they numerically solved it to evaluate the characters of the solution.

            In addition, they derived another mean-field equation for the gap-size PDF $P_t(g_k)$, where the gap-size is defined by $g_k:=\frac{p_{k+1}-p_k}{\Delta}$ with the $k$-th best ask price $p_k$ for $k\geq 1$ and $g_0$ is the spread $s$. They derived a recursive formula regarding the mean gap size $\la g_k\ra$, such that
                \begin{equation}\label{eq:main:Smith_gap_comp}
                    \begin{cases}
                        0=
                        \Le(\mu+v\Ri) \Le(\Le\langle g_1\Ri\rangle\Ri)
                        -\lambda\Le(\Le\langle g_0\Ri\rangle^2
                        -\Le\langle g_0\Ri\rangle\Ri)&\\
                        0=\{\mu+(k+1)v\}\langle g_{k+1}\rangle
                        -\lambda \langle g_{k}\rangle
                        \sum^{k}_{i=0}(\langle g_{i}\rangle-1) &(1\leq k)
                    \end{cases}.
                \end{equation}
            They evaluated the average spread via this equation. In other words, they considered the simultaneous system composed of the two mean-field equations for the LOB density profile $\psi(x)$ and the gap-size PDF $P_t(g_k)$. 

        \subsubsection{The goal of this paper}
            Let us describe our aim of this paper by pointing out several issues in E. Smith {\it et al.} Quantitative Finance 2003~\cite{ZIOB-QFin2003}:
            \begin{itemize}
                \item Point 1: The mean-field formulation in Ref.~\cite{ZIOB-QFin2003} is heuristic/phenomenological and lacks solid mathematical foundation. For example, is it true that their ``mean-field'' equations require only the mean-field and the small tick-size approximations, without introducing additional heuristic arguments? 
                \item Point 2: Logically, there are several candidates for dimensional analyses, different from the one proposed in Ref.~\cite{ZIOB-QFin2003}. As summarised in Eq.~\eqref{eq:Smith-characteristic-constants}, Ref.~\cite{ZIOB-QFin2003} assumed that the characteristic constants for price and time are given by $\mu/\lambda$ and $1/v$, respectively. However, there are other candidates for characteristic-constant selections, such as $v/\lambda$ for a characteristic price constant and $1/\mu$ for a characteristic time constant. It is generally difficult to uniquely identify the correct set of characteristic constants without theoretical background. Since Ref.~\cite{ZIOB-QFin2003} did not obtain explicit analytical solutions of their mean-field equations, there is no solid theoretical ground to support the correctness of their dimensional analysis. We will point out the specific point at which an error is located in their calculations.
            \end{itemize}
            To solve these points, we revisit the mean-field theory of the Santa Fe model by applying mathematical kinetic theory. Note that kinetic theory~\cite{Dorfman2021} has been mathematically generalised for interdisciplinary fields beyond conventional statistical physics, such as for active matter~\cite{Bertin2006}, traffic modelling~\cite{Schadschneider2010}, game theory (i.e., the mean-field games~\cite{Carmona2018}), and econophysics~\cite{Pareschi2013}. Indeed, Ref.~\cite{SocialPhysics,KanazawaPRL2018,KanazawaPRE2018,KanazawaJSP2023} formulated a kinetic framework to analytically study the dealer model---an agent-based LOB model describing individual traders' decision-making~\cite{Takayasu1992,Sato1998}, proposed earlier than the Santa Fe model. In this work, we generalise the technique in Ref.~\cite{SocialPhysics,KanazawaPRL2018,KanazawaPRE2018,KanazawaJSP2023} for the Santa Fe LOB model. Correspondingly to Points 1 and 2, our answers can be summarised as follow: 
            \begin{itemize}
                \item Our answer 1: We provide the solid mathematical ground for the mean-field theory in Ref.~\cite{ZIOB-QFin2003} by deriving an exact BBGKY hierarchical equation from an exact master equation. By applying only the mean-field and the small tick-size approximations, we derive the ``Boltzmann'' equation for the average LOB profile. This Boltzmann equation has an almost identical structure to the original mean-field equation in Ref.~\cite{ZIOB-QFin2003}, which implies that our framework serves as a mathematical foundation for the mean-field theory of the Santa Fe model based on an exact approach. 
                \item Our answer 2: In this paper, we explicitly solve the mean-field equations to uniquely identify scaling laws of various financial metrics. 
                Finally, we find that their original dimensional analysis is inconsistent with explicit analytical solutions of the mean-field equations. Specifically, for the asymptotic limit $\mu\gg v$, we show that the correct characteristic constants for price and time are given by $\mu/\lambda$ and $1/\mu$. Also, for the asymptotic limit $\mu\gg v$, the correct characteristic constants are shown to be given by $v/\lambda$ for price and $1/v$ for time. Our dimensional analysis is consistent with our explicit mean-field solutions for both $\mu\gg v$ and $\mu\gg v$. 
            \end{itemize}

\section{Formulation: state space and path-level stochastic dynamics}\label{sec:statespace-pathleveldynamics}
    In this section, we describe the stochastic dynamics of the Santa Fe model within the formulation of Markov processes. Particularly, we exactly derive the master equation---the time-evolution equation for the PDF of the phase-space point---, which is a starting point for our kinetic formulation. 

    \subsection{State space: fixed coordinate system and relative coordinate system}
        Let us first define the state space (or the ``phase space" in the word of analytical mechanics) for the description of the Santa Fe model as an infinite-dimensional Markov process. First, we introduce the {\it fixed-price coordinate} for the Santa Fe model. We next define the {\it relative-price coordinate} as a variable transformation from the fixed coordinate. We show that both price coordinates have identical information for uniquely identifying the state of the Santa Fe model. Finally, our kinetic framework is described based on the relative-price coordinate.
        
        \subsubsection{Fixed price coordinate system}
            \begin{figure}
                \centering
                \includegraphics[width=0.9\columnwidth]{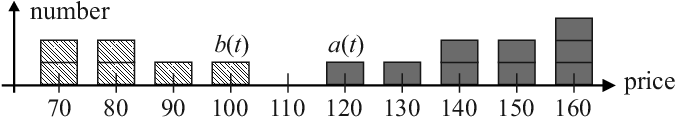}
                \caption{
                    Schematic of the fixed price coordinate with the tick size $\Delta=10$.
                }
                \label{fig:fixed-price-coordinates}
            \end{figure}  
            In the Santa Fe model, a trader submits their limit order at a discrete price with the lattice size $\Delta$. The number of ask orders at price~$x_i:= i\Delta$ at time $t$ is written as $\hat{N}^a_i(t)\in\mathbb{N}$, and the number of bid orders at price~$x_i$ at time $t$ is written as $\hat{N}^b_i(t)\in\mathbb{N}$. In this paper, the integer $i:=x_i/\Delta\in\mbb{Z}$ is referred to as the {\it price level}, instead of {\it price}, which is a real number. In the fixed-price coordinate (see Fig.~\ref{fig:fixed-price-coordinates} for a schematic), the state of the Santa Fe model is uniquely represented by a point in the state space $G_t$, 
            \eqsp{
                G_t:= \left(\hat{\mathbf{N}}^a(t),\hat{\mathbf{N}}^b(t)\right), \quad
                \hat{\mathbf{N}}^a(t):=\left\{\hat{N}^a_i(t)\right\}_{i\in \mathbb{Z}}\in\mathbb{N}^\infty, \quad
                \hat{\mathbf{N}}^b(t):=\left\{\hat{N}^b_i(t)\right\}_{i\in \mathbb{Z}}\in\mathbb{N}^\infty.
            }{}
            The best-bid and best-ask price levels are defined by 
                \eqsp{
                    A[\hat{\mathbf{N}}^a(t)]:=\text{min}[i\mid \hat{N}^a_i(t)\neq 0],\quad
                    B[\hat{\mathbf{N}}^b(t)]:=\text{max}[i\mid\hat{N}^b_i(t)\neq 0].
                }{}
            The {\it midprice level}~$\hat{m}_t$ is defined as
                \eq{
                    \hat{m}_t
                    :=
                    \frac{A[\hat{\mathbf{N}}^a(t)]+B[\hat{\mathbf{N}}^b(t)]}{2},
                }{}
            which is an element of $\mathbb{Z}_{1/2}$ (i.e., the set of integer and half-integer values).
    
        \subsubsection{Relative price coordinate system}        
            \begin{figure}
                \centering
                \includegraphics[width=0.9\columnwidth]{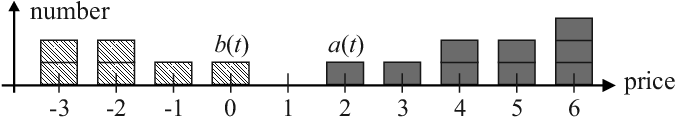}
                \caption{
                    Schematic of the relative ask price coordinate from the best bid price.
                }
                \label{fig:relative-price-coordinates}
            \end{figure}  
            We also introduce the relative price coordinate system, where the price level is measured relative to the opposite best price level. For a given price level $i$, the relative ask price level $i_a$ and the relative bid price level $i_b$ are defined as 
                \eqsp{
                    i_a:= i-B[\hat{\mathbf{N}}^b(t)]\in \mathbb{Z},\quad i_b:= i-A[\hat{\mathbf{N}}^a(t)]\in \mathbb{Z}.
                }{eq:coordinates_relative_fix}
            See Fig.~\ref{fig:relative-price-coordinates} for a schematic. Here, the price level on the fixed (relative) coordinate system is written in capital letters (lowercase letters), to stress the difference between the fixed price coordinate system and the relative price coordinate system.

            In the relative price coordinate system, the numbers of limit orders at the ask and bid price levels are written as $\hat{n}^a_{i_a}(t)$ and $\hat{n}^b_{i_b}(t)$, such that 
                \eqsp{
                    \hat{n}^a_{i_a}(t)
                    :=\hat{N}^a_{i_a+B[\hat{\mathbf{N}}^b(t)]}(t),\quad
                    \hat{n}^b_{i_b}(t)
                    :=\hat{N}^b_{i_b+A[\hat{\mathbf{N}}^a(t)]}(t)
                }{def:n_r}
            Trivially, there are no ask (bid) orders for $i_a\leq0$ ($i_b\geq0$). The sets for the number of ask and bid orders are written as
                \eqsp{
                    \hat{\mathbf{n}}^a_t:=\left\{\hat{n}^a_i(t)\right\}_{i\in \mathbb{Z}}\in\mathbb{N}^\infty,\quad
                    \hat{\mathbf{n}}^b_t:=\left\{\hat{n}^b_i(t)\right\}_{i\in \mathbb{Z}}\in\mathbb{N}^\infty.
                }{}    
            Furthermore, we define the best price level on the relative price coordinate system as 
                \eqsp{
                   a[\hat{\mbf{n}}^a_t]:=\min \{i\mid \hat{n}^a_i(t)\neq 0\},\quad
                   b[\hat{\mbf{n}}^a_t]:=\max \{i\mid \hat{n}^b_i(t)\neq 0\}.
                }{}
            Notably, several useful identities hold, such that 
                \eqsp{
                    a[\hat{\mbf{n}}^a_t]=A[\hat{\mathbf{N}}^a(t)]-B[\hat{\mathbf{N}}^b(t)],\quad
                    b[\hat{\mbf{n}}^b_t]=B[\hat{\mathbf{N}}^b(t)]-A[\hat{\mathbf{N}}^a(t)],\quad
                    a[\hat{\mbf{n}}^a_t]=-b[\hat{\mbf{n}}^b_t].
                }{eq:bestprice_f_and_r}
                
        \subsubsection{Preparation for kinetic theory: rewriting of the best prices based on Kronecker's delta}
            Let us rewrite the best prices by using Kronecker's delta for aiming at marginalisation (i.e., systematic integration of the PDFs), which will be applied for our kinetic formulation later. By using Eq.~\eqref{eq:best_price_f} in Appendix~\ref{subsec:I_gutairei}, the best prices on the fixed price coordinates system can be written by 
                \eqsp{
                A   [\hat{\mathbf{N}}^a(t)]
                    =
                    \sum_{i\in\mbb{Z}}i
                    (1-\delta_{\hat{N}^a_i(t),0})\prod^{i-1}_{j=-\infty}\delta_{\hat{N}^a_j(t),0}, \quad
                    B[\hat{\mathbf{N}}^b(t)]
                    =\sum_{i\in\mbb{Z}}i
                    (1-\delta_{\hat{N}^b_i(t),0})
                    \prod^{\infty}_{j=i+1}\delta_{\hat{N}^b_i(t),0},
                }{}
                where we use
                \eqsp{
                    \delta_{i,A[\hat{\mathbf{N}}^a(t)]}
                    =
                    (1-\delta_{\hat{N}^a_i(t),0})\prod^{i-1}_{j=-\infty}\delta_{\hat{N}^a_j(t),0},\quad
                    \delta_{i,B[\hat{\mathbf{N}}^a(t)]}
                    =
                    (1-\delta_{\hat{N}^b_i(t),0})
                    \prod^{\infty}_{j=i+1}\delta_{\hat{N}^b_i(t),0},
                }{}
                where we use the fact that there is no lower (upper) price than the best ask (bid) price. Likewise, by using Eq.~\eqref{eq:best_price_r} in Appendix~\ref{subsec:I_gutairei}, the best prices on the relative price coordinates system can be written as
                \eqsp{
                    a[\hat{\mathbf{n}}^a(t)] 
                    =
                    \sum_{i\in\mbb{Z}}i
                    (1-\delta_{\hat{n}^a_i(t),0})\prod^{i-1}_{j=-\infty}\delta_{\hat{n}^a_j(t),0}, \quad
                    b[\hat{\mathbf{n}}^b(t)]
                    =\sum_{i\in\mbb{Z}}i
                    (1-\delta_{\hat{n}^b_i(t),0})
                    \prod^{\infty}_{j=i+1}\delta_{\hat{n}^b_i(t),0},
                }{}
                where we use
                \eqsp{
                    \delta_{i,a[\hat{\mathbf{n}}^a(t)]}
                    =
                    (1-\delta_{\hat{n}^a_i(t),0})\prod^{i-1}_{j=-\infty}\delta_{\hat{n}^a_j(t),0},\quad
                    \delta_{i,b[\hat{\mathbf{n}}^b(t)]}
                    =
                    (1-\delta_{\hat{n}^b_i(t),0})
                    \prod^{\infty}_{j=i+1}\delta_{\hat{n}^b_i(t),0}.
                }{}

        \subsubsection{One-to-one correspondence between the fixed and relative price coordinate systems}
            In the fixed coordinate system, the state of the Santa Fe model is uniquely identified by $G_t:=(\hat{\mathbf{N}}^a(t),\hat{\mathbf{N}}^b(t))$. Correspondingly, in the relative coordinate system, the state of the Santa Fe model is uniquely identified by
            \begin{itembox}[l]{Definition: the state vector in the relative coordinate system}
            \vspace{-5mm}
            \eqsp{
                \Gamma_t := \left(\hat{m}_t,\hat{\mathbf{n}}^a_t,\hat{\mathbf{n}}^b_t\right) \in \mathcal{S}, \quad
                \mathcal{S}:= \mbb{Z}_{1/2}\times\mbb{N}^\infty\times\mbb{N}^\infty,
            }{}
            \end{itembox}
            where $\Gamma_t$ is the state vector in the state space $\mathcal{S}$. 

            Note that there is a one-to-one map between $G_t$ and $\Gamma_t$. Trivially, $\Gamma_t$ can be constructed by $G_t$. Inversely, $G_t$ can be constructed by $\Gamma_t$ as follows: Since we have the identities
            \eqsp{
                \hat{m}_t = \frac{A[\hat{\mathbf{N}}^a(t)]+B[\hat{\mathbf{N}}^b(t)]}{2}, \quad 
                a[\hat{\mbf{n}}^a_t] = A[\hat{\mathbf{N}}^a(t)]-B[\hat{\mathbf{N}}^b(t)],
            }{}
            we find the best price levels in the fixed coordinate systems, 
            \eqsp{
                A[\hat{\mathbf{N}}^a(t)]=\hat{m}_t+\frac{a[\hat{n}_t^a]}{2}, \quad 
                B[\hat{\mathbf{N}}^a(t)]=\hat{m}_t-\frac{a[\hat{n}_t^a]}{2}.
            }{}
            Finally, we can construct $\hat{\mathbf{n}}^a(t),\hat{\mathbf{n}}^b(t)$ by
            \eqsp{
                \hat{N}^a_{i}(t)=\hat{n}^a_{i-B[\hat{\mathbf{N}}^b(t)]}(t), \quad 
                \hat{N}^b_{i}(t)=\hat{n}^b_{i-A[\hat{\mathbf{N}}^a(t)]}(t).
            }{}

    \subsection{Stochastic dynamics}\label{subsec:dyn}
        Let us consider the time evolution of the state vector $\hat{\Gamma}_t:=(\hat{m}_t,\hat{\mathbf{n}}^a_t,\hat{\mathbf{n}}^b_t)$ during an infinitesimal time interval $[t,t+\dd t)$. We first assume that $L$ is finite and later takes the limit $L\to \infty$, recovering the original Santa Fe model without any cutoff. 

        \subsubsection{Total event intensity and its decomposition}
            Before formulating the stochastic time evolution of the state vector $\hat{\Gamma}_t$, we here evaluate the total intensity $\lambda_{\rm tot}$ of event occurrence: in other words, the probability $\mathcal{P}_{\rm all}= \lambda_{\rm tot}\dd t$ that some event occurs during $[t,t+\dd t)$ is given by the sum of the contribution of limit-order submission, cancellation, and market-order submission as 
            \eqsp{
                \mathcal{P}_{\rm tot} 
                =&\sum^L_{q=1}\Le(\lambda \Delta +v\hat{N}^a_{B[\hat{\mathbf{N}}^b(t)]+q}(t)\Ri)\mr{d}t
                +\sum^L_{q=1}\Le(\lambda \Delta+v\hat{N}^b_{A[\hat{\mathbf{N}}^a(t)]-q}(t)\Ri)\mr{d}t + 2\mu\dd t + o(\dd t)\\
                =&\left(\mathcal{P}_{\rm tot; S}^a + \mathcal{P}_{\rm tot; CM}^a\right)+ 
                  \left(\mathcal{P}_{\rm tot; S}^b + \mathcal{P}_{\rm tot; CM}^b\right) + o(\dd t),
            }{}
            where we consider the contribution up to the order $\dd t$. Here we define the total limit-order submission probability for the ask (bid) side $\mathcal{P}_{\rm tot; S}^a$ ($\mathcal{P}_{\rm tot; S}^b$) and the total cancellation and market-order probability for the ask (bid) side $\mathcal{P}_{\rm tot; CM}^a$ ($\mathcal{P}_{\rm tot; CM}^b$), which are defined by 
            \eqsp{
                &\mathcal{P}_{\rm tot; S}^a := \sum^L_{q=1} \lambda\Delta\dd t,\quad 
                \mathcal{P}_{\rm tot; CM}^a := \sum^L_{q=1} \big\{ v\hat{n}^a_{q} + \mu\delta_{q,a[\hat{\mathbf{n}}^a_t]}\big\}\dd t, \\
                &\mathcal{P}_{\rm tot; S}^b := \sum^L_{q=1} \lambda\Delta\dd t,\quad 
                \mathcal{P}_{\rm tot; CM}^b := \sum^L_{q=1} \{ v\hat{n}^b_{-q}(t) + \mu\delta_{-q,b[\hat{\mathbf{n}}^b_t]}\}\dd t.
            }{}
            In the following, we will further decompose these total probabilities into several contributions that are mutually exclusive by using the indicator function.

        \begin{figure}[t]
            \centering
            \includegraphics[width=0.9\columnwidth]{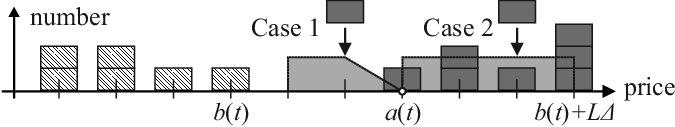}
            \caption{
                Case study of a new ask limit-order submission. 
                Case 1: the price of the new order is within $[1,a[\hat{\bm{n}}^a(t)])$ and the best ask price is changed. 
                Case 2: the price of the new order is within $[a[\hat{\bm{n}}^a(t)],L)$ and the best ask price is unchanged.
            }
            \label{fig:casestudy-askLOsubmission}
        \end{figure}  
        \subsubsection{Contribution of limit-order submission}
            Let us decompose the total limit-order submission intensity into four contributions that are mutually exclusive and collectively exhaustive, based on whether the submission price level is below (above) the best ask price level~$a[\hat{\mathbf{n}}^a(t)]$ (the best bid price level~$b[\hat{\mathbf{n}}^b(t)]$). 
                \begin{itemize}
                    \item Case 1: An ask limit order is submitted within the range $[1,a[\hat{\mathbf{n}}^a(t))$.
                    \item Case 2: An ask limit order is submitted within the range $[a[\hat{\mathbf{n}}^a(t)],L]$.
                    \item Case 3: A bid limit order is submitted within the range $(b[\hat{\mathbf{n}}^b(t)],1]$.
                    \item Case 4: A bid limit order is submitted within the range $[-L,b[\hat{\mathbf{n}}^b(t)]]$.
                \end{itemize}
            See Fig.~\ref{fig:casestudy-askLOsubmission} for a schematic. Each contribution can be organised by decomposing $\mathcal{P}_{\rm tot; S}^a$ and $\mathcal{P}_{\rm tot; S}^b$ as 
            \eqsp{
                \mathcal{P}_{\rm tot; S}^a &:= \sum^L_{q=1} \bigl\{
                    \underset{\text{Case 1}}{\underline{\lambda \Delta\dd t \mathbb{I}_{(0,a[\hat{\mathbf{n}}^a_t])}(q)}} + 
                    \underset{\text{Case 2}}{\underline{\lambda \Delta\dd t \mathbb{I}_{[a[\hat{\mathbf{n}}^a_t],L]}(q)}}
                \bigr\}, \\
                \mathcal{P}_{\rm tot; S}^b &:= \sum^L_{q=1} \bigl\{
                    \underset{\text{Case 3}}{\underline{\lambda \Delta\dd t \mathbb{I}_{(b[\hat{\mathbf{n}}^b_t],0)}(-q)}} + 
                    \underset{\text{Case 4}}{\underline{\lambda \Delta\dd t \mathbb{I}_{[-L,b[\hat{\mathbf{n}}^b_t]]}(-q)}}
                \bigr\}.
            }{eq:decompose:P_submission}
            For Case~1 and Case~3, there are changes in both best price and midprice. For Case~2 and Case~4, there is a change in neither the best price nor the midprice. In the following case-by-case analysis, we consider not only the path-level dynamics of $\hat{\Gamma}_t$ but also the corresponding probabilistic weight (PW) in the sum in Eq.~\eqref{eq:decompose:P_submission}.

            \begin{figure}[t]
                \centering
                \includegraphics[width=0.9\columnwidth]{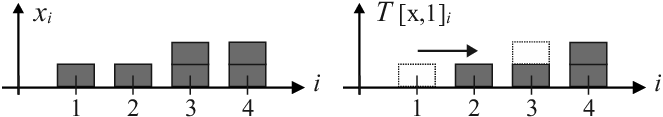}
                \caption{
                    Schematic of the shift operator $T[\mbf{x},a]$ when $a$ is $1$.
                }
                \label{fig:shift}
            \end{figure}

            Due to limit-order submissions, the path-level dynamics of the state vector $\hat{\Gamma}_t$ are summarised as follows:
            \begin{itembox}[l]{Dynamics of $\hat{\Gamma}_t$ due to limit-order submissions}
                \vspace{-5mm}
                \eqsp{
                    \hat{\Gamma}_{t+\mr{d}t}
                    =
                    \begin{cases}
                         (\hat{m}_t+\frac{b[\hat{\mathbf{n}}^b_t]+q}{2},\hat{\mathbf{n}}^a_t+\mathbf{e}^q,T[\hat{\mathbf{n}}^b_t,-b[\hat{\mathbf{n}}^b_t]-q])
                         & (\text{Case 1; }\text{PW}=\lambda\Delta\mr{d}t\mathbb{I}_{[1,a[\hat{\mathbf{n}}^a_t])}(q)) \\
                         (\hat{m}_{t},\hat{\mathbf{n}}^a_t+\mathbf{e}^q,\hat{\mathbf{n}}^b_t)
                         & (\text{Case 2; }\text{PW}=\lambda\Delta\mr{d}t\mathbb{I}_{[a[\hat{\mathbf{n}}^a_t],L]}(q))\\
                         (\hat{m}_{t}+\frac{a[\hat{\mathbf{n}}^a_t]-q}{2},T[\hat{\mathbf{n}}^a_t,-a[\hat{\mathbf{n}}^a_t]+q],\hat{\mathbf{n}}^b_t+\mathbf{e}^{-q})
                         & (\text{Case 3; }\text{PW}=\lambda\Delta\mr{d}t\mathbb{I}_{(b[\hat{\mathbf{n}}^b_t],-1]}(-q))\\
                         (\hat{m}_{t},\hat{\mathbf{n}}^a_t,\hat{\mathbf{n}}^b_t+\mathbf{e}^{-q})
                         & (\text{Case 4; }\text{PW}=\lambda\Delta\mr{d}t\mathbb{I}_{[-L,b[\hat{\mathbf{n}}^b_t]]}(-q))
                    \end{cases}.
                }{eq:dyn_sub}
            \end{itembox}
            The parameter $q$ represents the relative price level to the opposite best price level just before the order submission. This formula will be used to exactly evaluate the limit-order contributions in the master equation as shown in Sec.~\ref{subsec:master}. Here, $T[\mathbf{x},a]\in \mathbb{N}^{\infty}$ is the shift operator acting on $\mathbf{x}$ by $a\in\mathbb{Z}$ and $\mathbf{e}^q$ is the unit vector (see Fig.\ref{fig:shift} for a schematic), such that 
                \eq{
                    T[\mathbf{x},a]_i:= x_{i-a}, \quad 
                    \bm{e}^q:=(e^{q}_1,e^{q}_2,\dots) = (0,\dots,0,\underset{q\text{-th}}{\underline{1}},0,\dots), \quad
                    e^{q}_i=\delta_{i,q}\quad (i\in\mathbb{Z}).
                }{}
                \begin{proof}
                    Let us derive the result~\eqref{eq:dyn_sub} by a case-by-case analysis. For Case~1, the time evolutions of the LOB on the fixed coordinate systems $\hat{\mathbf{N}}^a(t),\hat{\mathbf{N}}^b(t)$ are given by 
                        \eqsp{
                            \hat{N}^a_i(t+\mr{d}t)
                            =\hat{N}^a_i(t)+\delta_{i,q+B[\hat{\mathbf{N}}^b(t)]},\quad
                            \hat{N}^b_i(t+\mr{d}t)
                            =\hat{N}^b_i(t) \quad \mbox{for all }i \in \mathbb{Z}.
                        }{eq:dyn_sub_case1_N}
                    The time evolution of the best price on the fixed coordinate system is given by 
                        \eqsp{
                            A[\hat{\mathbf{N}}^a(t+\mr{d}t)]=B[\hat{\mathbf{N}}^b(t)]+q=A[\hat{\mathbf{N}}^a(t)]+b[\hat{\mathbf{n}}^b_t]+q,\quad
                            B[\hat{\mathbf{N}}^b(t+\mr{d}t)]=B[\hat{\mathbf{N}}^b(t)].
                        }{eq:dyn_sub_case1_AB}
                    Correspondingly, the midprice after transaction $\hat{m}_{t\dd t}$ is given by 
                        \eq{
                           \hat{m}_{t+\mr{d}t}=\frac{A[\hat{\mathbf{N}}^a(t+\mr{d}t)]+B[\hat{\mathbf{N}}^b(t+\mr{d}t)]}{2}=\hat{m}_t+\frac{b[\hat{\mathbf{n}}^b_t]+q}{2}.
                        }{}
                    Then, the time evolutions of the LOBs on the relative coordinate system $\hat{\mathbf{n}}^a_t,\hat{\mathbf{n}}^b_t$ are given by
                        \eqsp{
                            \hat{n}^a_i(t+\mr{d}t)
                            &=\hat{N}^a_{i+B[\hat{\mathbf{N}}^b(t+\mr{d}t)]}(t+\mr{d}t)
                            =\hat{N}^a_{i+B[\hat{\mathbf{N}}^b(t)]}(t+\mr{d}t)
                            =\hat{N}^a_{i+B[\hat{\mathbf{N}}^b(t)]}(t)+\delta_{i,q}
                            =\hat{n}^a_i(t)+\delta_{i,q},\\
                            \hat{n}^b_i(t+\mr{d}t)
                            &=\hat{N}^b_{i+A[\hat{\mathbf{N}}^b(t+\mr{d}t)]}(t+\mr{d}t)
                            =\hat{N}^b_{i+A[\hat{\mathbf{N}}^b(t)]+b[\hat{\mathbf{n}}^b_t]+q}(t+\mr{d}t)
                            =\hat{N}^b_{i+A[\hat{\mathbf{N}}^b(t)]+b[\hat{\mathbf{n}}^b_t]+q}(t)
                            =\hat{n}^b_{i+b[\hat{\mathbf{n}}^b_t]+q}(t).
                        }{}
                    
                    For Case~2, the time evolutions of the LOB on the fixed coordinate systems $\hat{\mathbf{N}}^a(t),\hat{\mathbf{N}}^b(t)$ are given by 
                        \eqsp{
                            \hat{N}^a_i(t+\mr{d}t)
                            =\hat{N}^a_i(t)+\delta_{i,q+B[\hat{\mathbf{N}}^b(t)]},\quad
                            \hat{N}^b_i(t+\mr{d}t)
                            =\hat{N}^b_i(t).
                        }{eq:dyn_sub_case2_N}
                    There are no changes in the best price and the midprice:
                        \eqsp{
                            A[\hat{\mathbf{N}}^a(t+\mr{d}t)]=A[\hat{\mathbf{N}}^a(t)],\quad B[\hat{\mathbf{N}}^b(t+\mr{d}t)]=B[\hat{\mathbf{N}}^b(t)],\quad
                            \hat{m}_{t+\mr{d}t}=\hat{m}_t.
                        }{eq:dyn_sub_case2_AB}
                    Therefore, the time evolutions of the LOBs on the relative coordinate system $\hat{\mathbf{n}}^a_t,\hat{\mathbf{n}}^b_t$ are given by
                        \eqsp{
                            \hat{n}^a_i(t+\mr{d}t)
                            =&\hat{N}^a_{i+B[\hat{\mathbf{N}}^b(t+\mr{d}t)]}(t+\mr{d}t)
                            =\hat{N}^a_{i+B[\hat{\mathbf{N}}^b(t)]}(t+\mr{d}t)
                            =\hat{N}^a_{i+B[\hat{\mathbf{N}}^b(t)]}(t)+\delta_{i,q}
                            =\hat{n}^a_i(t)+\delta_{i,q}, \\
                            \hat{n}^b_i(t+\mr{d}t)
                            =&\hat{N}^b_{i+A[\hat{\mathbf{N}}^b(t+\mr{d}t)]}(t+\mr{d}t)
                            =\hat{N}^b_{i+A[\hat{\mathbf{N}}^b(t)]}(t+\mr{d}t)
                            =\hat{N}^b_{i+A[\hat{\mathbf{N}}^b(t)]}(t)
                            =\hat{n}^b_i(t).
                        }{}
                    Case~3 and Case~4 can be addressed likewise. Finally, we obtain Eq.~\eqref{eq:dyn_sub}. 
                \end{proof}
            
        \subsubsection{Contribution of cancellation and market order}
            \begin{figure}[t]
                \centering
                \includegraphics[width=0.9\columnwidth]{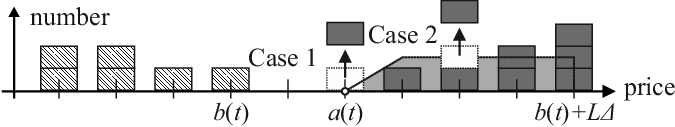}
                \caption{
                    Case study of an ask limit-order cancellation and a buy market order. 
                    Case 1: the best ask limit order is cancelled or a buy market order is issued. The best ask price is changed. 
                    Case 2: an ask limit order is cancelled beyond the best ask price. The best ask price is unchanged. 
                }
                \label{fig:casestudy-askCancel-MO}
            \end{figure}
            We next decompose the total cancellation intensity and the market order intensity.
            \eqsp{
                &\mathcal{P}_{\rm tot; CM}^a 
                = \sum^L_{q=1} \Big[
                    \underset{\text{Case 1}}{   \underline{  \big\{ \mu+v\hat{n}^a_{q} \big\}\delta_{q,a[\hat{\mathbf{n}}^a_t]}}  } +  \underset{\text{Case 2}}{  \underline{v\hat{n}^a_{q}\mathbb{I}_{(a[\hat{\mathbf{n}}^a_t],L]}(q)} } 
                \Big]\dd t, \\
                &\mathcal{P}_{\rm tot; CM}^b 
                = \sum^L_{q=1} \Big[
                    \underset{\text{Case 3}}{   \underline{  \big\{ \mu+v\hat{n}^b_{-q} \big\}\delta_{-q,b[\hat{\mathbf{n}}^b_t]}}  } +  
                    \underset{\text{Case 4}}{   \underline{v\hat{n}^b_{-q}\mathbb{I}_{[-L,b[\hat{\mathbf{n}}^b_t])}(-q) \Big]\dd t}  },
            }{}
            where each component can be interpreted as follows: 
                \begin{itemize}
                    \item Case 1: A best ask order is cancelled or executed at $q = a[\hat{\mathbf{n}}^a(t)]$.
                    \item Case 2: An ask order over the best ask price is cancelled within $(a[\hat{\mathbf{n}}^a(t)],L]$.
                    \item Case 3: A best bid order is cancelled or executed at $q = b[\hat{\mathbf{n}}^b(t)]$.
                    \item Case 4: A bid order under the best bid price is cancelled within $[-L,b[\hat{\mathbf{n}}^b(t)])$.
                \end{itemize}
            Note that there are changes in both best price and midprice for Case~1 and Case~3, whereas there is no change for Case~2 and Case~4. See Fig.~\ref{fig:casestudy-askCancel-MO} as a schematic. Finally, we obtain the path-level stochastic dynamics of $\hat{\Gamma}_t$ due to cancellation or market order submission, such that 
            \begin{itembox}[l]{Dynamics of $\hat{\Gamma}_t$ due to cancellation or market order}
                \vspace{-5mm}
                \eqsp{
                    \small
                    \hat{\Gamma}_{t+\mr{d}t}
                    =
                    \begin{cases}
                         (\hat{m}_t+\fr{a[\hat{\mbf{n}}^a_t-\mbf{e}^{q}]-q}{2},\hat{\mbf{n}}^a_t-\mbf{e}^q,T[\hat{\mbf{n}}^b_t,-a[\hat{\mbf{n}}^a_t-\mbf{e}^{q}]+q])
                         & (\text{Case 1; PW}=\Le\{\mu+v\hat{n}^a_{q}(t)\Ri\}\mr{d}t\delta_{q,a[\hat{\mathbf{n}}^a_t]})\\
                         (\hat{m}_t,\hat{\mbf{n}}^a_t-\mbf{e}^q,\hat{\mbf{n}}^b_t)
                         & (\text{Case 2; PW}=v\hat{n}^a_{q}(t)\mr{d}t\mathbb{I}_{(a[\hat{\mathbf{n}}^a_t],L]}(q)) \\
                         (\hat{m}_t+\fr{b[\hat{\mbf{n}}^b_t-\mbf{e}^{-q}]+q}{2},T[\hat{\mbf{n}}^a_t,-b[\hat{\mbf{n}}^b_t-\mbf{e}^{-q}]-q],\hat{\mbf{n}}^b_t-\mbf{e}^{-q})
                         & (\text{Case 3; PW}=\Le\{\mu+v\hat{n}^b_{-q}(t)\Ri\}\mr{d}t\delta_{-q,b[\hat{\mathbf{n}}^b_t]})\\
                         (\hat{m}_t,\hat{\mbf{n}}^a_t,\hat{\mbf{n}}^b_t-\mbf{e}^{-q})
                         & (\text{Case 4; PW}=v\hat{n}^b_{-q}(t)\mr{d}t\mathbb{I}_{[-L,b[\hat{\mathbf{n}}^b_t])}(-q))\\
                    \end{cases}.
                }{eq:dyn_can}
                \vspace{-5mm}
            \end{itembox}

            \begin{proof}
                We show Eq.~\eqref{eq:dyn_can} by considering a case-by-case analysis. 
                For Case~1, the time evolutions of $\hat{\mathbf{N}}^a(t),\hat{\mathbf{N}}^b(t)$ are given by 
                    \eqsp{
                        \hat{N}^a_i(t+\mr{d}t)
                        =\hat{N}^a_i(t)-\delta_{i,q+B[\hat{\mathbf{N}}^b(t)]},\quad
                        \hat{N}^b_i(t+\mr{d}t)
                        =\hat{N}^b_i(t).
                    }{}
                The time evolution of the best price is given by
                    \eqsp{
                        A[\hat{\mathbf{N}}^a(t+\mr{d}t)]
                        =A[\hat{\mathbf{N}}^a(t)-\mathbf{e}^{q+B[\hat{\mathbf{N}}^b(t)]}]
                        =B[\hat{\mathbf{N}}^b(t)]+a[\hat{\mathbf{n}}^a_t-\mathbf{e}^q],\quad
                        B[\hat{\mathbf{N}}^b(t+\mr{d}t)]=B[\hat{\mathbf{N}}^b(t)],
                    }{}
                which leads to the dynamics of the midprice $\hat{m}_t$ as 
                    \eq{
                        \hat{m}_{t+\mr{d}t}
                        =\frac{A[\hat{\mathbf{N}}^a(t+\mr{d}t)]+B[\hat{\mathbf{N}}^b(t+\mr{d}t)]}{2}
                        =\hat{m}_t+\frac{a[\hat{\mathbf{n}}^a_t-\mathbf{e}^q]-q}{2}
                    }{}
                Therefore, the time evolution of $\hat{\mathbf{n}}^a_t,\hat{\mathbf{n}}^b_t$ is
                    \eqsp{
                        \hat{n}^a_i(t+\mr{d}t)
                        =&\hat{N}^a_{i+B[\hat{\mathbf{N}}^b(t+\mr{d}t)]}(t+\mr{d}t)
                        =\hat{N}^a_{i+B[\hat{\mathbf{N}}^b(t)]}(t+\mr{d}t)\\
                        =&\hat{N}^a_{i+B[\hat{\mathbf{N}}^b(t)]}(t)-\delta_{i,q}
                        =\hat{n}^a_i(t)-\delta_{i,q},
                    }{}
                    \eqsp{
                        \hat{n}^b_i(t+\mr{d}t)
                        =&\hat{N}^b_{i+A[\hat{\mathbf{N}}^a(t+\mr{d}t)]}(t+\mr{d}t)
                        =\hat{N}^b_{i+A[\hat{\mathbf{N}}^a(t)]+a[\hat{\mathbf{n}}^a_t-\mathbf{e}^q]-q}(t+\mr{d}t)\\
                        =&\hat{N}^b_{i+A[\hat{\mathbf{N}}^a(t)]+a[\hat{\mathbf{n}}^a_t-\mathbf{e}^q]-q}(t)
                        =\hat{n}^b_{i+a[\hat{\mathbf{n}}^a_t-\mathbf{e}^q]-q}(t).
                    }{}

                For Case 2, The time evolutions of $\hat{\mathbf{N}}^a(t),\hat{\mathbf{N}}^b(t)$ are given by 
                    \eqsp{
                        \hat{N}^a_i(t+\mr{d}t)
                        =\hat{N}^a_i(t)-\delta_{i,q+B[\hat{\mathbf{N}}^b(t)]},\quad
                        \hat{N}^b_i(t+\mr{d}t)
                        =\hat{N}^b_i(t).
                    }{}
                Since there is no change in the best price and midprice, we have 
                    \eqsp{
                        A[\hat{\mathbf{N}}^a(t+\mr{d}t)]=A[\hat{\mathbf{N}}^a(t)],\quad
                        B[\hat{\mathbf{N}}^b(t+\mr{d}t)]=B[\hat{\mathbf{N}}^b(t)],\quad
                        \hat{m}_{t+\mr{d}t}=\hat{m}_t. 
                    }{}
                 The corresponding time evolutions of $\hat{\mathbf{n}}^a_t,\hat{\mathbf{n}}^b_t$ are given by 
                    \eqsp{
                        \hat{n}^a_i(t+\mr{d}t)
                        =&\hat{N}^a_{i+B[\hat{\mathbf{N}}^b(t+\mr{d}t)]}(t+\mr{d}t)
                        =\hat{N}^a_{i+B[\hat{\mathbf{N}}^b(t)]}(t+\mr{d}t)\\
                        =&\hat{N}^a_{i+B[\hat{\mathbf{N}}^b(t)]}(t)-\delta_{i,q}
                        =\hat{n}^a_i(t)-\delta_{i,q},
                    }{}
                    \eqsp{
                        \hat{n}^b_i(t+\mr{d}t)
                        =&\hat{N}^b_{i+A[\hat{\mathbf{N}}^b(t+\mr{d}t)]}(t+\mr{d}t)
                        =\hat{N}^b_{i+A[\hat{\mathbf{N}}^b(t)]}(t+\mr{d}t)\\
                        =&\hat{N}^b_{i+A[\hat{\mathbf{N}}^b(t)]}(t)
                        =\hat{n}^b_i(t).
                    }{}
                By parallel calculations, we can straightforwardly address Case~3 and Case~4. We thus obtain Eq.~\eqref{eq:dyn_can}.
            \end{proof}
    
        \subsubsection{Without events during $[t,t+\dd t)$}
            The last case is that there is no events---neither limit-order submissions, cancellations, nor market-order submission---during $[t,t+\dd t)$. For this case, the path-level stochastic dynamics of $\hat{\Gamma_t}$ is trivially given by $\hat{\Gamma}_{t+\mr{d}t}=\hat{\Gamma_t}$ with the total probability $1-\left\{2\lambda \Delta L -v\Le(\sum^{L}_{q=1}\hat{n}^a_q(t)+\sum^{-1}_{q=-L}\hat{n}^b_q(t)\Ri) - 2\mu\right\}\mr{d}t-o(\mr{d}t).$ 

        \subsection{Remark on the cutoff length $L$}
            The cutoff length $L$ is introduced to avoid the divergence of the total event intensity, such that $\lambda_{\rm tot}<\infty$ during the derivation of the master equation~\eqref{eq;master_eq}. After deriving the master equation~\eqref{eq;master_eq} and the BBGKY hierarchical equation~\eqref{eq:BBGKY} in the next section, we finally take the limit $L\to \infty$ because the precise value of $L$ does not impact the final results for a sufficiently large $L$. More technically, we take $L\to 0$ after taking $\dd t\to 0$, to keep the relation $L\dd t\to 0$.

\section{Exact kinetic theory: master equation and BBGKY hierarchical equation}\label{sec:kinetictheory-ME+BBGKY}
    \subsection{Exact master equation}\label{subsec:master}
        Let us formulate the mean-field theory of the Santa Fe LOB model within the framework of kinetic theory. The starting point of the kinetic theory is the exact master equation for the state vector---called the pseudo-Liouville equation in the conventional kinetic framework. By using the results in Sec.~\ref{subsec:dyn} and by taking the $L\to \infty$ limit, the exact master equation for the Santa Fe model is given by the following linear integro-differential equation:
            \begin{itembox}[l]{The exact master equation of the Santa Fe model}
                \vspace{-5mm}
                \eqsp{
                    \pdiff{P_t(\Gamma)}{t}
                    =&
                    \sum_{\hat{\Gamma}}\sum_{g, q,\dd m}
                    P_t(\hat{\Gamma})
                    \Le[w^{{\rm S};g}_{q,\dd m}[\hat{\mbf{n}}]
                    \left\{\delta_{\hat{\Gamma}^{{\rm S};g}_{q,\dd m},\Gamma}-\delta_{\hat{\Gamma},\Gamma}\right\} 
                    +
                    w^{{\rm CM};g}_{q,\dd m}[\hat{\mbf{n}}]
                    \left\{\delta_{\hat{\Gamma}^{{\rm CM};g}_{q,\dd m},\Gamma}-\delta_{\hat{\Gamma},\Gamma}\right\} \Ri]
                }{eq;master_eq}
                \vspace{-5mm}
            \end{itembox}
            where we define the order type $g\in \{a,b\}$ and the pre-jump state vectors
            \eqsp{
                \hat{\Gamma}^{{\rm S};a}_{q,\dd m}
                :=&(\hat{m}+\dd m,\hat{\mbf{n}}^a+\mbf{e}^q,T[\hat{\mbf{n}}^b,-2\dd m]),\quad
                \hat{\Gamma}^{{\rm S};b}_{q,\dd m}
                :=(\hat{m}+\dd m,T[\hat{\mbf{n}}^a,-2dm],\hat{\mbf{n}}^b+\mbf{e}^q),\\
                \hat{\Gamma}^{{\rm CM};a}_{q,\dd m}
                :=&(\hat{m}+\dd m,\hat{\mbf{n}}^a-\mbf{e}^q,T[\hat{\mbf{n}}^b,-2\dd m]),\quad
                \hat{\Gamma}^{{\rm CM};b}_{q,\dd m}
                :=(\hat{m}+\dd m,T[\hat{\mbf{n}}^a,-2\dd m],\hat{\mbf{n}}^b-\mbf{e}^q).
            }{}
            Also, the corresponding intensity weight functions are given by 
            \eqsp{
                w^{{\rm S};a}_{q,\dd m}[\hat{\mbf{n}}]
                := &
                \lambda\Delta\mbb{I}_{[1,\infty)}(q)\delta_{\dd m,-F(-\fr{b[\hat{\mbf{n}}^b]+q}{2})},\\
                w^{S;b}_{q,\dd m}[\hat{\mbf{n}}]
                := &
                \lambda\Delta\mbb{I}_{(-\infty,-1]}(q)\delta_{\dd m,F(\fr{-b[\hat{\mbf{n}}^b]+q}{2})}
                =
                \lambda\Delta\mbb{I}_{(-\infty,-1]}(q)\delta_{\dd m,F(\fr{a[\hat{\mbf{n}}^a]+q}{2})},\\
                w^{{\rm CM};a}_{q,\dd m}[\hat{\mbf{n}}]
                := &
                (\mu\delta_{q,a[\hat{\mbf{n}}^a]}+v\hat{n}^a_q)\mbb{I}_{[1,\infty)}(q)\delta_{\dd m,F(\fr{a[\hat{\mbf{n}}^a-\mbf{e}^q]-q}{2})},\\
                w^{{\rm CM};b}_{q,\dd m}[\hat{\mbf{n}}]
                := &
                (\mu\delta_{q,b[\hat{\mbf{n}}^b]}+v\hat{n}^b_q)\mbb{I}_{(-\infty,-1]}(q)\delta_{\dd m,-F(-\fr{b[\hat{\mbf{n}}^b-\mbf{e}^q]-q}{2})}
            }{}
            with the filter function $F(x):= x\mbb{I}_{(0,\infty)}(x)$. This is the first main result of this work.
        \begin{proof}
            Let us consider the time evolution $\dd f(\hat{\Gamma}_t) := f(\hat{\Gamma}_{t+\mr{d}t})-f(\hat{\Gamma}_t)$ for an arbitrary function $f(\hat{\Gamma}_t)$ during an infinitesimal interval $[t,t+\dd t)$. On average, the time evolution of $\dd f(\hat{\Gamma}_t)$ is given by 
                \eqsp{
                    \Le\langle \mr{d}f(\hat{\Gamma}_t)\Ri\rangle
                    =&
                    \lambda\Delta\mr{d}t\sum_{\hat{\Gamma}}\sum_{q,\dd m}
                    P_t(\hat{\Gamma})
                    \mbb{I}_{[1,L]}(q)\delta_{\dd m,-F(-\fr{b[\hat{\mbf{n}}^b]+q}{2})}
                    \left\{f(\hat{\Gamma}^{{\rm S};a}_{q,\dd m})-f(\hat{\Gamma})\right\} \\
                    &+
                    \lambda\Delta\mr{d}t\sum_{\hat{\Gamma}}\sum_{q,\dd m}
                    P_t(\hat{\Gamma})
                    \mbb{I}_{[-L,-1]}(q)\delta_{\dd m,F(\fr{a[\hat{\mbf{n}}^a]+q}{2})}
                    \left\{f(\hat{\Gamma}^{{\rm S};b}_{q,\dd m})-f(\hat{\Gamma})\right\}\\
                    &+
                    \mr{d}t\sum_{\hat{\Gamma}}\sum_{q,\dd m}
                    P_t(\hat{\Gamma})
                    (\mu\delta_{q,a[\hat{\mbf{n}}^a]}+v\hat{n}^a_q)\mbb{I}_{[1,L]}(q)\delta_{\dd m,F(\fr{a[\hat{\mbf{n}}^a-\mbf{e}^q]-q}{2})}
                    \left\{f(\hat{\Gamma}^{{\rm CM};a}_{q,\dd m})-f(\hat{\Gamma})\right\} \\
                    &+
                    \mr{d}t\sum_{\hat{\Gamma}}\sum_{q,\dd m}
                    P_t(\hat{\Gamma})
                    (\mu\delta_{q,b[\hat{\mbf{n}}^b]}+v\hat{n}^b_q)\mbb{I}_{[-L,-1]}(q)\delta_{\dd m,-F(-\fr{b[\hat{\mbf{n}}^b-\mbf{e}^q]-q}{2})}
                    \left\{f(\hat{\Gamma}^{{\rm CM};b}_{q,\dd m})-f(\hat{\Gamma})\right\}
                }{eq:f_time_ev}
            up to $o(\dd t)$. See Appendix~\ref{App:master_cal} for the detailed derivation.
            
            We next substitute $\delta_{\hat{\Gamma},\Gamma}:=\delta_{\hat{m},m}\prod_{i\in\mbb{Z}}\delta_{\hat{n}^a_{i},n^a_{i}}\prod_{j\in\mbb{Z}}\delta_{\hat{n}^b_{j},n^b_{j}}$ for $f(\hat{\Gamma})$ to obtain the master equation. The left-hand side (LHS) can be rewritten by 
                \eqsp{
                    \left\langle \mr{d}f(\hat{\Gamma}_t)\right\rangle
                    =
                    \left\langle f(\hat{\Gamma}_{t+\mr{d}t})-f(\hat{\Gamma}_t)\right\rangle
                    =
                    \sum_{\hat{\Gamma}_{t+\mr{d}t}}P_{t+\mr{d}t}(\hat{\Gamma}_{t+\mr{d}t})f(\hat{\Gamma}_{t+\mr{d}t})
                    -
                    \sum_{\hat{\Gamma}_{t}}P_{t}(\hat{\Gamma}_{t})f(\hat{\Gamma}_{t})
                    =
                    P_{t+\mr{d}t}(\Gamma)-P_{t}(\Gamma).
                }{}
            Therefore, we obtain the exact master equation Eq.~\eqref{eq;master_eq} by taking the limit $\dd t\to 0$ and $L \to \infty$. Note that the $L\to \infty$ limit is taken after the $\dd t\to 0$ limit, to keep the relation $L\dd t \to 0$. 
        \end{proof}

    \subsection{Exact BBGKY  hierarchical equation}
        The master equation~\eqref{eq;master_eq} is exact but cannot be solved explictly. In the standard statistical-physics program of kinetic theory, we study the BBGKY hierarchy---a series of exact equations between multi-body PDFs---and then proceed with the Boltzmann/Vlasov equations as a mean-field theory. In this subsection, we derive the first-order BBGKY hierarchical equation as an exact identity. 

        Let us consider the one-body distribution function $P_t^a(n,k)$ for the ask side as $\text{Prob}[\hat{n}^a_k(t)=n]$. In other words, we are interested in the following quantity:         
            \eqsp{
                P^a_t(n,k):=\text{Prob}[\hat{n}^a_k(t)=n]=\sum_{m\in\mbb{Z}_{1/2}}\sum_{\mathbf{n}\in\mbb{N}^\infty}\delta_{n^a_k,n}P_t(\Gamma).
            }{}
        for any $k\in[1,\infty)$ and $n\in\mbb{N}$. Note that $P^a_t(n,k)$ trivially satisfies the normalisation condition $\sum_{n}P^a_t(n,k)=1$. 
        By exactly integrating both sides of the master equation~\eqref{eq;master_eq}, we obtain the first-order BBGKY hierarchical equation as an identity:
        \begin{itembox}[l]{The exact first-order BBGKY hierarchical equation}
            \vspace{-5mm}
            \eqsp{
                &\pdiff{P^a_t(n,k)}{t}
                =
                \lambda\Delta 
                \Le\{P^a_t(n-1,k)-P^a_t(n,k)\Ri\}
                +
                v\Le\{(n+1)P^a_t(n+1,k)-nP^a_t(n,k)\Ri\}\\
                &+
                \mu\sum_{\hat{\Gamma}}P_t(\hat{\Gamma})
                \delta_{k,a[\hat{\mbf{n}}^a]}\Le(\delta_{\hat{n}^a_k-1,n}-\delta_{\hat{n}_k^a,n}\Ri)            
                +\sum_{\hat{\Gamma}}\sum_{q,\dd m}P_t(\hat{\Gamma})
                    \Le(w^{{\rm S};b}_{q,\dd m}[\hat{\mbf{n}}]+w^{{\rm CM};b}_{q,\dd m}[\hat{\mbf{n}}]\Ri)\Le(\delta_{\hat{n}^a_{k+2\dd m},n}-\delta_{\hat{n}^a_k,n}\Ri)
            }{eq:BBGKY}
            \vspace{-5mm}
        \end{itembox}
        This equation is the second main result of this work. On the RHS, the first and second terms represent the limit-order submission and cancellation at the ask $k$-th relative price level and the third term represents the buy market-order submission. The fourth term represents all bid-side events causing changes in the best bid price. 
        
        \begin{proof}
            By integrating the LHS of the master equation~\eqref{eq;master_eq}, we obtain 
                \eqsp{
                    \sum_{m}\sum_{\mathbf{n}}\delta_{n^a_k,n}\pdiff{P_t(\Gamma)}{t}
                    =
                    \pdiff{P^a_t(n,k)}{t}.
                }{}
            Here, let us consider the following identities related to the Kronecker delta:
                \eqsp{
                    \sum_{m}\sum_{\mathbf{n}}\delta_{n^a_k,n}\delta_{\hat{\Gamma}^{{\rm S};a}_{q,\dd m},\Gamma}
                    =&
                    \delta_{\hat{n}^a_k+\delta_{q,k},n},\quad
                    \sum_{m}\sum_{\mathbf{n}}\delta_{n^a_k,n}\delta_{\hat{\Gamma}^{{\rm S};b}_{q,\dd m},\Gamma}
                    =
                    \delta_{\hat{n}^a_{k+2\dd m},n},\\
                    \sum_{m}\sum_{\mathbf{n}}\delta_{n^a_k,n}\delta_{\hat{\Gamma}^{{\rm CM};a}_{q,\dd m},\Gamma}
                    =&
                    \delta_{\hat{n}^a_k-\delta_{q,k},n},\quad
                    \sum_{m}\sum_{\mathbf{n}}\delta_{n^a_k,n}\delta_{\hat{\Gamma}^{{\rm CM};b}_{q,\dd m},\Gamma}
                    =
                    \delta_{\hat{n}^a_{k+2\dd m},n},\quad 
                    \sum_{m}\sum_{\mathbf{n}}\delta_{n_k^a,n}\delta_{\hat{\Gamma},\Gamma}
                    =
                    \delta_{\hat{n}_k^a,n}.
                }{}
            By integrating both sides of the master equation~\eqref{eq;master_eq}, we obtain the time-evolution equation for the one-body distribution $P_t(n,k)$ as 
                \eqsp{
                    \pdiff{P^a_t(n,k)}{t}
                    =&
                    \sum_{\hat{\Gamma}}\sum_{q,\dd m}P_t(\hat{\Gamma})
                    \Le\{w^{{\rm S};a}_{q,\dd m}[\hat{\mbf{n}}]\Le(\delta_{\hat{n}^a_k+\delta_{q,k},n}-\delta_{\hat{n}^a_k,n}\Ri)
                    +w^{{\rm S};b}_{q,\dd m}[\hat{\mbf{n}}]\Le(\delta_{\hat{n}^a_{k+2\dd m},n}-\delta_{\hat{n}^a_k,n}\Ri)\Ri\}\\
                    &+
                    \sum_{\hat{\Gamma}}\sum_{q,\dd m}P_t(\hat{\Gamma})
                    \Le\{w^{{\rm CM};a}_{q,\dd m}[\hat{\mbf{n}}]\Le(\delta_{\hat{n}^a_k-\delta_{q,k},n}-\delta_{\hat{n}^a_k,n}\Ri)
                    +w^{{\rm CM};b}_{q,\dd m}[\hat{\mbf{n}}]\Le(\delta_{\hat{n}^a_{k+2\dd m},n}-\delta_{\hat{n}^a_k,n}\Ri)\Ri\}\\
                    =&
                    \sum_{\hat{\Gamma}}\sum_{\dd m}P_t(\hat{\Gamma})
                    w^{{\rm S};a}_{k,\dd m}[\hat{\mbf{n}}]\Le(\delta_{\hat{n}^a_k+1,n}-\delta_{\hat{n}_k^a,n}\Ri)
                    +
                    \sum_{\hat{\Gamma}}\sum_{\dd m}P_t(\hat{\Gamma})
                    w^{{\rm CM};a}_{k,\dd m}[\hat{\mbf{n}}]\Le(\delta_{\hat{n}^a_k-1,n}-\delta_{\hat{n}_k^a,n}\Ri)\\
                    &+\sum_{\hat{\Gamma}}\sum_{q,\dd m}P_t(\hat{\Gamma})
                    \Le\{w^{{\rm S};b}_{q,\dd m}[\hat{\mbf{n}}]\Le(\delta_{\hat{n}^a_{k+2\dd m},n}-\delta_{\hat{n}_k^a,n}\Ri)
                    +w^{{\rm CM};b}_{q,\dd m}[\hat{\mbf{n}}]\Le(\delta_{\hat{n}^a_{k+2\dd m},n}-\delta_{\hat{n}_k^a,n}\Ri)\Ri\}.
                }{}
            On the RHS of the third and fourth lines, the ask-side events are described by the first and second terms: i.e., the first term represents the order-number increase due to ask limit-order submission, and the second term represents the order-number decrease due to ask limit-order cancellations or buy market-order submissions. The third term represents all events for the bid side causing changes in the best bid price: e.g., an arrival of a new bid limit order revising the best bid price, a cancellation of the best bid price order, and a sell market-order clearing the best bid price. 

            By using two identities
                \eqsp{
                    \sum_{\dd m}w^{{\rm S};a}_{k,\dd m}[\hat{\mbf{n}}]=\lambda\Delta\mbb{I}_{[1,\infty)}(k)=\lambda\Delta,\quad
                    \sum_{\dd m}w^{{\rm CM};a}_{k,\dd m}[\hat{\mbf{n}}]=\mu\delta_{k,a[\hat{\mbf{n}}^a]}+v\hat{n}^a_k,
                }{}
            we finally obtain the first-order BBGKY hierarchical equation,
                \eqsp{
                    \pdiff{P^a_t(n,k)}{t}
                    =&
                    \lambda\Delta\sum_{\hat{\Gamma}}P_t(\hat{\Gamma})
                    \Le(\delta_{\hat{n}^a_k+1,n}-\delta_{\hat{n}^a_k,n}\Ri)
                    +
                    \sum_{\hat{\Gamma}}P_t(\hat{\Gamma})
                    (\mu\delta_{k,a[\hat{\mbf{n}}^a]}+v\hat{n}^a_k)\Le(\delta_{\hat{n}^a_k-1,n}-\delta_{\hat{n}^a_k,n}\Ri)\\
                    &+\sum_{\hat{\Gamma}}\sum_{q,\dd m}P_t(\hat{\Gamma})
                    \Le\{w^{{\rm S};b}_{q,\dd m}[\hat{\mbf{n}}]\Le(\delta_{\hat{n}^a_{k+2\dd m},n}-\delta_{\hat{n}^a_k,n}\Ri)
                    +w^{{\rm CM};b}_{q,\dd m}[\hat{\mbf{n}}]\Le(\delta_{\hat{n}^a_{k+2\dd m},n}-\delta_{\hat{n}^a_k,n}\Ri)\Ri\}\\
                    =&
                    \lambda\Delta 
                    \Le\{P^a_t(n-1,k)-P^a_t(n,k)\Ri\}
                    +
                    v\Le\{(n+1)P^a_t(n+1,k)-nP^a_t(n,k)\Ri\}\\
                    &+
                    \mu\sum_{\hat{\Gamma}}P_t(\hat{\Gamma})
                    \delta_{k,a[\hat{\mbf{n}}^a]}\Le(\delta_{\hat{n}^a_k-1,n}-\delta_{\hat{n}^a_k,n}\Ri)\\
                    &+\sum_{\hat{\Gamma}}\sum_{q,\dd m}P_t(\hat{\Gamma})
                    \Le(w^{{\rm S};b}_{q,\dd m}[\hat{\mbf{n}}]+w^{{\rm CM};b}_{q,\dd m}[\hat{\mbf{n}}]\Ri)\Le(\delta_{\hat{n}^a_{k+2\dd m},n}-\delta_{\hat{n}^a_k,n}\Ri).\\
                }{}
            This BBGKY hierarchical equation corresponds to an exact form of the master equation for $P_t(n,k)$ in E. Smith {\it et al.} 2003.
        \end{proof}

\section{Mean-field theory: Boltzmann equation and its explicit steady solution}\label{sec:BoltzmannEq}
    The aim of this section is to obtain explicit analytical formulas for the Santa Fe model within the mean-field theory. In the previous section, we have derived the master equation~\eqref{eq;master_eq} and the first-order BBGKY hierarchical equation~\eqref{eq:BBGKY}. While these equations are exact without any approximation, their explicit solutions cannot be solved due to the genuinely many-body nature of the Santa Fe model. Generally, truly many-body problems with nonlinearity require some approximation in obtaining their explicit solutions. The aim of this section is to revisit a mean-field theory, which was heuristically proposed by E. Smith {\it et al.} 2003, through the lens of kinetic theory, particularly the BBGKY hierarchy and the Boltzmann equation. 

    \subsection{Boltzmann equation}
        In this subsection, we derive the Boltzmann equation for the Santa Fe model, which turns out to be a systematic counterpart of the mean-field equation for the LOB profile, heuristically derived by E. Smith {\it et al.} 2003. Our mathematical starting point is the first-order BBGKY hierarchical equation~\eqref{eq;master_eq}, where we particularly consider a specific form of the PDF $P_t^a(1,k)$ by setting $n=1$. We then apply two systematic approximations: 
        \begin{itemize}
            \item Step 1: The mean-field approximation for the many-body PDF $P_t(\Gamma)$
            \item Step 2: The small tick-size limit (or, equivalently, the sparse order-book limit)
        \end{itemize}
        After these approximations, we obtain a Boltzmann-like equation for the order-density profile, 
            \begin{itembox}[l]{Boltzmann equation for the order-density profile}
                \vspace{-5mm}
                    \eqsp{
                        \pdiff{\rho^a_t(r)}{t}
                        \simeq&\lambda-\rho^a_t(r)(v+\mu e^{-\INT{0}{r}{x}\rho^a_t(x)})
                        +
                        \INT{-\infty}{\infty}{y}\Tilde{W}(y)\Le\{\rho^a_t(r-y)-\rho^a_t(r)\Ri\}\\
                    }{eq:Boltzmann}
                where the jump term $\Tilde{W}(y)$ is given by
                    \eqsp{
                        \tilde{W}(y)
                        =&
                        \mbb{I}_{(-\infty,0)}(y)\lambda\INT{0}{\infty}{z}
                        \rho^a_t(z-y)e^{-\INT{0}{z-y}{x}\rho^a_t(x)}\\
                        &+\mbb{I}_{(0,\infty)}(y)(\mu+v)\INT{0}{\infty}{z}
                        \rho^a_t(z)\rho^a_t(z+y)e^{-\INT{0}{z+y}{x}\rho^a_t(x)}\\
                    }{eq:Boltzmann_jump}
                \vspace{-5mm}
            \end{itembox}
            This Boltzmann equation is the third main result of this work and corresponds to the mean-field density profile equation in E.~Smith {\it et al.} 2003. 
            
            Technically, there are minor differences between E.~Smith {\it et al.} 2003 and ours. First, our Boltzmann equation is based on the coordinate relative to the opposite best price, while the mean-field equation in E.~Smith {\it et al.} 2003 was based on the coordinate relative to the midprice. Second, E.~Smith {\it et al.} 2003 assumed that the order-submission contribution is almost equal to the cancellation and market-order contribution and that the overall diffusion constant is roughly estimated as twice the diffusion constant originating from order submissions. We evaluated their contributions systematically.

        \subsubsection{Step 1: Mean-field approximation}
            Let us systematically derive this Boltzmann equation~\eqref{eq:Boltzmann}. As the first step, we apply a mean-field approximation---the assumption of independence between the random variables: 
            \begin{itembox}[l]{Mean-field approximation}
                \vspace{-5mm}
                \eqsp{
                    P_t(\hat{\Gamma})\simeq P_t(\hat{m})\prod_{i}P^a_t(\hat{n}^a_i,i)\prod_{j}P^b_t(\hat{n}^b_{j},j).
                }{} 
                \vspace{-6mm}
            \end{itembox}
            By focusing on $P^a_t(n=1,k)$, we can rewrite \eqref{eq:BBGKY} as 
                \eqsp{
                    \pdiff{P^a_t(1,k)}{t}
                    =&
                    \lambda\Delta 
                    \Le\{P^a_t(0,k)-P^a_t(1,k)\Ri\}
                    +
                    v\Le\{2P^a_t(2,k)-P^a_t(1,k)\Ri\}
                    +
                    \mu\sum_{\hat{\Gamma}}P_t(\hat{\Gamma})
                    \delta_{k,a[\hat{\mbf{n}}^a]}\Le(\delta_{\hat{n}^a_k,2}-\delta_{\hat{n}^a_k,1}\Ri)\\
                    &+\sum_{\hat{\Gamma}}\sum_{q,\dd m}P_t(\hat{\Gamma})
                    \Le(w^{{\rm S};b}_{q,\dd m}[\hat{\mbf{n}}]+w^{{\rm CM};b}_{q,\dd m}[\hat{\mbf{n}}]\Ri)
                    \Le(\delta_{\hat{n}^a_{k+2\dd m},1}-\delta_{\hat{n}^a_k,1}\Ri).
                }{eq:BBGKY_n=1_mod}
            Then, we apply the mean-field approximation to \eqref{eq:BBGKY_n=1_mod} to obtain
                \eqsp{
                    &\pdiff{P^a_t(1,k)}{t}
                    =
                    \lambda\Delta 
                    \Le\{P^a_t(0,k)-P^a_t(1,k)\Ri\}
                    +
                    v\Le\{2P^a_t(2,k)-P^a_t(1,k)\Ri\}\\
                    &
                    +\mu\sum_{\hat{\mbf{n}}^a}\Le\{\prod_{i}P^a_t(\hat{n}^a_i,i)\Ri\}
                    \delta_{k,a[\hat{\mbf{n}}^a]}\Le(\delta_{\hat{n}^a_k,2}-\delta_{\hat{n}^a_k,1}\Ri)
                    +\sum_{h}W(h)\Le\{P^a_{t}(1,k-h)-P^a_{t}(1,k)\Ri\}
                }{eq:MF_after}
            where $W(h)$ is given by
                \eqsp{
                    W(h)&
                    :=\sum_{\hat{\mbf{n}}^b}\sum_{q}\Le\{\prod_{i}P^b_t(\hat{n}^b_{i},i)\Ri\}
                    \lambda\Delta\mbb{I}_{(-\infty,-1]}(q)\mbb{I}_{(-\infty,q)}(b[\hat{\mbf{n}}^b])\delta_{h,b[\hat{\mbf{n}}^b]-q}\\
                    &+\sum_{\hat{\mbf{n}}^b}\sum_{q}\Le\{\prod_{i}P^b_t(\hat{n}^b_{i},i)\Ri\}
                    (\mu\delta_{q,b[\hat{\mbf{n}}^b]}+v\hat{n}^b_q)\mbb{I}_{(-\infty,-1]}(q)\mbb{I}_{(-\infty,q)}(b[\hat{\mbf{n}}^b-\mbf{e}^q])\delta_{h,q-b[\hat{\mbf{n}}^b-\mbf{e}^q]}.\\
                }{}
            and we convert $\dd m$ to $h:=-2\dd m\in\mbb{Z}$.

    \subsubsection{Step 2: Small tick-size limit}
        As the second step, we take the small tick-size limit $\Delta \to 0$. This limit can be interpreted as the continuous price limit because the ``lattice size'' of the price axis is taken as infinitesimal. Also, this limit can be interpreted as the sparse LOB limit, where each queue is expected to have only a single or zero order. In other words, for the small tick-size limit $\Delta \to 0$, we assume a $\Delta$ scaling for the PDF $P_t^a(n,k)$,
            \eqsp{
                P^a_t(n,k)= O(\Delta^n),\quad 
                P^b_t(n,k)= O(\Delta^n),
            }{}
        implying that both $P^a_t(n,k)$ and $P^b_t(n,k)$ can be negligible for $n\geq 2$. We therefore define the average density profile at the relative continuous price $r:=k\Delta \in \mathbb{R}$, corresponding to the relative integer price level $k\in \mathbb{Z}$ as 
        \begin{itembox}[l]{Small tick-size assumption for the density profile}
            \vspace{-5mm}
            \eqsp{
                \rho^a_t(r):=
                \frac{\sum_{n}nP^a_t(n,k)}{\Delta}\simeq\frac{P^a_t(1,k)}{\Delta}\sim O(1), \quad 
                \rho^b_t(r):=
                \frac{\sum_{n}nP^b_t(n,k)}{\Delta}\simeq\frac{P^b_t(1,k)}{\Delta}\sim O(1).
            }{}
        \end{itembox}
        We can assume their symmetry without losing generality, such that 
            \eqsp{
                \rho^a_t(r)=\rho^b_t(-r).
            }{}
        Let us rewrite $P^a_t(1,k)$ by using $\rho_t^a(r)$ with $r:=k\Delta$ as 
            \eqsp{
                P^a_t(n,k)
                =\delta_{n,0}(1-P^a_t(1,k))+\delta_{n,1}P^a_t(1,k) + O(\Delta^2)
                =\delta_{n,0}(1-\rho^a_t(r)\Delta)
                +\delta_{n,1}\rho^a_t(r)\Delta + O(\Delta^2).
            }{}
        Therefore, while the Santa Fe model is an infinite-dimensional Markovian stochastic process, needing the many-body joint PDF $P_t(\Gamma)$ as an exact description, its approximate dynamics can be described by the one-body distribution $\rho^a_t(r\Delta)$ under the mean-field approximation for the small tick-size limit. Finally, we derive the mean-field density equation~\eqref{eq:Boltzmann} from Eq.~\eqref{eq:MF_after} (see Appendix~\ref{section:App_density_eq} for its more detailed derivation).

    \subsection{Approximate stationary solution solution}\label{sec:bound_theory}
     Here we apply a diffusive approximation for the Boltzmann equation~\eqref{eq:Boltzmann} in the stationary state under a boundary condition: 
        \begin{itembox}[l]{Exact boundary condition}
            \vspace{-5mm}
            \eqsp{
                \rho^a_t(0)=\frac{\lambda}{v+\mu}.
            }{eq:boundary_condition_secmain}
        \end{itembox}
    This boundary condition is exactly derived for the small tick-size limit, without needing the mean-field approximation (see Appendix~\ref{sec:App_bound_condition} for its derivation). Finally, we derive an analytical approximate formula: 
        \begin{itembox}[l]{Approximate stationary solution of $\rho^a_{st}(r)$}
            \vspace{-5mm}
            \eqsp{
                \rho^a_{\rm st}(r)
                =
                \Le\{\frac{\lambda}{v+\mu}-\fr{\lambda}{v}\Ri\}\exp{\Le(-\sqrt{\fr{v}{D}}r\Ri)}+\frac{\lambda}{v}.
            }{eq:sol_Boltzmann_in}
        \end{itembox}

        \begin{proof}
            Let us introduce a typical jump size in the jump term in the Boltzmann equation~\eqref{eq:Boltzmann}
                \eqsp{
                    \epsilon := \frac{v+\mu}{\lambda}=\frac{1}{\rho^a_t(0)},
                }{eq:epsilon_def}
            which is assumed to be small in the continuous asymptotic limit $\epsilon\to 0$. We then apply the Kramers-Moyal expansion to the jump term in the Boltzmann equation~\eqref{eq:Boltzmann}: 
            \begin{equation}
                \INT{-\infty}{\infty}{y}\Tilde{W}(y)\Le\{\rho^a_t(r-y)-\rho^a_t(r)\Ri\}
                =\sum_{k=1}^\infty \frac{(-1)^k}{k!}\left(\int_{-\infty}^\infty y^k \Tilde{W}(y)\dd y\right)\frac{\partial^k}{\partial r^{k}}\rho_t^{a}(r).
            \end{equation}
            By truncating the Kramers-Moyal series up to the second order, the Boltzmann equation~\eqref{eq:Boltzmann} reduces to a diffusive partial differential equation for the region $r\gg\epsilon$,
                \eqsp{
                    \pdiff{\rho^a_t(r)}{t}
                    \simeq&\lambda-v\rho^a_t(r)
                    +
                    A\fr{\partial}{\partial r}\rho^a_t(r)
                    +
                    D\fr{\partial^2}{\partial r^2}\rho^a_t(r),
                }{eq:explicit-sol-rho-trans1}
            where the coefficients $A$ and $D$ are defined as  
                \eqsp{
                    A:=\INT{-\infty}{\infty}{y}y\Tilde{W}(y) \approx 0,\quad
                    D:=\fr{1}{2}\INT{-\infty}{\infty}{y}y^2\Tilde{W}(y) \approx 2\frac{(v+\mu)^3}{\lambda^2}.
                }{}
            See Appendix.~\ref{app:1st-KM-calc} for the derivation of $A=0$ and for Appendix.~\ref{app:diffusion-const-calc} the derivation of $D=2(v+\mu)^3/\lambda^2$. Note that the nonlinear term $\mu e^{-\int_0^r\dd x\rho_t^a(x)}\approx \mu e^{-r/\epsilon}$ is negligible for $r\gg \epsilon$ under the boundary condition~\eqref{eq:boundary_condition_secmain}. Technical remarks are made in Sec.~\ref{sec:remark-SSE-KMtruncation} for the validity of this second-order truncation approximation. Solving this linear differential equation~\eqref{eq:explicit-sol-rho-trans1} yields the following general solution $\rho^a_{\rm st}(r)$:
                \eqsp{
                    \rho^a_{\rm st}(r)
                    =C_1\exp{\Le(\sqrt{\fr{v}{D}}r\Ri)}+C_2\exp{\Le(-\sqrt{\fr{v}{D}}r\Ri)}+\frac{\lambda}{v}.
                }{eq:sol_Boltzmann_out}
            The coefficients $C_1$ and $C_2$ are fixed by the boundary condition 
                \eqsp{
                    \lim_{r\to0}\rho^a_{\rm st}(r)=&\fr{\lambda}{v+\mu},\quad
                    \lim_{r\to\infty}\rho^a_{\rm st}(r)=\fr{\lambda}{v}.
                }{}
            As a result, we obtain the explicit formula of the stationary solution~\eqref{eq:sol_Boltzmann_in}.
        \end{proof}
        \subsubsection{Technical remarks about the approximation}\label{sec:remark-SSE-KMtruncation}
            There are two technical remarks on the above approximation. First, the KM expansion is valid only for $r\gg \epsilon$, and the general solution~\eqref{eq:sol_Boltzmann_out} is also expected to be valid for $r\gg \epsilon$. Imposing the boundary condition~\eqref{eq:boundary_condition_secmain} at $r=0$ on the solution~\eqref{eq:sol_Boltzmann_out} is not technically exact even for the $\eps\to 0$ limit; it should be regarded as an approximation. Further investigation of this technical point would be an interesting work for future. 

            In addition, our second-order truncation is formally validated when we focus on the diffusive limit: 
            \begin{equation}
                \eps=\frac{v+\mu}{\lambda}\to 0, \quad v+\mu \to \infty, \quad \mbox{s.t.}\quad D:= (v+\mu)\eps^2 = {\rm const}.
            \end{equation}
            For instance, taking a limit $v+\mu \to \infty$ by assuming a scaling $\lambda = c(v+\mu)^{3/2}$ is an easiest example with a constant $c=O((v+\mu)^0)$. This scaling is essentially equivalent to the system-size expansion~\cite{VanKampenB}---a traditional framework in statistical physics for a microscopic derivation of the Langevin equation for kinetic theory. For this diffusive limit, the Kramers-Moyal coefficients have the following asymptotic scaling, 
            \begin{equation}
                \int_{-\infty}^\infty y^n \Tilde{W}(y)\dd y \propto (v+\mu)\eps^n = D\eps^{n-2}.
            \end{equation}
            Thus, the overall contribution above the second order is negligible for this limit. 
        
    \subsection{The ``method of image'' solution for $\mu\gg v$}
        Let us consider the asymptotic limit $\mu\gg v$. The mean-field stationary order-book formula reduces to the following formula:
        \begin{itembox}[l]{The ``method of image'' solution of $\rho^a_{st}(r)$ for $\mu\gg v$}
            \vspace{-5mm}
            \eqsp{
                \rho^{a}_{\rm st}(r) =\frac{\lambda}{v}\left[1-\exp{\Le(-\sqrt{\fr{v}{D}}r\Ri)}\right].
            }{}
        \end{itembox}
        This formula is equivalent to the method of image solution\footnote{
            Our exact boundary condition~\eqref{eq:boundary_condition_secmain} is inconsistent with the absorbing boundary condition $\rho_{\rm st}^{a}(0)=0$, which was adopted in Ref.~\cite{ZIOB-Bouchaud2002} in deriving the method of image solution. However, our boundary condition~\eqref{eq:boundary_condition_secmain} effectively reduces to the absorbing boundary condition for large $\mu$ because $\lim_{\mu \to \infty}\rho_{\rm st}^a(0)/ \rho_{\rm st}^a(\infty) = 0$. 
        }, which was proposed by Bouchaud-M\'ezard-Potters in Quantitative Finance 2002~\cite{ZIOB-Bouchaud2002} (see also Chapter 8.8 of Ref.~\cite{BouchaudB}). Thus, our mean-field framework provides a mathematical ground of their heuristic arguments from the microscopic dynamics of the Santa Fe model. 
        
\section{Summary of predicted scaling formulas}\label{sec:summary-scaling}
    Since we obtain the explicit stationary order-book profile formula $\rho_{\rm st}^a(r)$, we can derive the scaling laws for financial metrics. Particularly, we focus on the following quantities:
    \begin{itemize}
        \item Spread $s$: the average difference between the best ask price $a$ and the best bid price $b$, such that $s:=\la a-b\ra$.
        \item Diffusion constant $D:= \lim_{\tau\to \infty}\la[p(t+\tau)-p(t)]^2\ra/\tau$, proportional to the volatility. 
        \item Instantaneous price impact $\la\varepsilon \dd m \ra_{\text{in}}$: the midprice change just after a market order of unit size. A positive order sign $\varepsilon=+1$ represents a buy market order and a negative order sign $\varepsilon=-1$ represents a sell market order. The instantaneous price impact is defined as the expected instantaneous midprice change $\dd m$ multiplied by the order sign $\varepsilon$.
    \end{itemize}
    Within the mean-field approximation, these quantities can be analytically evaluated from the stationary LOB distribution $\rho_{\rm st}^a(r)$, explicitly given by Eq.~\eqref{eq:sol_Boltzmann_in}. Our overall results are summarized in a table \ref{tab:scaling}. 
    
    We next compare our results with the previous results in Refs.~\cite{ZIOB-PRL2003,ZIOB-QFin2003}. Particularly, we find that our explicit solution is inconsistent with the scaling for the diffusion constant $D$ in Refs.~\cite{ZIOB-PRL2003,ZIOB-QFin2003}. We identify a misspecification of a characteristic constant in the dimensional analysis proposed in Ref.~\cite{ZIOB-QFin2003}. Finally, simulations are performed and compared with the theoretical analysis results, illustrating limitations of the mean-field approximation.

    \begin{table}
        \centering
            \begin{tabular}{|c|c|c|c|c|}
                \hline
                & Spread $s$ & Diffusion constant $D$ & \makecell{Instantaneous price impact\\ $\la\varepsilon\dd m \ra_{\text{in}}$} & Numerical validity\\
                \hline
                \makecell{This work \\for $\mu\ll v$} & $\frac{v}{\lambda}$ & $\frac{2v^3}{\lambda^2}$ & $\frac{v}{2\lambda}$ & Very well\\
                \hline
                \makecell{This work \\ for $\mu\gg v$} & $\frac{\mu}{\lambda}$ & $\frac{2\mu^3}{\lambda^2}$ & $\frac{\mu}{2\lambda}$  & \makecell{Limited}\\
                \hline
                Refs.~\cite{ZIOB-PRL2003,ZIOB-QFin2003} & $\frac{\mu}{\lambda}$ & $\frac{v\mu^2}{\lambda^2}$ & Not reported & Limited\\
                \hline
            \end{tabular}
        \caption{
            Summary of the mean-field scalings for financial metrics.
        }
        \label{tab:scaling}
    \end{table}

    \subsection{Asymptotic results for $\mu \ll v$}
        For $\mu\ll v$, we obtain the following asymptotic results for the financial metrics:
            \begin{itembox}[l]{financial metrics in the case of $\mu\ll v$}
                \vspace{-5mm}
                \eqsp{
                    s\simeq\frac{v}{\lambda},\quad
                    D\simeq 2\fr{v^3}{\lambda^2},\quad
                    \langle \varepsilon\mr{d}m\rangle_{\text{in}}\simeq\fr{v}{2\lambda}.
                }{eq:financial metrics_v}
            \end{itembox}
        This scaling is very natural and can be intuitively understood: For $\mu\ll v$, the characteristic time constant should be given by $1/v$, instead of $1/\mu$. Therefore, the dimensional analysis should be based on 
            \eqsp{
                \frac{v}{\lambda}\sim\text{[dimension of price]}, \quad 
                \frac{1}{v}\sim\text{[dimension of time]}.
            }{}
        Based on this scenario,  the spread $s$ and the diffusion constant $D$ should have the following scalings, 
            \eqsp{
                s \propto \frac{v}{\lambda}\sim\text{[dimension of price]}, \quad
                D\propto \frac{v^3}{\lambda^2}\sim\text{[price}^2\text{ per time]}.
            }{}
        Our mean-field formulas perfectly agree with this dimensional analysis.

    \subsection{Asymptotic results for $\mu\gg v$}
        For $v\ll\mu$, we obtain the following asymptotic results for the financial metrics:
            \begin{itembox}[l]{financial metrics in the case of $\mu\gg v$}
                \vspace{-5mm}
                \eqsp{
                    s\simeq\frac{\mu}{\lambda},\quad
                     D\simeq 2\fr{\mu^3}{\lambda^2},\quad
                     \langle \varepsilon\mr{d}m\rangle_{\text{in}}\simeq\fr{\mu}{2\lambda}.
                }{eq:financial_metrics_exe}
            \end{itembox}
        Again, this scaling is very natural: For $v\ll\mu$, the characteristic time constant should be given by $1/\mu$, instead of $1/v$. Therefore, the dimensional analysis should be based on 
            \eqsp{
                \frac{\mu}{\lambda}\sim\text{[dimension of price]}, \quad 
                \frac{1}{\mu}\sim\text{[dimension of time]}.
            }{}
        Based on this assumption, the scalings of the spread $s$ and the diffusion constant $D$ should be estimated as 
            \eqsp{
                s \propto \frac{\mu}{\lambda}\sim\text{[dimension of price]}, \quad
                D\propto \frac{\mu^3}{\lambda^2}\sim\text{[price}^2\text{ per time]}.
            }{}

    \subsection{Comparison 1 with E. Smith et al. 2003: dimensional analysis}
        Here, our scaling for $D\propto \mu^3/\lambda^2$ for $\mu \gg v$ is inconsistent with the E. Smith et. al. 2003, claiming $D\propto v\mu^2/\lambda^2$. In our view, this inconsistency stems from their misspecification of characteristic constants for $\mu \gg v$, formulated in Eq.~\eqref{eq:Smith-characteristic-constants}, particularly because they did not solve their mean-field equations explictly. Indeed, our explicit solution~\eqref{eq:sol_Boltzmann_in} of the mean-field equation is inconsistent with their estimation regarding $D$.

        As an additional explanation, we directly show where E. Smith's paper has an error in dimensional arguments for their mean-field equation. In Ref.~\cite{ZIOB-QFin2003}, they derived a non-dimensional mean-field equation (i.e., Eq.~{(37)} on the page 500),
            \eqsp{
                1+\varphi(\hat{p})=-\Le[\fr{d\varphi(\hat{p})}{d\hat{p}}+\tilde{\epsilon}\Le(1-\beta\fr{d^2}{d\hat{p}^2}\Ri)\fr{d\log\varphi(\hat{p})}{d\hat{p}}\Ri],
            }{eq:Smith_density}
        where $\hat{p}$ is a non-dimensional price and $\varphi(\hat{p})$ is the exponentially-cumulative density:
            \eqsp{            
                \hat{p}:=\frac{r\lambda}{\mu},\quad 
                \varphi(\hat{p})=e^{-\INT{0}{\fr{\mu}{\lambda}\hat{p}}{r}\rho^a_{st}(r)}.
            }{}
        In addition, $\tilde{\epsilon}$, and $\beta$ are non-dimensional constants, defined by 
            \eqsp{
                \tilde{\epsilon}:=\fr{v}{\mu}, \quad 
                \beta:=\frac{D\lambda^2}{v\mu^2}.
            }{}
        Ref.~\cite{ZIOB-QFin2003} regarded $\beta$ as a non-dimensional diffusion constant based on their dimensional analysis $D\propto v\mu^2/\lambda^2$. They also provided an analytical formula 
        \eqsp{
                \beta=\fr{4}{\tilde{\epsilon}}\INT{0}{\infty}{\hat{p}}\hat{p}^2\varphi(\hat{p}).
            }{}
        In Ref.~\cite{ZIOB-QFin2003}'s framework, $\varphi(\hat{p})$ is a non-dimensional quantity for the limit $\tilde{\epsilon}\to 0$ because their mean-field equation~\eqref{eq:Smith_density} has no parameter for $\tilde{\epsilon}\to 0$. This means that the integral $\int_{0}^\infty \hat{p}^2\varphi(\hat{p})\dd \hat{p}$ converges to a non-dimensional real number. However, at the same time, one can notice that $\beta$ does not converge to a finite real number as it is proportional to $\eps^{-1}$. Clearly, this is erroneous as a dimensional analysis. Instead, the correct non-dimensionalisation of the diffusion constant is given by 
            \eqsp{
                \tilde{\beta}:=\tilde{\epsilon}\beta=4\INT{0}{\infty}{\hat{p}}\hat{p}^2\varphi(\hat{p})=O(1),
            }{}
        which converges to a finite real number for $\tilde{\epsilon}\to 0$. Finally, the assumption of $\tilde{\beta}=O(1)$ implies our scaling formula $D\propto \mu^3/\lambda^2$, even based on the original mean-field framework by Ref.~\cite{ZIOB-QFin2003}. 

    \subsection{Comparison 2 with E. Smith et al. 2003: mean-field gap equation}
        In Ref.~\cite{ZIOB-QFin2003}, they evaluated the spread via another mean-field equation~\eqref{eq:main:Smith_gap_comp} regarding the average gap size $\la g_k\ra$. However, we have evaluated the average spread $s$ and the price impacts only via the mean-field equation for the density profile; the mean-field equation for the average gap size is redundant and unnecessary for deriving the financial metrics addressed in this work. Note that the mean-field equaton for the average gap size can be also derived from the BBGKY hierarchical equation, as shown in Appendix~\ref{sec:app:GapMeanField}.

    \subsection{Numerical simulation}
        In this subsection, we show the scaling laws of financial metrics for the Santa Fe model based on the Monte Carlo simulation. Our numerical scheme is essentially based on the description in Sec.~\ref{sec:model}. 

        \subsubsection{Spread}
            \begin{figure}
                \centering
                \includegraphics[width=0.9\columnwidth]{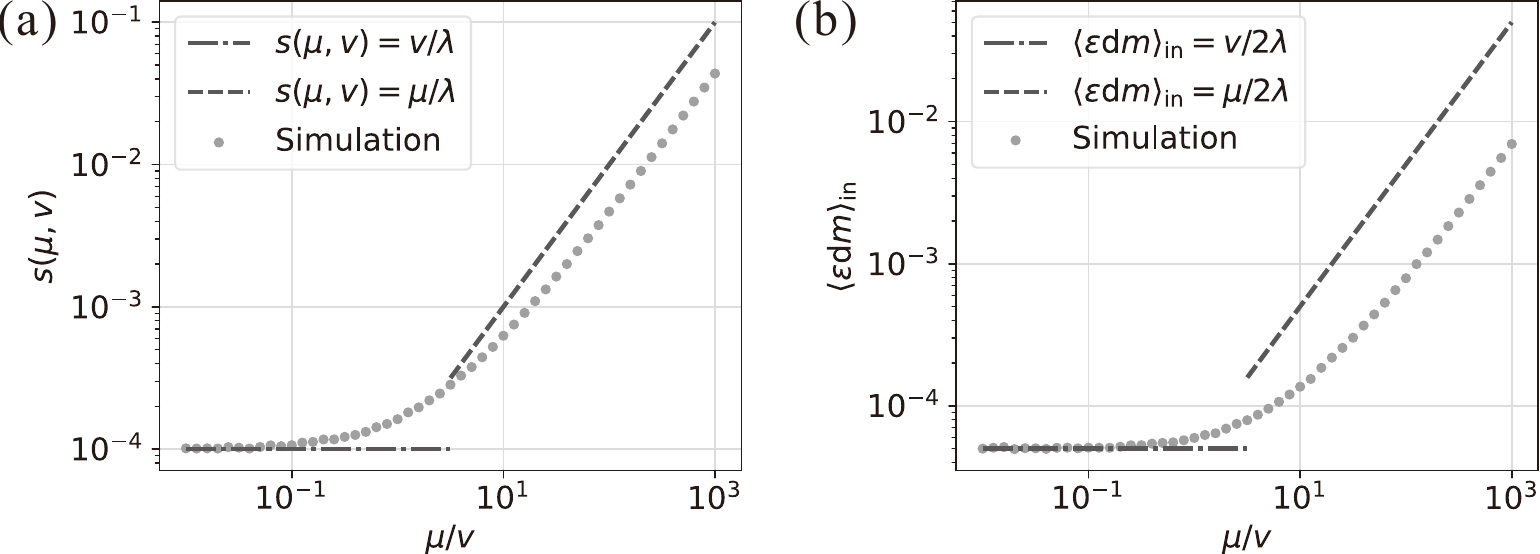}
                \caption{
                    Comparison between the numerical simulation and the mean-field scalings as a function of the market-order intensity $\mu \in [10^{-2},10^3]$. The figure~(a) plots the average spread, and the figure~(b) plots the instantaneous price impact. The simulation parameters are given by $\lambda=10000.0$, $v=1.0$, $\Delta=10^{-6}$, and $L=10^6$.
                }
                \label{fig:spreadandPI}
            \end{figure}
            Figure~\ref{fig:spreadandPI}(a) presents the comparison between our numerical results and the mean-field predictions. The figure shows two points: (i)~For small $\mu$, the mean-field prediction~\eqref{eq:financial metrics_v} works almost perfectly, including the coefficient beyond the scaling law. (ii)~For large $\mu$, the mean-field scaling~\eqref{eq:financial_metrics_exe} works well, while its coefficient is not overestimated. Anyway, we conclude that the mean-field scaling laws nicely work for both $\mu\to0$ and $\mu\to\infty$ limits. 

        \subsubsection{Instantaneous price impact}        
            Figure~\ref{fig:spreadandPI}(b) shows the numerical instantaneous price impact as a function of $\mu$. For small $\mu$, the mean-field prediction~\eqref{eq:financial metrics_v} perfectly works, even at the coefficient level. For large $\mu$, the mean-field scaling~\eqref{eq:financial_metrics_exe} works as $\langle \varepsilon\mr{d}m\rangle_{\text{in}}\propto \mu/\lambda$, while there is a discrepancy at the coefficient level. Again, the mean-field theory works nicely in predicting the scaling laws for the instantaneous price impact. Note that the permanent price impact behaves differently from the instantaneous price impact, as will be discussed in Sec.~\ref{sec:why-MFdoesNotWork?}

        \subsubsection{Diffusion constant}
             \begin{figure}
                \centering
                \includegraphics[width=0.9\columnwidth]{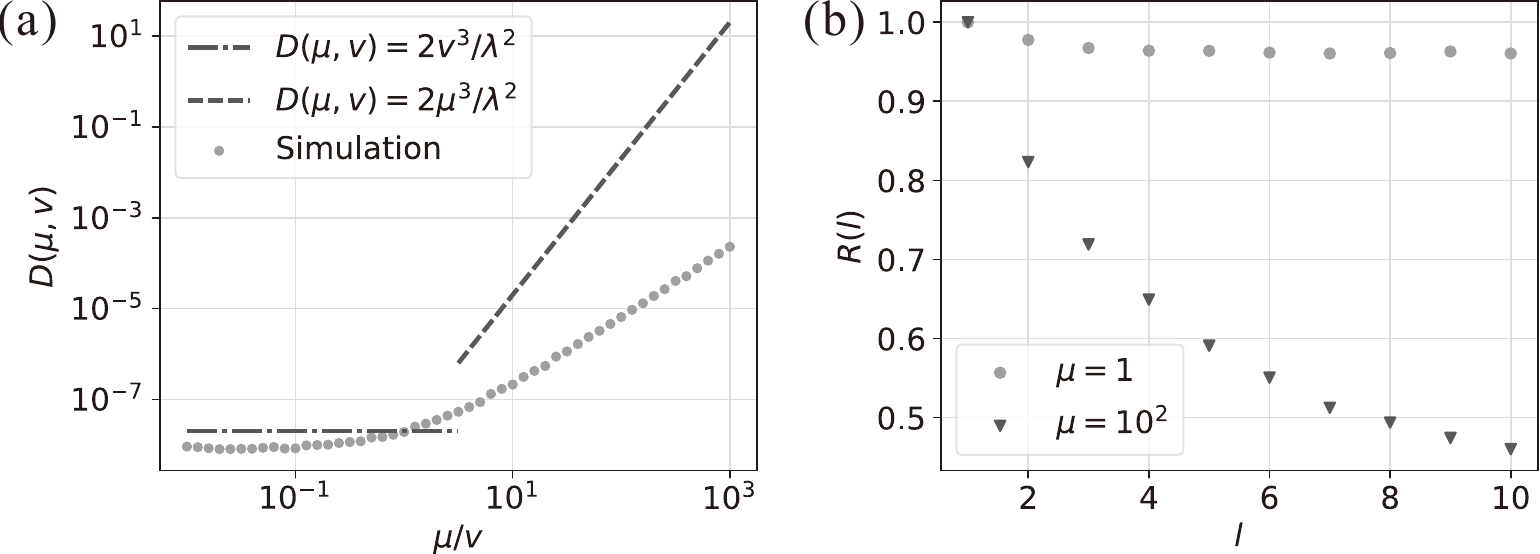}
                \caption{
                    The figure~(a) plots the diffusion constant for the numerical simulation and the mean-field scaling as a function of the market-order intensity $\mu \in [10^{-2},10^3]$. The simulation parameters are given by $\lambda=10000.0$, $v=1.0$, $\Delta=10^{-6}$, and $L=10^6$. The figure~(b) is the numerical plot of $R(l):=\la\varepsilon \dd m(l)\ra/\la\varepsilon \dd m(1)\ra$, i.e., the ratio of the $l$-lag to the one-lag price impacts for $\mu=1$ and $\mu=100$. The simulation parameters are given by $\lambda=10000.0$, $v=1.0$, $\Delta=10^{-6}$, and $L=10^6$.
                }
                \label{fig:diffandlagPI}
            \end{figure}            
            Figure~\ref{fig:diffandlagPI}(a) shows the numerical result on the diffusion constant. For small $\mu$, the numerical diffusion constant is consistent with the mean-field prediction~\eqref{eq:financial metrics_v}, while we find a small discrepancy regarding a numerical factor. For large $\mu$, the numerical diffusion constant behaves very differently from the mean-field scaling~\eqref{eq:financial_metrics_exe}. 

            This numerical result is consistent with the previous report~\cite{ZIOB-QFin2003}. Indeed, Ref.~\cite{ZIOB-QFin2003} numerically reported $D\propto (\mu v)^{3/2}/\lambda^2$. Note that the origin of this mysterious scaling has not been theoretically resolved to the best of our knowledge\footnote{
                In Ref.~\cite{FarmerPNAS2005}, J.~D. Farmer states that this exponent cannot be derived from their mean-field theory: ``The dependence on order size and the values of the scaling exponents are not so obvious. It has so far not been possible to derive this formula from theoretical considerations (although dimensional analysis was essential for guessing this functional form)''.
            }. Thus, there fact illustrates a clear limitation of the mean-field theory, particularly for the diffusion constant. 

        \subsubsection{Why does the mean-field theory have limitation?}\label{sec:why-MFdoesNotWork?}
            We observe the limitation of the mean-field theory for the Santa Fe model. Why is the usefulness of the mean-field theory limited for the Santa Fe model. While we have no clear answer yet, we conjecture that this comes from the non-Markovian effect in the Santa Fe model for large $\mu$. Indeed, it is known that the Santa Fe model exhibits mean-reverting behaviour in its price dynamics for $\mu$, which is actually non-Markovian. 

            To illustrate this point, Let us consider the $l$-lag price impact $\la \varepsilon \dd m(l)\ra$. We plot the ratio
                \eqsp{
                    R(l):= \fr{\la \varepsilon \dd m(l)\ra}{\la \varepsilon \dd m(1)\ra}
                }{}
            in Figure~\ref{fig:diffandlagPI}(b) for two different $\mu$: $\mu=1$ as a small parameter and $\mu=100$ as a large parameter. For the small $\mu$ case, there is no decay in the price impact, and the instantaneous price impact is almost equal to the permanent price impact: 
            \begin{equation}
                \mbox{[Permanent price impact]} \simeq \mbox{[Instantaneous price impact]} \propto \frac{v}{\lambda} \mbox{ for small }\mu.
            \end{equation}
            For such a regime, the Markovian mean-field approximation should apply, and the diffusion constant is estimated by 
            \begin{equation}
                D \propto \mbox{[Intensity]} \times \mbox{[Permanent price impact]}^2 \simeq v \left(\frac{v}{\lambda}\right)^2 = \frac{v^3}{\lambda^2},
            \end{equation}
            where we use the approximate fact that the permanent price impact is equal to the instantaneous price impact. On the other hand, for large $\mu$, we observe a significant decay of the price impact as the lag $l$ increases. This implies the strong mean-reverting behaviour of the price dynamics in the Santa Fe model. Due to this non-Markovian effect, we cannot assume any more that the permanent price impact is approximately equal to the instantaneous price impact: 
            \begin{equation}
                \mbox{[Permanent price impact]} \not\simeq \mbox{[Instantaneous price impact] for large }\mu. 
            \end{equation}
            Thus, the naive scaling estimation based on the mean-field approximation does not work even qualitatively. 

            If the above scenario is correct, it will be necessary to develop a different approximation that can account for the non-Markovian mean-reverting nature of the Santa Fe model for the large $\mu$ limit.
           
\section{Concluding remarks}\label{sec:conclusion}
    In this work, we revisit the mean-field theory of the Santa Fe model through the lens of kinetic theory. While the original work~\cite{ZIOB-PRL2003,ZIOB-QFin2003} depends on heuristic reasoning based on dimensional analysis, we derive the Boltzmann-like equation via the BBGKY hierarchical equation. Our Boltzmann equation corresponds to the mean-field equation for the LOB profile in Ref.~\cite{ZIOB-PRL2003,ZIOB-QFin2003}, providing its solid mathematical foundation. By explicitly solving the Boltzmann equation, we derive the mean-field analytical formulas for the average order-book profile and various financial metrics (i.e., the average spread, the instantaneous price impact, and diffusion constant) for the small tick-size limit. Our explicit solution shows that (i)~the mean-field theory works for small $\mu$, but (ii)~its utility is limited for large $\mu$. We find that the dimension analysis in the original work~\cite{ZIOB-PRL2003,ZIOB-QFin2003} has errors inconsistent with the explicit mean-field solutions, which are corrected in our formulation. For large $\mu$, our order-book formula coincides with the method of image order-book formula proposed in Ref.~\cite{ZIOB-Bouchaud2002}. Thus, our work supports their heuristic solution within the mean-field approximation via a systematic calculation. 

    In this paper, we study the most basic version of the Santa Fe model with spatially uniform submission intensity $\lambda$. Extension of our approach to more advanced Santa Fe models is left as a future topic. For example, it is an important generalisation to consider the spatially inhomogeneity of the submission intensity and cancellation rate, such that $\lambda \to \lambda(r)$ and $v\to v(r)$ by introducing the $r$ dependence. This generalisation will be necessary in replicating realistic hump shapes~\cite{ZIOB-Bouchaud2002,PotterPhysA2003} of the average LOB profile. Also, it would be important to include the long memory of market-order flow, which is empirically modelled by the Lillo-Mike-Farmer model~\cite{LMF_PRE2005,TothJEDC2015,SatoPRL2023,SatoPRR2023}. Indeed, the long-ranged correlation was the key in updating the zero-intelligence model to the ``$\epsilon$-intelligence'' model in Ref.~\cite{TothPRX2011} for studying the nonlinearity of price impact (called the square-root law~\cite{BouchaudB,TothPRX2011,SatoKz2025}; see Ref.~\cite{TothPRX2011,Donier2015} also for the latent order-book model as a potential candidate model for its microscopic description). 

\section*{Acknowledgements}
    TW was supported by JST SPRING (Grant Number~JPMJSP2110). KK was supported by JSPS KAKENHI (Grant Number~22H01141).
 
\appendix
\section{Summary of the indicator functions and its logical structure}\label{sec:App_indicator}
    In this work, how to technically use the indicator functions is the key to our kinetic formulation, particularly for case-by-case analyses. In this Appendix, we will summarise the indicator function and its logical structure regarding mathematical statement, which are exploited for the derivation of the BBGKY hierarchical equations.

    \subsection{Formulas for the indicator functions}\label{subsec:i_pro}
        Let us $\mathcal{P}(x)$ a mathematical statement that depends on some variable $\bm{x}$, that can be a vector argument. For the statement $\mathcal{P}(\bm{x})$, we define $\mbf{1}_{\mathcal{P}(\bm{x})}$ as the indicator function, such that 
        \begin{equation}
            \mbf{1}_{\mathcal{P}(\bm{x})}
            :=
            \begin{cases}
                1&(\text{$\mathcal{P}(x)$ is true})\\
                0&(\text{$\mathcal{P}(x)$ is false})
            \end{cases}
        \end{equation}
        For example, if $\mc{P}(x)$ is the statement that ``$x$ is an even integer'', then $\mc{P}(5)$ is a false statement and $\mbf{1}_{\mathcal{P}(5)}=0$; inversely, $\mc{P}(2)$ is a true statement and $\mbf{1}_{\mathcal{P}(2)}=0$. For this indicator function, the following formulas hold as identities: 
        \begin{enumerate}
            \item  
            If $\mathcal{P}(\bm{x})\Leftrightarrow\mathcal{Q}(\bm{x})$, we have an identity, 
                \eqsp{
                    \mbf{1}_{\mathcal{P}(\bm{x})}=\mbf{1}_{\mathcal{Q}(\bm{x})}
                }{eq:pro_i_1}
            \item
            If $\mathcal{P}(x)\Rightarrow\mathcal{Q}(\bm{x})$, we have an identity, 
                \eqsp{
                    \mbf{1}_{\mathcal{P}(\bm{x})}=\mbf{1}_{\mathcal{P}(\bm{x})}\mbf{1}_{\mathcal{Q}(\bm{x})}
                }{}
            \item Let $\neg\mathcal{P}(\bm{x})$ be the negation of $\mathcal{P}(\bm{x})$. We have an identity,
                \eqsp{
                    \mbf{1}_{\neg\mathcal{P}(\bm{x})}=1-\mbf{1}_{\mathcal{P}(\bm{x})}
                }{}
            \item
            Let ``$\land$'' be the logical AND (conjunction) symbol\footnote{
                Consider a two-dimensional vector argument $\bm{x}:=(x_1,x_2)$. Let $\mc{P}_1(\bm{x})$ be the statement that ``$x_1$ is an even integer'' and $\mc{P}_2(\bm{x})$ be the statement that ``$x_2$ is zero''. Their logical conjunction $\mc{P}_1(\bm{x})\land \mc{P}_2(\bm{x})$ represents the statement that ``$x_1$ is an even integer and $x_2$ is zero''. Trivially, $\mc{P}_1(\bm{x})\land \mc{P}_2(\bm{x})$ is false for $\bm{x}=(2,2)$ and is true for $\bm{x}=(4,0)$.
            }. The indicator function $\mbf{1}_{\land^n_{i=1}\mathcal{P}_i(\bm{x})}$ for a multiple-conjunction statement $\land^n_{i=1}\mathcal{P}_i(\bm{x})$ satisfies an identity,
                \eqsp{
                    \mbf{1}_{\land^n_{i=1}\mathcal{P}_i(\bm{x})}
                    =\prod^n_{i=1}\mbf{1}_{\mathcal{P}_i(\bm{x})}
                }{}
            \item
            Let ``$\lor$'' be the logical OR (disjunction) symbol\footnote{
                Consider a two-dimensional vector argument $\bm{x}:=(x_1,x_2)$. 
                Let $\mc{P}_1(\bm{x})$ be the statement that ``$x_1$ is an even integer'' and $\mc{P}_2(\bm{x})$ be the statement that ``$x_2$ is one''. Their logical disjunction $\mc{P}_1(\bm{x})\lor \mc{P}_2(\bm{x})$ represents the statement that ``$x_1$ is an even integer or $x_2$ is one''. Trivially, $\mc{P}_1(\bm{x})\lor \mc{P}_2(\bm{x})$ is false for $\bm{x}=(2,2)$ and is true for $\bm{x}=(1,1)$.
            }. Also, let $\mathcal{P}_1(\bm{x}),\cdots,\mathcal{P}_n(\bm{x})$ be statements that are mutually exclusive\footnote{If $\mc{P}_k$ is true for some $k$, all the other statements are false, such that $\mc{P}_l$ is false for all $l\neq k$.}. For the multiple-disjunction statement $\lor^n_{i=1}\mathcal{P}_i(\bm{x})$, we have an identity, 
                \eqsp{
                    \mbf{1}_{\lor^n_{i=1}\mathcal{P}_i(\bm{x})}
                    =
                    \sum^n_{i=1}\mbf{1}_{\mathcal{P}_i(\bm{x})}.
                }{}
            Particularly, if $\mathcal{P}_1(\bm{x}),\cdots,\mathcal{P}_n(\bm{x})$ are not only mutually exclusive but also collectively exhaustive, we have an identity,
                \eqsp{
                    1= \mbf{1}_{\lor^n_{i=1}\mathcal{P}_i(\bm{x})} = \sum^n_{i=1}\mbf{1}_{\mathcal{P}_i(\bm{x})}.
                }{eq:i_all_1}
        \end{enumerate}

    \subsection{Application to represent the best price levels with the indicator function}\label{subsec:I_gutairei}
        The above formulas are useful for us to represent the best price levels by using the indicator functions: 
        \begin{itembox}[l]{Best price levels on the fixed coordinate system}
            \vspace{-5mm}
            \eqsp{            
                \delta_{i,A[\hat{\mathbf{N}}^a(t)]}
                =
                (1-\delta_{\hat{N}^a_i(t),0})\prod^{i-1}_{j=-\infty}\delta_{\hat{N}^a_j(t),0}, \quad                
                A[\hat{\mathbf{N}}^a(t)]
                =
                \sum_{i\in\mbb{Z}}i
                (1-\delta_{\hat{N}^a_i(t),0})\prod^{i-1}_{j=-\infty}\delta_{\hat{N}^a_j(t),0},\\    
                \delta_{i,B[\hat{\mathbf{N}}^a(t)]}
                =
                (1-\delta_{\hat{N}^b_i(t),0})
                \prod^{\infty}_{j=i+1}\delta_{\hat{N}^b_j(t),0},\quad 
                B[\hat{\mathbf{N}}^b(t)]
                =\sum_{i\in\mbb{Z}}i
                (1-\delta_{\hat{N}^b_i(t),0})
                \prod^{\infty}_{j=i+1}\delta_{\hat{N}^b_j(t),0}.
            }{eq:best_price_f}
        \end{itembox}
        \begin{proof}
            By writing the statement that ``$\hat{N}^a_{j}(t)=0$'' as $\mathcal{E}(\hat{N}^a_{j}(t))$, the following statement holds true:  
                \eqsp{
                    i=A[\hat{\mathbf{N}}^a(t)]
                    \Leftrightarrow
                    \neg\mathcal{E}(\hat{N}^a_{i}(t))\wedge\Le\{\bigwedge^{i-1}_{j=-\infty}\mathcal{E}(\hat{N}^a_{j}(t))\Ri\},
                }{}
            because there is no ask orders below the best price. Using the identities in Appendix~\ref{subsec:i_pro}, we have an identity, 
                \eqsp{
                    \mbf{1}_{i=A[\hat{\mathbf{N}}^a(t)]}
                    =(1-\mbf{1}_{\mathcal{E}(\hat{N}^a_{i}(t))})\prod^{i-1}_{j=-\infty}\mbf{1}_{\mathcal{E}(\hat{N}^a_{j}(t))}.
                }{}
            In other words, by using the Kronecker delta, we obtain  
                \eq{
                    \delta_{i,A[\hat{\mathbf{N}}^a(t)]}
                    =
                    (1-\delta_{\hat{N}^a_i(t),0})\prod^{i-1}_{j=-\infty}\delta_{\hat{N}^a_j(t),0}.
                }{}

            Also, let $\mc{P}_i[\hat{\mathbf{N}}^a(t)]$ be the statement that ``$i=A[\hat{\mathbf{N}}^a(t)]$''. Since the statements $\{\mc{P}_i[\hat{\mathbf{N}}^a(t)]\}_{i\in\mbb{Z}}$ are mutually exclusive and collectively exhaustive, we have an identity
                \eq{
                    1=
                    \sum_{i\in\mbb{Z}}\delta_{i,A[\hat{\mathbf{N}}^a(t)]}
                    =
                    \sum_{i\in\mbb{Z}}(1-\delta_{\hat{N}^a_i(t),0})\prod^{i-1}_{j=-\infty}\delta_{\hat{N}^a_j(t),0}
                }{}
            by using Eq.~\eqref{eq:i_all_1}. Thus, $A[\hat{\mathbf{N}}^a(t)]$ can be rewritten as
                \eq{
                    A[\hat{\mathbf{N}}^a(t)]
                    =
                    \sum_{i\in\mbb{Z}}i\delta_{i,A[\hat{\mathbf{N}}^a(t)]}
                    =
                    \sum_{i\in\mbb{Z}}i
                    (1-\delta_{\hat{N}^a_i(t),0})\prod^{i-1}_{j=-\infty}\delta_{\hat{N}^a_j(t),0},
                }{}
            which is an explicit function of $\hat{\mbf{N}}^a(t)$. In a parallel calculation, we have identities 
                \eqsp{
                    \delta_{i,B[\hat{\mathbf{N}}^a(t)]}
                    =
                    (1-\delta_{\hat{N}^b_i(t),0})
                    \prod^{\infty}_{j=i+1}\delta_{\hat{N}^b_j(t),0},\quad
                    B[\hat{\mathbf{N}}^b(t)]
                    =\sum_{i\in\mbb{Z}}i
                    (1-\delta_{\hat{N}^b_i(t),0})
                    \prod^{\infty}_{j=i+1}\delta_{\hat{N}^b_j(t),0},
                }{}
            because there are no buy orders over the best bid price level. 
        \end{proof}

        By applying parallel calculations for $a[\hat{\mbf{n}}^a_t],b[\hat{\mbf{n}}^b_t]$ as functions with respect to $\hat{\mbf{n}}^a_t,\hat{\mbf{n}}^b_t$, we obtain 
        \begin{itembox}[l]{Best price levels on the relative coordinate system}
            \vspace{-5mm}
            \eqsp{
                \delta_{i,a[\hat{\mathbf{n}}^a(t)]}
                =
                (1-\delta_{\hat{n}^a_i(t),0})\prod^{i-1}_{j=-\infty}\delta_{\hat{n}^a_j(t),0},\quad
                a[\hat{\mathbf{n}}^a(t)]
                =
                \sum_{i\in\mbb{Z}}i
                (1-\delta_{\hat{n}^a_i(t),0})\prod^{i-1}_{j=-\infty}\delta_{\hat{n}^a_j(t),0}\\
                \delta_{i,b[\hat{\mathbf{n}}^b(t)]}
                =
                (1-\delta_{\hat{n}^b_i(t),0})
                \prod^{\infty}_{j=i+1}\delta_{\hat{n}^b_j(t),0},\quad
                b[\hat{\mathbf{n}}^b(t)]
                =\sum_{i\in\mbb{Z}}i
                (1-\delta_{\hat{n}^b_i(t),0})
                \prod^{\infty}_{j=i+1}\delta_{\hat{n}^b_j(t),0}.
            }{eq:best_price_r}
        \end{itembox}

\section{Extreme value statistics based on the mean-field approximation}\label{sec:App_kyoku}
    Let us organise our results regarding extreme value statistics based on the mean-field approximation. We consider a PDF of the best ask price
        \eqsp{
            \text{Prob}[i=a[\hat{\mbf{n}}^a]]
            =\sum_{\hat{\mbf{n}}^a}\delta_{i,a[\hat{\mbf{n}}^a]}P^a_t[\hat{\mbf{n}}^a]
            \simeq \sum_{\hat{\mbf{n}}^a}\delta_{i,a[\hat{\mbf{n}}^a]}\prod_{r}P^a_t(\hat{n}^a_r,r),            
        }{}
    where the mean-field approximation $P^a_t[\hat{\mbf{n}}^a] \simeq \prod_{r}P^a_t(\hat{n}^a_r,r)$ is applied. A parallel calculation can be performed even for the bid side. By using the results in Appendix~\ref{subsec:i_pro}, we can transform $\text{Prob}[i=a[\hat{\mbf{n}}^a]]$ as 
        \eqsp{
            \text{Prob}[i=a[\hat{\mbf{n}}^a]]
            \simeq\sum_{\hat{\mbf{n}}^a}\prod_{r}P^a_t(\hat{n}^a_r,r)
            (1-\delta_{\hat{n}^a_i,0})\prod^{i-1}_{j=1}\delta_{\hat{n}^a_j,0}
            \simeq(1-P^a_t(0,i))\prod^{i-1}_{j=1}P^a_t(0,j).
        }{}
    For the small tick-size limit $\Delta \to 0$ (or equivalently the assumption of the sparsely-distributed LOBs), we obtain 
        \eqsp{
           P^a_t(0,i)\simeq1-P^a_t(1,i)\simeq 1-\rho^a_t(i\Delta)\Delta.
        }{}
    Thus, we can rewrite $\text{Prob}[i=a[\hat{\mbf{n}}^a]]$ as 
        \eqsp{
            \text{Prob}[i=a[\hat{\mbf{n}}^a]]
            \simeq \rho^a_t(i\Delta)\Delta \prod^{i-1}_{j=1}\Le(1-\rho^a_t(j\Delta)\Delta\Ri)
            \simeq \rho^a_t(i\Delta)\Delta e^{-\Delta\sum^{i-1}_{j=1}\rho^a_t(j\Delta)}
            \simeq \rho^a_t(x)\Delta e^{-\int_{0}^x \dd y\rho^a_t(y)}
        }{eq:best_kyoku}
    for small $\Delta$, where the real numbers $x:=i\Delta$ and $y:=j\Delta$ represent continuous price, instead of integer price levels.

    Also, the following forms of integrals often appear in our calculations for the spread and diffusion constant, such as 
        \eqsp{
            \sum^{\infty}_{i=1}g(i\Delta)\text{Prob}[i=a[\hat{\mbf{n}}^a]]
            \simeq
            \sum^{\infty}_{i=1}g(i\Delta)\rho^a_t(i\Delta)\Delta e^{-\Delta\sum^{i-1}_{j=1}\rho^a_t(j\Delta)}
            \xrightarrow{\text{spase limit}}
            \INT{0}{\infty}{x}g(x)\rho^a_t(x)e^{-\INT{0}{x}{y}\rho^a_t(y)}
        }{}
   Because the contribution in the integral exponentially decays for large $x$, we apply the following approximation, 
        \eqsp{
            \INT{0}{\infty}{x}g(x)\rho^a_t(x)e^{-\INT{0}{x}{y}\rho^a_t(y)}
            \simeq
            g(0)\INT{0}{\infty}{x}\rho^a_t(x) e^{-\INT{0}{x}{y}\rho^a_t(y)},
        }{eq:approxi_kyokuti}
    where we focus on the contribution around $x\sim 0$. 

\section{Detailed derivation of the exact master equation~\eqref{eq;master_eq}}\label{App:master_cal}
    In this Appendix, we provide the detailed calculations to derive the exact master equation~\eqref{eq;master_eq}. First, let us decompose $\left\langle \dd f(\hat{\Gamma}_t)\right\rangle $ into two types of contributions: one from a new order arrival and the other from an order cancellation or market order.
        \eqsp{
            \left\langle \mr{d}f(\hat{\Gamma}_t)\right\rangle
            =&
            \dd t\sum_{\hat{\Gamma}_t}\sum_{q\in\mbb{Z}}
            P_t(\hat{m}_t,\hat{\mbf{n}}_t)
            \Le[\dd f^{{\rm S};q}(\hat{\Gamma}_t)
            +
            \dd f^{{\rm CM};q}(\hat{\Gamma}_t)\Ri],
        }{}
    where $\dd f^{s;q}(\hat{\Gamma}_t)$ and $\dd f^{c,e;q}(\hat{\Gamma}_t)$ are defined by
        \eqsp{
            \dd f^{{\rm S};q}(\hat{\Gamma}_t)
            :=&
            \lambda\Delta\mbb{I}_{[1,a[\hat{\mbf{n}}^a_t])}(q)
            \mbb{I}_{[1,L]}(q)
            \Le\{f\Le(\hat{m}_t+\fr{b[\hat{\mbf{n}}^b_t]+q}{2},\hat{\mbf{n}}^a_t+\mbf{e}^q,T[\hat{\mbf{n}}^b_t,-b[\hat{\mbf{n}}^b_t]-q]\Ri)-f(\hat{m}_t,\hat{\mbf{n}}^a_t,\hat{\mbf{n}}^b_t)\Ri\}\\
            &+
            \lambda\Delta\mbb{I}_{[a[\hat{\mbf{n}}^a_t],L]}(q)\mbb{I}_{[1,L]}(q)
            \Le\{f(\hat{m}_t,\hat{\mbf{n}}^a_t+\mbf{e}^q,\hat{\mbf{n}}^b_t)
            -f(\hat{m}_t,\hat{\mbf{n}}^a_t,\hat{\mbf{n}}^b_t)\Ri\}\\
            &+
            \lambda\Delta\mbb{I}_{(b[\hat{\mbf{n}}^b_t],-1]}(q)\mbb{I}_{[-L,-1]}(q)
            \Le\{f\Le(\hat{m}_t+\frac{a[\hat{\mbf{n}}^a_t]+q}{2},T[\hat{\mbf{n}}^a_t,-a[\hat{\mbf{n}}^a_t]-q],\hat{\mbf{n}}^b_t+\mbf{e}^q\Ri)
            -f(\hat{m}_t,\hat{\mbf{n}}^a_t,\hat{\mbf{n}}^b_t)\Ri\}\\
            &+
            \lambda\Delta\mbb{I}_{[-L,b[\hat{\mbf{n}}^b_t]]}(q)\mbb{I}_{[-L,-1]}(q)
            \Le\{f(\hat{m}_t,\hat{\mbf{n}}^a_t,\hat{\mbf{n}}^b_t+\mbf{e}^q)
            -f(\hat{m}_t,\hat{\mbf{n}}^a_t,\hat{\mbf{n}}^b_t)\Ri\},
        }{}
        \eqsp{
            &\dd f^{{\rm CM};q}(\hat{\Gamma}_t)
            :=
            v\hat{n}^a_q(t)\mbb{I}_{(a[\hat{\mbf{n}}^a_t],L]}(q)\mbb{I}_{[1,L]}(q)
            \Le\{f\Le(\hat{m}_t,\hat{\mbf{n}}^a_t-\mbf{e}^q,\hat{\mbf{n}}^b_t\Ri)-f(\hat{m}_t,\hat{\mbf{n}}^a_t,\hat{\mbf{n}}^b_t)\Ri\}\\
            &+
            (\mu+v\hat{n}^a_q(t))\delta_{q,a[\hat{\mbf{n}}^a_t]}\mbb{I}_{[1,L]}(q)
            \Le\{f\Le(\hat{m}_t+\fr{a[\hat{\mbf{n}}^a_t-\mbf{e}^q]-q}{2},\hat{\mbf{n}}^a_t-\mbf{e}^q,T[\hat{\mbf{n}}^b_t,-a[\hat{\mbf{n}}^a_t-\mbf{e}^q]+q]\Ri)
            -f(\hat{m}_t,\hat{\mbf{n}}^a_t,\hat{\mbf{n}}^b_t)\Ri\}\\
            &+
            v\hat{n}^b_q(t)\mbb{I}_{[-L,b[\hat{\mbf{n}}^b_t])}(q)\mbb{I}_{[-L,-1]}(q)
            \Le\{f\Le(\hat{m}_t,\hat{\mbf{n}}^a_t,\hat{\mbf{n}}^b_t-\mbf{e}^q\Ri)-f(\hat{m}_t,\hat{\mbf{n}}^a_t,\hat{\mbf{n}}^b_t)\Ri\}\\
            &+
            (\mu+v\hat{n}^b_q(t))\delta_{q,b[\hat{\mbf{n}}^b_t]}\mbb{I}_{[-L,-1]}(q)
            \Le\{f\Le(\hat{m}_t+\fr{b[\hat{\mbf{n}}^a_t-\mbf{e}^q]-q}{2},T[\hat{\mbf{n}}^a_t,-b[\hat{\mbf{n}}^b_t-\mbf{e}^q]+q],\hat{\mbf{n}}^b_t-\mbf{e}^q\Ri)
            -f(\hat{m}_t,\hat{\mbf{n}}^a_t,\hat{\mbf{n}}^b_t)\Ri\}.
        }{}
    These contributions can be rewritten by using Kronecker's delta, such that
        \eqsp{
            \dd f^{{\rm S};q}
            =&
            \lambda\Delta\mbb{I}_{[1,a[\hat{\mbf{n}}^a_t])}(q)\mbb{I}_{[1,L]}(q)
            \sum_{\dd m}\delta_{\dd m,-F(-\fr{b[\hat{\mbf{n}}^b_t]+q}{2})}
            \Le\{f\Le(\hat{m}_t+\dd m,\hat{\mbf{n}}^a_t+\mbf{e}^q,T[\hat{\mbf{n}}^b_t,-2\dd m]\Ri)-f(\hat{m}_t,\hat{\mbf{n}}^a_t,\hat{\mbf{n}}^b_t)\Ri\}\\
            &+
            \lambda\Delta\mbb{I}_{[a[\hat{\mbf{n}}^a_t],L]}(q)\mbb{I}_{[1,L]}(q)
            \sum_{\dd m}\delta_{\dd m,-F(-\fr{b[\hat{\mbf{n}}^b_t]+q}{2})}
            \Le\{f\Le(\hat{m}_t+\dd m,\hat{\mbf{n}}^a_t+\mbf{e}^q,T[\hat{\mbf{n}}^b_t,-2\dd m]\Ri)-f(\hat{m}_t,\hat{\mbf{n}}^a_t,\hat{\mbf{n}}^b_t)\Ri\}\\
            &+
            \lambda\Delta\mbb{I}_{(b[\hat{\mbf{n}}^b_t],-1]}(q)\mbb{I}_{[-L,-1]}(q)
            \sum_{\dd m}\delta_{\dd m,F(\fr{a[\hat{\mbf{n}}^a_t]+q}{2})}
            \Le\{f\Le(\hat{m}_t+\dd m,T[\hat{\mbf{n}}^a_t,-2\dd m],\hat{\mbf{n}}^b_t+\mbf{e}^q\Ri)
            -f(\hat{m}_t,\hat{\mbf{n}}^a_t,\hat{\mbf{n}}^b_t)\Ri\}\\
            &+
            \lambda\Delta\mbb{I}_{[-L,b[\hat{\mbf{n}}^b_t]]}(q)\mbb{I}_{[-L,-1]}(q)
            \sum_{\dd m}\delta_{\dd m,F(\fr{a[\hat{\mbf{n}}^a_t]+q}{2})}
            \Le\{f\Le(\hat{m}_t+\dd m,T[\hat{\mbf{n}}^a_t,-2\dd m],\hat{\mbf{n}}^b_t+\mbf{e}^q\Ri)
            -f(\hat{m}_t,\hat{\mbf{n}}^a_t,\hat{\mbf{n}}^b_t)\Ri\}\\
            =&
            \lambda\Delta\mbb{I}_{[1,L]}(q)
            \sum_{\dd m}\delta_{\dd m,-F(-\fr{b[\hat{\mbf{n}}^b_t]+q}{2})}
            \Le\{f\Le(\hat{m}_t+\dd m,\hat{\mbf{n}}^a_t+\mbf{e}^q,T[\hat{\mbf{n}}^b_t,-2\dd m]\Ri)-f(\hat{m}_t,\hat{\mbf{n}}^a_t,\hat{\mbf{n}}^b_t)\Ri\}\\
            &+
            \lambda\Delta\mbb{I}_{[-L,-1]}(q)
            \sum_{\dd m}\delta_{\dd m,F(\fr{a[\hat{\mbf{n}}^a_t]+q}{2})}
            \Le\{f\Le(\hat{m}_t+\dd m,T[\hat{\mbf{n}}^a_t,-2\dd m],\hat{\mbf{n}}^b_t+\mbf{e}^q\Ri)
            -f(\hat{m}_t,\hat{\mbf{n}}^a_t,\hat{\mbf{n}}^b_t)\Ri\},
        }{}
        \eqsp{
            &\dd f^{{\rm CM};q}
            =
            v\hat{n}^a_q(t)\mbb{I}_{(a[\hat{\mbf{n}}^a_t],L]}(q)\mbb{I}_{[1,L]}(q)
            \sum_{\dd m}\delta_{\dd m,F(\fr{a[\hat{\mbf{n}}^a_t-\mbf{e}^q]-q}{2})}
            \Le\{f\Le(\hat{m}_t+\dd m,\hat{\mbf{n}}^a_t-\mbf{e}^q,T[\hat{\mbf{n}}^b_t,-2\dd m]\Ri)
            -f(\hat{m}_t,\hat{\mbf{n}}^a_t,\hat{\mbf{n}}^b_t)\Ri\}\\
            &+
            (\mu+v\hat{n}^a_q(t))\delta_{q,a[\hat{\mbf{n}}^a_t]}\mbb{I}_{[1,L]}(q)
            \sum_{\dd m}\delta_{\dd m,F(\fr{a[\hat{\mbf{n}}^a_t-\mbf{e}^q]-q}{2})}
            \Le\{f\Le(\hat{m}_t+\dd m,\hat{\mbf{n}}^a_t-\mbf{e}^q,T[\hat{\mbf{n}}^b_t,-2\dd m]\Ri)
            -f(\hat{m}_t,\hat{\mbf{n}}^a_t,\hat{\mbf{n}}^b_t)\Ri\}\\
            &+
            v\hat{n}^b_q(t)\mbb{I}_{[-L,b[\hat{\mbf{n}}^b_t])}(q)\mbb{I}_{[-L,-1]}(q)
            \sum_{\dd m}\delta_{\dd m,-F(-\fr{b[\hat{\mbf{n}}^b_t-\mbf{e}^q]-q}{2})}
            \Le\{f\Le(\hat{m}_t+\dd m,T[\hat{\mbf{n}}^a_t,-2\dd m],\hat{\mbf{n}}^b_t-\mbf{e}^q\Ri)
            -f(\hat{m}_t,\hat{\mbf{n}}^a_t,\hat{\mbf{n}}^b_t)\Ri\}\\
            &+
            (\mu+v\hat{n}^b_q(t))\delta_{q,b[\hat{\mbf{n}}^b_t]}\mbb{I}_{[-L,-1]}(q)
            \sum_{\dd m}\delta_{\dd m,-F(-\fr{b[\hat{\mbf{n}}^b_t-\mbf{e}^q]-q}{2})}
            \Le\{f\Le(\hat{m}_t+\dd m,T[\hat{\mbf{n}}^a_t,-2\dd m],\hat{\mbf{n}}^b_t-\mbf{e}^q\Ri)
            -f(\hat{m}_t,\hat{\mbf{n}}^a_t,\hat{\mbf{n}}^b_t)\Ri\}\\
            =&
            (\mu\delta_{q,a[\hat{\mbf{n}}^a_t]}+v\hat{n}^a_q(t))\mbb{I}_{[1,L]}(q)
            \sum_{\dd m}\delta_{\dd m,F(\fr{a[\hat{\mbf{n}}^a_t-\mbf{e}^q]-q}{2})}
            \Le\{f\Le(\hat{m}_t+\dd m,\hat{\mbf{n}}^a_t-\mbf{e}^q,T[\hat{\mbf{n}}^b_t,-2\dd m]\Ri)
            -f(\hat{m}_t,\hat{\mbf{n}}^a_t,\hat{\mbf{n}}^b_t)\Ri\}\\
            &+
            (\mu\delta_{q,b[\hat{\mbf{n}}^b_t]}+v\hat{n}^b_q(t))\mbb{I}_{[-L,-1]}(q)
            \sum_{\dd m}\delta_{\dd m,-F(-\fr{b[\hat{\mbf{n}}^b_t-\mbf{e}^q]-q}{2})}
            \Le\{f\Le(\hat{m}_t+\dd m,T[\hat{\mbf{n}}^a_t,-2\dd m],\hat{\mbf{n}}^b_t-\mbf{e}^q\Ri)
            -f(\hat{m}_t,\hat{\mbf{n}}^a_t,\hat{\mbf{n}}^b_t)\Ri\},
        }{}
    where we defined the filter function $F(x)$
    \eqsp{
            F(x):=
            x\mbb{I}_{(0,\infty)}(x)
            =
            \begin{cases}
                x&(x>0)\\
                0&(x\leq0)
            \end{cases}
        }{}
    and used the following relation
        \eqsp{
            (\mu\delta_{q,a[\hat{\mbf{n}}^a_t]}+v\hat{n}^a_q(t))\mbb{I}_{[1,a[\hat{\mbf{n}}^a_t])}(q)=0,\quad
            (\mu\delta_{q,b[\hat{\mbf{n}}^b_t]}+v\hat{n}^b_q(t))\mbb{I}_{(b[\hat{\mbf{n}}^a_t],-1]}(q)=0.
        }{} 
    We then replace the dummy variable $\hat{\Gamma}_t$ with $\hat{\Gamma}_t\to \hat{\Gamma}:=(\hat{m},\hat{\mbf{n}}^a,\hat{\mbf{n}}^b)$. Also, we define 
        \eqsp{
            \hat{\Gamma}^{{\rm S};a}_{q,\dd m}
            :=&(\hat{m}+\dd m,\hat{\mbf{n}}^a+\mbf{e}^q,T[\hat{\mbf{n}}^b,-2\dd m]),\quad
            \hat{\Gamma}^{{\rm S};b}_{q,\dd m}
            :=(\hat{m}+\dd m,T[\hat{\mbf{n}}^a,-2\dd m],\hat{\mbf{n}}^b+\mbf{e}^q)\\
            \hat{\Gamma}^{{\rm CM};a}_{q,\dd m}
            :=&(\hat{m}+\dd m,\hat{\mbf{n}}^a-\mbf{e}^q,T[\hat{\mbf{n}}^b,-2\dd m]),\quad
            \hat{\Gamma}^{{\rm CM};b}_{q,\dd m}
            :=(\hat{m}+\dd m,T[\hat{\mbf{n}}^a,-2\dd m],\hat{\mbf{n}}^b-\mbf{e}^q).
        }{}
    Finally, we obtain Eq.~\eqref{eq:f_time_ev}.

\section{Exact boundary condition}\label{sec:App_bound_condition}
    We derive the boundary condition~\eqref{eq:boundary_condition_secmain} at $r\to0$ from the BBGKY hierarchical equation \eqref{eq:BBGKY}. Let us start our calculation from 
        \eqsp{
            0
            =&
            \lambda\Delta 
            \Le\{P^a_{\rm st}(0,1)-P^a_{\rm st}(1,1)\Ri\}
            +
            v\Le\{2P^a_{\rm st}(2,1)-P^a_{\rm st}(1,1)\Ri\}
            +
            \mu\sum_{\hat{\Gamma}}P_{\rm st}(\hat{\Gamma})
            \delta_{1,a[\hat{\mbf{n}}^a]}\Le(\delta_{\hat{n}^a_1,2}-\delta_{\hat{n}^a_1,1}\Ri)\\
            &+\sum_{\hat{\Gamma}}\sum_{q,\dd m}P_{\rm st}(\hat{\Gamma})
            \Le\{w^{{\rm S};b}_{q,\dd m}[\hat{\mbf{n}}]\Le(\delta_{\hat{n}^a_{1+2\dd m},1}-\delta_{\hat{n}^a_1,1}\Ri)
            +w^{{\rm CM};b}_{q,\dd m}[\hat{\mbf{n}}]\Le(\delta_{\hat{n}^a_{1+2\dd m},1}-\delta_{\hat{n}^a_1,1}\Ri)\Ri\},\\
        }{}
    which is the BBGKY hierarchical equation in the steady state with $n=1$ and $k=1$. We evaluate each term one by one. 

    First, we evaluate the term $\mu\sum_{\hat{\Gamma}}P_{\rm st}(\hat{\Gamma})\delta_{1,a[\hat{\mbf{n}}^a]}\Le(\delta_{\hat{n}^a_1,2}-\delta_{\hat{n}^a_1,1}\Ri)$. Trivially, the relation $\hat{n}^a_1\neq 0\Rightarrow \delta_{1,a[\hat{\mbf{n}}^a]}=1$ holds because the best ask price exists at $r=1$ if $\hat{n}^a_1\neq 0$. We then have a simplified relation, 
        \eqsp{
            \mu\sum_{\hat{\Gamma}}P_{\rm st}(\hat{\Gamma})\delta_{1,a[\hat{\mbf{n}}^a]}\Le(\delta_{\hat{n}^a_1,2}-\delta_{\hat{n}^a_1,1}\Ri)
            =\mu\sum_{\hat{\Gamma}}P_{\rm st}(\hat{\Gamma})\Le(\delta_{\hat{n}^a_1,2}-\delta_{\hat{n}^a_1,1}\Ri)
            =\mu\Le\{P^a_{\rm st}(2,1)-P^a_{\rm st}(1,1)\Ri\}.
        }{}        
    We next evaluate the sum regarding $w^{{\rm S};b}_{q,\dd m}[\hat{\mbf{n}}]$ as follows:
        \eqsp{
            &\sum_{\hat{\Gamma}}\sum_{q,\dd m}P_{\rm st}(\hat{\Gamma})
            w^{{\rm S};b}_{q,\dd m}[\hat{\mbf{n}}]\Le(\delta_{\hat{n}^a_{1+2\dd m},1}-\delta_{\hat{n}^a_1,1}\Ri)\\
            =&\sum_{\hat{\Gamma}}\sum_{q,\dd m}P_{\rm st}(\hat{\Gamma})
            \lambda\Delta\mbb{I}_{(-\infty,-1]}(q)\delta_{\dd m,F(\fr{a[\hat{\mbf{n}}^a]+q}{2})}\Le(\delta_{\hat{n}^a_{1+2\dd m},1}-\delta_{\hat{n}^a_1,1}\Ri)\\
            =&\lambda\Delta\sum_{\hat{\Gamma}}\sum_{\dd m}P_{\rm st}(\hat{\Gamma})\sum_{q\in\mbb{Z}}\mbb{I}_{(-a[\hat{\mbf{n}}^a],-1]}(q)
            \delta_{\dd m,\fr{a[\hat{\mbf{n}}^a]+q}{2}}\Le(\delta_{\hat{n}^a_{1+2\dd m},1}-\delta_{\hat{n}^a_1,1}\Ri)\\
            =&\lambda\Delta\sum_{\hat{\Gamma}}P_{\rm st}(\hat{\Gamma})
            \Le(\sum_{q\in\mbb{Z}}\mbb{I}_{(-a[\hat{\mbf{n}}^a],-1]}(q)\delta_{\hat{n}^a_{1+a[\hat{\mbf{n}}^a]+q},1}-\delta_{\hat{n}^a_1,1}\sum_{q\in\mbb{Z}}\mbb{I}_{(-a[\hat{\mbf{n}}^a],-1]}(q)\Ri)\\
            =&\lambda\Delta\sum_{\hat{\Gamma}}P_{\rm st}(\hat{\Gamma})
            \sum_{q\in\mbb{Z}}\mbb{I}_{(-a[\hat{\mbf{n}}^a],-1]}(q)\delta_{\hat{n}^a_{1+a[\hat{\mbf{n}}^a]+q},1},
        }{}
    where we use $\delta_{\hat{n}^a_1,1}\mbb{I}_{(-a[\hat{\mbf{n}}^a],-1]}(q)=0$ in the fourth line because $a[\hat{\mbf{n}}^a]=1$ if $\hat{n}^a_1=1$. Here, the following identity\footnote{ 
        For the case with $a[\hat{\mbf{n}}^a]=1$, this identity holds because $\mbb{I}_{(-a[\hat{\mbf{n}}^a],-1]}(q)$ is zero. For the case with $a[\hat{\mbf{n}}^a]\neq 1$, we obtain
            \eqsp{
                \sum_{q\in\mbb{Z}}\mbb{I}_{(-a[\hat{\mbf{n}}^a],-1]}(q)\delta_{\hat{n}^a_{1+a[\hat{\mbf{n}}^a]+q},1}
                =
                \underbrace{\delta_{n^a_{2},1}+\dots +\delta_{n^a_{a[\hat{\mbf{n}}^a]-1},1}}_{=0}+\delta_{n^a_{a[\hat{\mbf{n}}^a]},1}
                = \delta_{\hat{n}^a_{a[\hat{\mbf{n}}^a]},1},
            }{}    
        because there is no ask orders below the best ask price, such that $i<a[\hat{\mbf{n}}^a]\Rightarrow \hat{n}^a_i=0$. Also, we introduce an indicator function $\delta_{\hat{n}^a_1,0}$ based on the following fact: $a[\hat{\mbf{n}}^a]\neq1 \Leftrightarrow \hat{n}^a_1=0$.
    } holds
        \eqsp{
            \sum_{q\in\mbb{Z}}\mbb{I}_{(-a[\hat{\mbf{n}}^a],-1]}(q)\delta_{\hat{n}^a_{1+a[\hat{\mbf{n}}^a]+q},1}
            =
            \begin{cases}
                0 & (a[\hat{\mbf{n}}^a]=1)\\
                \delta_{\hat{n}^a_{a[\hat{\mbf{n}}^a]},1} & (a[\hat{\mbf{n}}^a]\neq1)
            \end{cases}
            =\delta_{\hat{n}^a_1,0}\delta_{\hat{n}^a_{a[\hat{\mbf{n}}^a]},1}.
        }{}
    Therefore, we obtain 
        \eqsp{
            \sum_{\hat{\Gamma}}\sum_{q,\dd m}P_{\rm st}(\hat{\Gamma})
            w^{{\rm S};b}_{q,\dd m}[\hat{\mbf{n}}]\Le(\delta_{\hat{n}^a_{1+2\dd m},1}-\delta_{\hat{n}^a_1,1}\Ri) = \lambda\Delta\sum_{\hat{\Gamma}}P_{\rm st}(\hat{\Gamma})\delta_{\hat{n}^a_1,0}\delta_{\hat{n}^a_{a[\hat{\mbf{n}}^a]},1}.
        }{}

    We also evaluate the sum regarding $w^{{\rm CM};b}_{q,\dd m}[\hat{\mbf{n}}]$ as follows:
        \eqsp{
            &\sum_{\hat{\Gamma}}\sum_{q,\dd m}P_{\rm st}(\hat{\Gamma})
            w^{{\rm CM};b}_{q,\dd m}[\hat{\mbf{n}}]\Le(\delta_{\hat{n}^a_{1+2\dd m},1}-\delta_{\hat{n}^a_1,1}\Ri)\\
            =&
            \sum_{\hat{\Gamma}}\sum_{q,\dd m}P_{\rm st}(\hat{\Gamma})
            (\mu\delta_{q,b[\hat{\mbf{n}}^b]}+v\hat{n}^b_q)\mbb{I}_{(-\infty,-1]}(q)\delta_{\dd m,-F(-\fr{b[\hat{\mbf{n}}^b-\mbf{e}^q]-q}{2})}\Le(\delta_{\hat{n}^a_{1+2\dd m},1}-\delta_{\hat{n}^a_1,1}\Ri)\\
            =&
            \sum_{\hat{\Gamma}}\sum_{q,\dd m}P_{\rm st}(\hat{\Gamma})
            (\mu\delta_{q,b[\hat{\mbf{n}}^b]}+v\hat{n}^b_q)\mbb{I}_{(b[\hat{\mbf{n}}^b-\mbf{e}^q],-1]}(q)\delta_{\dd m,\fr{b[\hat{\mbf{n}}^b-\mbf{e}^q]-q}{2}}\Le(\delta_{\hat{n}^a_{1+2\dd m},1}-\delta_{\hat{n}^a_1,1}\Ri)\\
            =&
            \sum_{\hat{\Gamma}}\sum_{q}P_{\rm st}(\hat{\Gamma})
            (\mu\delta_{q,b[\hat{\mbf{n}}^b]}+v\hat{n}^b_q)\mbb{I}_{(b[\hat{\mbf{n}}^b-\mbf{e}^q],-1]}(q)\Le(\delta_{\hat{n}^a_{1+b[\hat{\mbf{n}}^b-\mbf{e}^q]-q},1}-\delta_{\hat{n}^a_1,1}\Ri)\\
            =&
            -\sum_{\hat{\Gamma}}\sum_{q}P_{\rm st}(\hat{\Gamma})
            (\mu\delta_{q,b[\hat{\mbf{n}}^b]}+v\hat{n}^b_q)\mbb{I}_{(b[\hat{\mbf{n}}^b-\mbf{e}^q],b[\hat{\mbf{n}}^b]]}(q)\delta_{\hat{n}^a_1,1}\\
            =&
            -\sum_{\hat{\Gamma}}P_{\rm st}(\hat{\Gamma})
            (\mu+v\hat{n}^b_{b[\hat{\mbf{n}}^b]})\delta_{\hat{n}^a_1,1}.
        }{}
    From the second to the third line, we remove the contribution for $q\in (-\infty, b[\hat{\mbf{n}}^b-\mbf{e}^q]]$ because $F(-(b[\hat{\mbf{n}}^b-\mbf{e}^q]-q)/2)=0$ for $q<b[\hat{\mbf{n}}^b-\mbf{e}^q]$ and $\delta_{\hat{n}^a_{1+2\dd m},1}$ is equal to $\delta_{\hat{n}^a_1,1}=0$ for $\dd m=0$. From the fourth to the fifth line, we use two relations: $\mbb{I}_{(b[\hat{\mbf{n}}^b-\mbf{e}^q],-1]}(q)\delta_{\hat{n}^a_{1+b[\hat{\mbf{n}}^b-\mbf{e}^q]-q},1}=0$ because $\hat{n}^a_{1+b[\hat{\mbf{n}}^b-\mbf{e}^q]-q}=1$ only for $q\leq b[\hat{\mbf{n}}^b-\mbf{e}^q]$. Also, $\mbb{I}_{(b[\hat{\mbf{n}}^b-\mbf{e}^q],-1]}(q)\delta_{\hat{n}^a_1,1}=\mbb{I}_{(b[\hat{\mbf{n}}^b-\mbf{e}^q],b[\hat{\mbf{n}}^b]]}(q)\delta_{\hat{n}^a_1,1}$ because $b[\hat{\mbf{n}}^b]=-a[\hat{\mbf{n}}^b]=-1$ if $\hat{n}^a_1=1$.

    In summary, we obtain an exact boundary condition as 
        \eqsp{
            0
            =&
            \lambda\Delta 
            \Le\{P^a_{\rm st}(0,1)-P^a_{\rm st}(1,1)\Ri\}
            +
            v\Le\{2P^a_{\rm st}(2,1)-P^a_{\rm st}(1,1)\Ri\}
            +
            \mu\Le\{P^a_{\rm st}(2,1)-P^a_{\rm st}(1,1)\Ri\}\\
            &+\lambda\Delta \sum_{\hat{\Gamma}}P_{\rm st}(\hat{\Gamma})\delta_{\hat{n}^a_1,0}\delta_{\hat{n}^a_{a[\hat{\mbf{n}}^a]},1}
            -\sum_{\hat{\Gamma}}P_{\rm st}(\hat{\Gamma})
            (\mu+v\hat{n}^b_{b[\hat{\mbf{n}}^b]})\delta_{\hat{n}^a_1,1}.\\
        }{eq:rigorous_bound_small_tick_off}
    For the small tick-size limit, the total number of orders at each price is either zero or one. This implies $\hat{n}^a_{a[\hat{\mbf{n}}^a]}=1$ and $\hat{n}^b_{b[\hat{\mbf{n}}^b]}=1$ always for $\Delta \to 0$. Thus, Eq.~\eqref{eq:rigorous_bound_small_tick_off} is rewritten as 
        \eqsp{
            0
            =&
            \lambda\Delta 
            \Le\{P^a_{\rm st}(0,1)-P^a_{\rm st}(1,1)\Ri\}
            -
            (v+\mu)P^a_{\rm st}(1,1)+\lambda\Delta P^a_{\rm st}(0,1)-P^a_{\rm st}(1,1)(\mu+v)
            +o(\Delta)\\
            =&2\lambda\Delta
            -2\rho^a_{\rm st}(\Delta)(\mu+v)\Delta+o(\Delta),
        }{}
    which implies the exact boundary condition~\eqref{eq:boundary_condition_secmain} for the limit $\Delta\to 0$.

\section{Derivation of the mean-field equation for the order-book profile \eqref{eq:Boltzmann}}\label{section:App_density_eq}
    In this appendix, we derive the Boltzmann equation for the order-book profile as the mean-field approximation and by taking the small tick-size limit.
    \subsection{Evaluation of the market-order contribution}
        Let us evaluate the contribution by market orders 
        \eqsp{
            \mu\sum_{\hat{\mbf{n}}^a}\Le\{\prod_{i}P^a_t(\hat{n}^a_i,i)\Ri\}
            \delta_{k,a[\hat{\mbf{n}}^a]}\Le(\delta_{\hat{n}^a_k,2}-\delta_{\hat{n}^a_k,1}\Ri).
        }{eq:exe_term}
        In the small tick-size limit, we can drop $\delta_{\hat{n}^a_k,2}$ because the probability that  $\hat{n}^a_k=2$ is negligible. According to Appendix~\ref{subsec:I_gutairei}, $\delta_{k,a[\hat{\mbf{n}}^a]}$ can be rewritten as 
            \eqsp{
                \delta_{k,a[\hat{\mbf{n}}^a]}
                =(1-\delta_{\hat{n}^a_k,0})\prod^{k-1}_{j=-\infty}\delta_{\hat{n}^a_j,0}
                \simeq \delta_{\hat{n}^a_k,1}\prod^{k-1}_{j=-\infty}\delta_{\hat{n}^a_j,0}.
            }{}
        Therefore, the contribution by market orders is evaluated as
            \eqsp{
                \mu\sum_{\hat{\mbf{n}}^a}\Le\{\prod_{i}P^a_t(\hat{n}^a_i,i)\Ri\}
                \delta_{k,a[\hat{\mbf{n}}^a]}\Le(\delta_{\hat{n}^a_k,2}-\delta_{\hat{n}^a_k,1}\Ri)
                \simeq &
                -\mu\sum_{\hat{\mbf{n}}^a}\Le\{\prod_{i}P^a_t(\hat{n}^a_i,i)\Ri\}
                \delta_{\hat{n}^a_k,1}\prod^{k-1}_{j=-\infty}\delta_{\hat{n}^a_j,0}\\
                \simeq&
                -\mu \rho^a_t(k\Delta)\Delta \prod^{k-1}_{j=1}(1-\rho^a_t(j\Delta)\Delta)\\
                \simeq&-\mu \rho^a_t(k\Delta)\Delta e^{-\Delta\sum^{k-1}_{j=1}\rho^a_t(j\Delta)},
            }{}
        where we use the techniques for extreme value statistics in the Appendix~\ref{sec:App_kyoku}.
        
    \subsection{Evaluation of the jump term $W(h)$}
        Let us rewrite $W(h)$ using the indicator functions $\delta_{s,b[\hat{\mbf{n}}^b]}$ and $\delta_{s,b[\hat{\mbf{n}}^b-\mbf{e}^q]}$:
            \eqsp{
                W(h)&
                =\sum_{\hat{\mbf{n}}^b}\sum_{q,s}\left\{\prod_{i}P^b_t(\hat{n}^b_{i},i)\right\}
                \lambda\Delta\mbb{I}_{(-\infty,-1]}(q)\mbb{I}_{(-\infty,q)}(s)\delta_{h,s-q}\delta_{s,b[\hat{\mbf{n}}^b]}\\
                &+\sum_{\hat{\mbf{n}}^b}\sum_{q,s}\left\{\prod_{i}P^b_t(\hat{n}^b_{i},i)\right\}
                (\mu\delta_{q,b[\hat{\mbf{n}}^b]}+v\hat{n}^b_q)\mbb{I}_{(-\infty,-1]}(q)\mbb{I}_{(-\infty,q)}(s)\delta_{h,q-s}\delta_{s,b[\hat{\mbf{n}}^b-\mbf{e}^q]}.
            }{}
        By the way, the following identity holds\footnote{
            For the case with $\hat{n}^b_q=0$, both sides are zero. Also, for the case with $\hat{n}^b_q\neq 0$, the proposition holds true
            \eqsp{
                \hat{n}^b_q\neq 0,\quad s<q\quad \text{and}\quad s=b[\hat{\mbf{n}}^b-\mbf{e}^q]
                \Rightarrow
                q=b[\hat{\mbf{n}}^b].
            }{}
        }
            \eqsp{
                \hat{n}^b_q\mbb{I}_{(-\infty,q)}(s)\delta_{s,b[\hat{\mbf{n}}^b-\mbf{e}^q]}
                =
                \delta_{q,b[\hat{\mbf{n}}^b]}\hat{n}^b_q\mbb{I}_{(-\infty,q)}(s)\delta_{s,b[\hat{\mbf{n}}^b-\mbf{e}^q]}.
            }{} 
            Therefore, $W(h)$ can be rewritten as 
            \eqsp{
                W(h)&
                =\sum_{\hat{\mbf{n}}^b}\sum_{q,s}\left\{\prod_{i}P^b_t(\hat{n}^b_{i},i)\right\}
                \lambda\Delta\mbb{I}_{(-\infty,-1]}(q)\mbb{I}_{(-\infty,q)}(s)\delta_{h,s-q}\delta_{s,b[\hat{\mbf{n}}^b]}\\
                &+\sum_{\hat{\mbf{n}}^b}\sum_{q,s}\left\{\prod_{i}P^b_t(\hat{n}^b_{i},i)\right\}
                (\mu+v\hat{n}^b_q)\mbb{I}_{(-\infty,-1]}(q)\mbb{I}_{(-\infty,q)}(s)\delta_{h,q-s}\delta_{q,b[\hat{\mbf{n}}^b]}\delta_{s,b[\hat{\mbf{n}}^b-\mbf{e}^q]}.\\
            }{}
        In the small tick-size limit, $\delta_{s,b[\hat{\mbf{n}}^b]}$ and $\delta_{q,b[\hat{\mbf{n}}^b]}\delta_{s,b[\hat{\mbf{n}}^b-\mbf{e}^q]}$ can be rewritten as 
            \eqsp{
                \delta_{s,b[\hat{\mbf{n}}^b]}
                =(1-\delta_{\hat{n}^b_s,0})\prod^{-1}_{l=s+1}\delta_{\hat{n}^b_l,0}
                \simeq \delta_{\hat{n}^b_s,1}\prod^{-1}_{l=s+1}\delta_{\hat{n}^b_l,0},
            }{eq:ind_best}
            \eqsp{
                \delta_{q,b[\hat{\mbf{n}}^b]}\delta_{s,b[\hat{\mbf{n}}^b-\mbf{e}^q]}
                =(1-\delta_{\hat{n}^b_q,0})(1-\delta_{\hat{n}^b_s,1})
                \left\{\prod^{-1}_{l=q+1}\delta_{\hat{n}^b_l,0}\right\}\prod^{q-1}_{l'=s+1}\delta_{\hat{n}^b_{l'},0}
                \simeq \delta_{\hat{n}^b_q,1}\delta_{\hat{n}^b_s,1}
                \prod^{-1}_{\substack{l=s+1 \\ l\neq q}}\delta_{\hat{n}^b_{l},0}
            }{eq:ind_best_nextbest}
        Therefore, we calculate $W(h)$ using the small tick limit and, rewrite using $\rho^b_t(q\Delta)$.
            \eqsp{
                W(h)
                =&\sum_{\hat{\mbf{n}}^b}\sum_{q,s}\left\{\prod_{i}P^b_t(\hat{n}^b_{i},i)\right\}\lambda\Delta
                 \mbb{I}_{(-\infty,-1]}(q)\mbb{I}_{(-\infty,q)}(s) \delta_{h,s-q}\delta_{\hat{n}^b_s,1}\prod^{-1}_{l=s+1}\delta_{\hat{n}^b_l,0}\\
                &+\sum_{\hat{\mbf{n}}^b}\sum_{q,s}\left\{\prod_{i}P^b_t(\hat{n}^b_{i},i)\right\}
                (\mu+v\hat{n}^b_q)\mbb{I}_{(-\infty,-1]}(q)\mbb{I}_{(-\infty,q)}(s)\delta_{h,q-s}\delta_{\hat{n}^b_q,1}\delta_{\hat{n}^b_s,1}\prod^{-1}_{\substack{l=s+1 \\ l\neq q}}\delta_{\hat{n}^b_{l},0}\\
                \simeq &\sum_{q,s}
                \lambda\Delta\delta_{h,s-q}\rho^b_t(s\Delta)\Delta \mbb{I}_{(-\infty,-1]}(q)\mbb{I}_{(-\infty,q)}(s)\prod^{-1}_{l=s+1}(1-\rho^b_t(l\Delta)\Delta)\\
                &+\sum_{q,s}
                (\mu+v)\mbb{I}_{(-\infty,-1]}(q)\mbb{I}_{(-\infty,q)}(s)\delta_{h,q-s}\rho^b_t(q\Delta)\Delta\rho^b_t(s\Delta)\Delta\prod^{-1}_{\substack{l=s+1 \\ l\neq q}}(1-\rho^b_t(l\Delta)\Delta).
            }{}
        By using the techniques in extreme value statistics as described in \eqref{eq:best_kyoku} in the Appendix~\ref{sec:App_kyoku}, we obtain 
            \eqsp{
                W(h)
                \simeq&\sum_{q,s}
                \lambda\Delta^2\delta_{h,s-q}\rho^b_t(s\Delta)e^{-\Delta\sum^{-1}_{l=s+1}\rho^b_t(l\Delta)}\mbb{I}_{(-\infty,-1]}(q)\mbb{I}_{(-\infty,q)}(s)\\
                &+\sum_{q,s}
                (\mu+v)\Delta^2\delta_{h,q-s}\rho^b_t(q\Delta)\rho^b_t(s\Delta)e^{-\Delta\sum^{-1}_{l=q+1}\rho^b_t(l\Delta)-\Delta\sum^{q-1}_{l'=s+1}\rho^b_t(l'\Delta)}\mbb{I}_{(-\infty,-1]}(q)\mbb{I}_{(-\infty,q)}(s).\\
            }{}
        By applying the variable transformation $q\to -q$ and $s\to -s$ and by using the symmetry $\rho^a_t(x)=\rho^b_t(-x)$, $W(h)$ can be rewritten as 
            \eqsp{
                W(h)
                \simeq&\sum_{q,s}
                \lambda\Delta^2\delta_{h,q-s}\rho^a_t(s\Delta)e^{-\Delta\sum^{s-1}_{l=1}\rho^a_t(l\Delta)}\mbb{I}_{[1,\infty)}(q)\mbb{I}_{(q,\infty)}(s)\\
                &+\sum_{q,s}
                (\mu+v)\Delta^2\delta_{h,s-q}\rho^a_t(q\Delta)\rho^a_t(s\Delta)e^{-\Delta\sum^{q-1}_{l=1}\rho^a_t(l\Delta)-\Delta\sum^{s-1}_{l'=q+1}\rho^a_t(l'\Delta)}\mbb{I}_{[1,\infty)}(q)\mbb{I}_{(q,\infty)}(s)\\
                =&\lambda\Delta^2\sum_{q}
                \rho^a_t(q\Delta-h\Delta)e^{-\Delta\sum^{q-h-1}_{l=1}\rho^a_t(l\Delta)}\mbb{I}_{[1,\infty)}(q)\mbb{I}_{(-\infty,-1]}(h)\\
                &+(\mu+v)\Delta^2\sum_{q}
                \rho^a_t(q\Delta)\rho^a_t(q\Delta+h\Delta)e^{-\Delta\sum^{q-1}_{l=1}\rho^a_t(l\Delta)-\Delta\sum^{q+h-1}_{l'=q+1}\rho^a_t(l'\Delta)}\mbb{I}_{[1,\infty)}(q)\mbb{I}_{[1,\infty)}(h).
            }{}

    \subsection{Continuum limit}
        In summary, by dividing both hand sides by $\Delta$, the Boltzmann equation is simplified as
            \eqsp{
                \pdiff{\rho^a_t(k\Delta)}{t}
                =&
                \lambda 
                \Le\{1-2\rho^a_t(k\Delta)\Delta\Ri\}
                -
                v\rho^a_t(k\Delta)
                -\mu \rho^a_t(k\Delta)e^{-\Delta\sum^{k-1}_{j=1}\rho^a_t(j\Delta)}
                +\sum_{h}W(h)\Le\{\rho^a_t(k\Delta-h\Delta)-\rho^a_t(k\Delta)\Ri\}.
            }{}
        By taking the continuous limit $\Delta\to 0$, we replace the integer labels $k$ and $h$ with the continuous labels $r:= k\Delta$ and $y:=h\Delta$ (i.e., the relative continuous price and price jump size). Also, we replace the discrete sum with the integral: 
            \eqsp{
                \Delta\sum^k_{l=1}\rho^a_t(l\Delta)\to\INT{0}{r}{x}\rho^a_t(x),\quad 
                \Delta\sum_{h} [\dots ] \to\INT{-\infty}{\infty}{y} [\dots].
            }{}
        Therefore, the continuous form of the Boltzmann equation is given by 
            \eqsp{
                \pdiff{\rho^a_t(r)}{t}
                \simeq&\lambda-\rho^a_t(r)(v+\mu e^{-\INT{0}{r}{x}\rho^a_t(x)})
                +
                \INT{-\infty}{\infty}{y}\Tilde{W}(y)\Le\{\rho^a_t(r-y)-\rho^a_t(r)\Ri\}\\
            }{}
        with the jump intensity density $\Tilde{W}(y)$ defined by 
            \eqsp{
                \tilde{W}(y)
                =&
                \mbb{I}_{(-\infty,0)}(y)\lambda\INT{0}{\infty}{z}
                \rho^a_t(z-y)e^{-\INT{0}{z-y}{x}\rho^a_t(x)}
                +\mbb{I}_{(0,\infty)}(y)(\mu+v)\INT{0}{\infty}{z}
                \rho^a_t(z)\rho^a_t(z+y)e^{-\INT{0}{z+y}{x}\rho^a_t(x)}.
            }{}

    \subsection{Kramers-Moyal coefficients}
        We next study the explicit analytical forms of the Kramers-Moyal coefficients. 
        \subsubsection{First-order Kramers-Moyal coefficient}\label{app:1st-KM-calc}
                Here we explicitly show $A:=\int_{-\infty}^\infty y\Tilde{W}(y)\dd y = 0$ as follows:
                \eqsp{
                    A
                    =&\lambda\INT{-\infty}{0}{y}y
                    \INT{0}{\infty}{z}
                    \rho^a_{st}(z-y)e^{-\INT{0}{z-y}{x}\rho^a_{st}(x)}
                    +(\mu+v)\INT{0}{\infty}{y}y
                    \INT{0}{\infty}{z}\rho^a_{st}(z)\rho^a_{st}(z+y)e^{-\INT{0}{z+y}{x}\rho^a_{st}(x)}\\
                    =&-\lambda\INT{-\infty}{0}{y}y
                    \INT{0}{\infty}{z}
                    \fr{\partial}{\partial z}e^{-\INT{0}{z-y}{x}\rho^a_{st}(x)}
                    -(\mu+v)\INT{0}{\infty}{y}y
                    \INT{0}{\infty}{z}
                    \rho^a_{st}(z)
                    \fr{\partial}{\partial y}e^{-\INT{0}{z+y}{x}\rho^a_{st}(x)}\\
                    =&
                    -\lambda\INT{0}{\infty}{y}y
                    e^{-\INT{0}{y}{x}\rho^a_{st}(x)}
                    +(\mu+v)\INT{0}{\infty}{y}
                    \INT{0}{\infty}{z}
                    \rho^a_{st}(z)
                    e^{-\INT{0}{z+y}{x}\rho^a_{st}(x)}\\
                    \simeq&
                    -\lambda\INT{0}{\infty}{y}y
                    e^{-\rho^a_{st}(0)y}
                    +(\mu+v)\rho^a_{st}(0)\INT{0}{\infty}{y}
                    \INT{0}{\infty}{z}
                    e^{-\rho^{a}_{st}(0)(z+y)}
                    =-\fr{\lambda}{\rho^a_{st}(0)^2}+\fr{\mu+v}{\rho^a_{st}(0)}=0.
                }{}

        \subsubsection{Second-order Kramers-Moyal coefficient}\label{app:diffusion-const-calc}
            We also study the second-order Kramers-Moyal coefficient, or equivalently the diffusion constant:
               \eqsp{
                    D
                    :=&
                    \fr{1}{2}\INT{-\infty}{\infty}{y}y^2\Tilde{W}(y\mid\infty)\\
                    =&
                    \fr{\lambda}{2}\INT{-\infty}{0}{y}y^2\INT{0}{\infty}{z}
                    \rho^a_t(z-y)e^{-\INT{0}{z-y}{x}\rho^a_t(x)}
                    +\fr{\mu+v}{2}\INT{0}{\infty}{y}y^2
                    \INT{0}{\infty}{z}
                    \rho^a_t(z)\rho^a_t(z+y)e^{-\INT{0}{z+y}{x}\rho^a_t(x)}.
                }{}
            We use $\rho^a_{st}(x)$ for the calculation.
                \eqsp{
                    D
                    \simeq&
                    -\fr{\lambda}{2}\INT{-\infty}{0}{y}y^2\INT{0}{\infty}{z}
                    \fr{\partial}{\partial z}e^{-\INT{0}{z-y}{x}\rho^a_{st}(x)}
                    -\fr{\mu+v}{2}\INT{0}{\infty}{y}y^2
                    \INT{0}{\infty}{z}\rho^a_{st}(z)
                    \fr{\partial}{\partial y}e^{-\INT{0}{z+y}{x}\rho^a_{st}(x)}\\
                    \simeq&
                    \fr{\lambda}{2}\INT{0}{\infty}{y}y^2
                    e^{-\INT{0}{y}{x}\rho^a_{st}(x)}
                    +
                    (\mu+v)\INT{0}{\infty}{y}y
                    \INT{0}{\infty}{z}\rho^a_{st}(z)
                    e^{-\INT{0}{z+y}{x}\rho^a_{st}(x)}\\
                    \simeq&
                    \fr{\lambda}{2}\INT{0}{\infty}{y}y^2
                    e^{-\rho^a_{st}(0)y}
                    +
                    (\mu+v)\rho^a_{st}(0)\INT{0}{\infty}{y}y
                    \INT{0}{\infty}{z}
                    e^{-\rho^a_{st}(0)(z+y)}\\
                    =&
                    \fr{\lambda}{\rho^{in}_{st}(0)^3}+\fr{\mu+v}{\rho^{in}_{st}(0)^2}
                    =2\fr{(v+\mu)^3}{\lambda^2}.
                }{}

    \section{Detailed derivation of financial metrics}
        In this Appendix, we show the detailed calculations for two financial metrics, such as the spread $s$ and the instantaneous price impact $\la\varepsilon\dd m\ra_{\rm in}$. 
        \subsection{Average spread}
            \begin{itembox}[l]{Average spread~$s$}
                \vspace{-5mm}
                \eqsp{
                    s\simeq\epsilon=\frac{\mu+v}{\lambda}
                }{eq:spread}
            \end{itembox}
            \begin{proof}
                The average spread is equal to the average of $a[\hat{\mbf{n}}^a]\Delta $ based on the relative coordinate. We thus obtain 
                    \eqsp{
                        s
                        = 
                        \la a[\hat{\mbf{n}}^a]\Delta \ra
                        = 
                        \sum^{\infty}_{i=1}i\Delta\text{Prob}[i=a[\hat{\mbf{n}}^a]]
                        \simeq
                        \sum^{\infty}_{i=1}i\Delta\rho^a_{\rm st}(i\Delta)\Delta e^{-\Delta\sum^{i-1}_{j=1}\rho^a_{\rm st}(j\Delta)}
                        \simeq 
                        \INT{0}{\infty}{z}z\rho^a_{\rm st}(z)e^{-\INT{0}{z}{x}\rho^a_{\rm st}(x)},
                    }{eq:spread_cal}
                by using the techniques in extreme value theory \eqref{eq:best_kyoku}. By considering an identity $\rho^a_{\rm st}(z)e^{-\INT{0}{z}{x}\rho^a_{\rm st}(x)} = -(\partial/\partial z)e^{-\INT{0}{z}{x}\rho^a_{\rm st}(x)}$, we obtain 
                    \eqsp{
                        s\simeq&
                        -\INT{0}{\infty}{z}z\fr{\partial}{\partial z}e^{-\INT{0}{z}{x}\rho^a_{\rm st}(x)}
                        =
                        \INT{0}{\infty}{z}e^{-\INT{0}{z}{x}\rho^a_{\rm st}(x)}
                        \simeq
                        \INT{0}{\infty}{z}e^{-z\rho^a_{\rm st}(0)}
                        =\epsilon,
                    }{}
                where we use $\rho^a_{\rm st}(0):=\eps^{-1}$.
            \end{proof}

    \subsection{Instantaneous price impact}
         The instantaneous price impact is the price change in the midprice immediately after a market order. The instantaneous price impact is equal to half of the first price gap (i.e., the difference between the best price and the second best price), which is given by the following formula: 
        \begin{itembox}[l]{Instantaneous price impact~$\langle \mr{d}m\rangle$}
            \vspace{-5mm}
            \eqsp{
                \langle \varepsilon\mr{d}m\rangle_{\text{in}}
                = 
                \fr{1}{2}\Le\langle\text{First price gap}\Ri\rangle 
                \simeq
                \fr{\epsilon}{2}
                =\fr{\mu+v}{2\lambda}
            }{eq:Instant_priceimpact}
        \end{itembox}
        \begin{proof}
            In a parallel calculation Eq.~\eqref{eq:ind_best_nextbest} in the Appendix~\ref{section:App_density_eq}, we obtain 
                \eqsp{
                    \text{Prob}[i=a[\mbf{\hat{n}}^a]\cap j=a[\mbf{\hat{n}}^a-\mbf{e}^i]]
                    =\langle\delta_{i,a[\mbf{\hat{n}}^a]}\delta_{j,a[\mbf{\hat{n}}^a-\mbf{e}^i]}\rangle.
                }{}
            Using this formula, the first price gap can be evaluated as 
                \eqsp{
                    \Le\langle\text{First price gap}\Ri\rangle   &=\sum_{i,j}(j-i)\Delta\text{Prob}[i=a[\mbf{\hat{n}}^a]\cap j=a[\mbf{\hat{n}}^a-\mbf{e}^i]]\\
                    &\simeq
                    \sum^{\infty}_{i=1}\sum^{\infty}_{j=i+1}(j-i)\Delta\rho^a_{st}(i\Delta)\Delta\rho^a_{st}(j\Delta)\Delta e^{-\Delta\sum^{i-1}_{l=1}\rho^a_{st}(l\Delta)-\Delta\sum^{j-1}_{l'=i+1}\rho^a_{st}(l'\Delta)}\\
                    &\simeq
                    \INT{0}{\infty}{z}\INT{z}{\infty}{y}(y-z)\rho^a_{st}(z)\rho^a_{st}(y)e^{-\INT{0}{y}{x}\rho^a_{st}(x)}
                }{}
            By using the partial integration, we obtain the formula for the instantaneous price impact as 
                \eqsp{
                    \langle \varepsilon\mr{d}m\rangle_{\text{in}}
                    \simeq&
                    -\fr{1}{2}\INT{0}{\infty}{z}\rho^a_{st}(z)\INT{z}{\infty}{y}y\fr{\partial}{\partial y}e^{-\INT{0}{y}{x}\rho^a_{st}(x)}
                    +
                    \fr{1}{2}\INT{0}{\infty}{z}z\rho^a_{st}(z)\INT{z}{\infty}{y}\fr{\partial}{\partial y}e^{-\INT{0}{y}{x}\rho^a_{st}(x)}\\
                    =&
                    \fr{1}{2}\INT{0}{\infty}{z}\rho^a_{st}(z)\INT{z}{\infty}{y}e^{-\INT{0}{y}{x}\rho^a_{st}(x)}
                    \simeq
                    \fr{1}{2}\INT{0}{\infty}{z}\rho^a_{st}(z)\INT{z}{\infty}{y}e^{-\rho^a_{st}(0)y}\\
                    =&
                    \fr{1}{2\rho^a_{st}(0)}\INT{0}{\infty}{z}\rho^a_{st}(z)e^{-\rho^a_{st}(0)z}
                    \simeq
                    \fr{1}{2\rho^a_{st}(0)}\INT{0}{\infty}{z}\rho^a_{st}(0)e^{-\rho^a_{st}(0)z}
                    =\fr{\epsilon}{2}.
                }{}
        \end{proof}

\section{systematic derivation of the gap-type mean-field equation via kinetic theory}\label{sec:app:GapMeanField}
    In this Appendix, we systematically derive another mean-field equation for price gaps via kinetic theory. 
\subsection{Definition of the gap space}
    While we formally consider discrete price levels, we assume that the tick size $\Delta$ is sufficiently small and that each price level has only zero or one orders. For positive integer $i$, the $i$-th lowest ask price at time $t$ is represented by $\hat{p}_i$ and the $i$-th highest bid price is represented by $\hat{p}_{-i}$. The spread at time $t$ is defined by     
        \eq{
            \hat{g}_0(t):=\frac{\hat{p}_1(t)-\hat{p}_{-1}(t)}{\Delta}.
        }{}
    We also define the $i$-th gap at the ask side $g_i(t)$ and the $i$-th gap at the bid side $g_{-i}(t)$ as 
        \eq{
            \hat{g}_i(t):=\frac{\hat{p}_{i+1}(t)-\hat{p}_{i}(t)}{\Delta}, \quad 
            \hat{g}_{-i}(t):=\frac{\hat{p}_{-i}(t)-\hat{p}_{-i-1}(t)}{\Delta},
        }{}
    where $g_0$ and $g_i$ are natural numbers.
    The state of the Santa Fe model is characterised by the state vector $\mathbf{\hat{g}}(t):=\left\{\hat{g}_i(t)\right\}^{\infty}_{i=-\infty}\in \mbb{N}^\infty$, which is called the {\it gap vector}. In this Appendix, we refer to this state space as {\it gap space}. The gap vector is an element of the gap space. 

\subsection{Dynamics of the gap vector}
    The stochastic dynamics of the gap is given by four patterns:
    \begin{itemize}
        \item Case 1: With $i\geq 1$, the ask price $\hat{p}_{i}$ disappears due to cancellation or a buy market order\\
              \eq{
                  \hat{g}_{i-1}(t+\mr{d}t)=\hat{g}_{i-1}(t)+\hat{g}_i(t),\quad \hat{g}_{j}(t+\mr{d}t)=\hat{g}_{j+1}(t)\quad (j\geq i).
              }{}
        \item Case 2: With $i\leq -1$, the bid price $\hat{p}_{i}$ disappears due to cancellation or a sell market order\\
              \eq{
                  \hat{g}_{i+1}(t+\mr{d}t)=\hat{g}_{i+1}(t)+\hat{g}_i(t),\quad \hat{g}_{j}(t+\mr{d}t)=\hat{g}_{j-1}(t)\quad(j\leq i).
              }{}
        \item Case 3: With $i\geq 0$, a new ask limit order is submitted within the $i$-th ask gap. Its ask price is written as $\hat{p}_{i+1}-x \in (\hat{p}_{i}, \hat{p}_{i+1})$ with $x \in (0,\hat{g}_i)$. The updated gaps are given by 
              \eq{
                  \hat{g}_i(t+\mr{d}t)=\hat{g}_{i}(t)-x,\quad 
                  \hat{g}_{i+1}(t+\mr{d}t)=x,\quad
                  \hat{g}_{j}(t+\mr{d}t)=\hat{g}_{j-1}(t)\quad (j\geq i+2).
              }{}
        \item Case 4: With $i\leq 0$, a new bid limit order is submitted within the $i$-th bid gap. Its ask price is written as $\hat{p}_{i-1}+x \in (\hat{p}_{i-1}, \hat{p}_{i1})$ with $x \in (0,\hat{g}_i)$. The updated gaps are given by 
              \eq{
                    \hat{g}_i(t+\mr{d}t)=\hat{g}_{i}(t)-x,\quad 
                    \hat{g}_{i-1}(t+\mr{d}t)=x,\quad 
                    \hat{g}_{j}(t+\mr{d}t)=\hat{g}_{j+1}(t)\quad (j\leq i-2).
              }{}
    \end{itemize}

\subsection{Exact master equation}    
    The master equation for the gap vector $\mathbf{\hat{g}}(t)$ is given by 
    \begin{itembox}[l]{Exact master equation on the gap space}
        \vspace{-5mm}
        \eqsp{
            \pdiff{}{t}P_t(\mathbf{g})
            =&
            \sum_{i\geq 1}(v+\mu\delta_{i,1})
            \Le\{\sum^{g_{i-1}-1}_{y=1}P_t(\cdots,\underset{(i-1)\text{-th}}{\underline{g_{i-1}-y}},\underset{i\text{-th}}{\underline{y}},\underset{(i+1)\text{-th}}{\underline{g_{i}}},\cdots)
            -
            P_t(\mathbf{g})\Ri\}\\
            &+
            \sum_{i\leq -1}(v+\mu\delta_{i,-1})
            \Le\{\sum^{g_{i+1}-1}_{y=1}
            P_t(\cdots,\underset{(i-1)\text{-th}}{\underline{g_{i}}},\underset{i\text{-th}}{\underline{y}},\underset{(i+1)\text{-th}}{\underline{g_{i+1}-y}},\cdots)
            -
            P_t(\mathbf{g})\Ri\}
            \\
            &+
            \lambda\Delta
            \sum_{i\geq0}
            \left\{
            P_t(\cdots,\underset{i\text{-th}}{\underline{g_{i}+g_{i+1}}},\underset{(i+1)\text{-th}}{\underline{g_{i+2}}},\cdots)-
            (g_i-1)
            P_t(\mathbf{g})\right\}\\
            &+
            \lambda\Delta
            \sum_{i\leq 0}
            \left\{P_t(\cdots,\underset{(i-1)\text{-th}}{\underline{g_{i-2}}},\underset{i\text{-th}}{\underline{g_i+g_{i-1}}},\cdots)-(g_i-1)P_t(\mbf{g})\right\}.
        }{eq:app:ME-gap}
        \vspace{-5mm}
    \end{itembox}
        
\subsubsection{Derivation}
    Its derivation is parallel to that for Eq.~\eqref{eq;master_eq}: for an arbitrary function $f(\mathbf{\hat{g}}(t))$, let us consider its time evolution $\dd f:=(\mathbf{\hat{g}}(t+\dd t))-(\mathbf{\hat{g}}(t))$ by taking into account the four cases, 
    \eq{
        \mr{d}f(\hat{\mathbf{g}}(t))
        =
        \begin{cases}
            \mr{d}f(\hat{\mathbf{g}}(t))^{{\rm CM};a}
            :=f(\cdots,\underset{(i-1)\text{-th}}{\underline{\hat{g}_{i-1}(t)+\hat{g}_{i}(t)}},\underset{i\text{-th}}{\underline{\hat{g}_{i+1}(t)}},\cdots)-f(\hat{\mathbf{g}}(t))&(\text{Case 1})\\
            \mr{d}f(\hat{\mathbf{g}}(t))^{{\rm CM};b}
            :=f(\cdots,\underset{i\text{-th}}{\underline{\hat{g}_{i-1}(t)}},\underset{(i+1)\text{-th}}{\underline{\hat{g}_{i+1}(t)+\hat{g}_{i}(t)}},\cdots)-f(\hat{\mathbf{g}}(t))&(\text{Case 2})\\
            \mr{d}f(\hat{\mathbf{g}}(t))^{{\rm S};a}
            :=f(\cdots,\underset{i\text{-th}}{\underline{\hat{g}_{i}(t)-x}},\underset{(i+1)\text{-th}}{\underline{x}},\underset{(i+2)\text{-th}}{\underline{\hat{g}_{i+1}(t)}},\cdots)-f(\hat{\mathbf{g}}(t))&(\text{Case 3})\\
            \mr{d}f(\hat{\mathbf{g}}(t))^{{\rm S};b}
            :=f(\cdots,\underset{(i-2)\text{-th}}{\underline{\hat{g}_{i-1}(t)}},\underset{(i-1)\text{-th}}{\underline{x}},\underset{i\text{-th}}{\underline{\hat{g}_{i}(t)-x}},\cdots)-f(\hat{\mathbf{g}}(t))&(\text{Case 4})
        \end{cases}
    }{}
    where $\mr{d}f(\hat{\mathbf{g}}(t))^{{\rm CM};a}$, $\mr{d}f(\hat{\mathbf{g}}(t))^{{\rm CM};b}$, $\mr{d}f(\hat{\mathbf{g}}(t))^{{\rm S};a}$, and $\mr{d}f(\hat{\mathbf{g}}(t))^{{\rm S};b}$ correspond to Case 1, 2, 3, and 4, respectively. By taking the ensemble average of both hand sides, we obtain 
    \eq{
        \left\langle \mr{d}f(\hat{\mathbf{g}}(t))\right\rangle
        =\left\langle \mr{d}f(\hat{\mathbf{g}}(t))\right\rangle^{{\rm CM};a}
        +\left\langle \mr{d}f(\hat{\mathbf{g}}(t))\right\rangle^{{\rm CM};b}
        +\left\langle \mr{d}f(\hat{\mathbf{g}}(t))\right\rangle^{{\rm S};a}
        +\left\langle \mr{d}f(\hat{\mathbf{g}}(t))\right\rangle^{{\rm S};b}.
    }{eq:mean_time_ev_gap}    

    We next evaluate the average contributions for Case 1-4 explicitly. The contribution for Case 1 is evaluated as follow: 
        \eqsp{
            \langle\mr{d}f(\hat{\mbf{g}}(t))\rangle^{\text{CM};a}
            :=&\mr{d}t\sum_{i\geq 1}(v+\mu\delta_{i,1})
            \sum_{\hat{\mathbf{g}}(t)}P_t(\hat{\mathbf{g}}(t))\Le\{f(\cdots,\underset{(i-1)\text{-th}}{\underline{\hat{g}_{i-1}(t)+\hat{g}_{i}(t)}},\underset{i\text{-th}}{\underline{\hat{g}_{i+1}(t)}},\cdots)
            -
            f(\hat{\mathbf{g}}(t))\Ri\}.
        }{}
    The contribution for Case 2 is given by 
        \eqsp{
            \left\langle \mr{d}f(\hat{\mbf{g}}(t))\right\rangle^{{\rm CM};b}
            :=\mr{d}t\sum_{i\leq -1}(v+\mu\delta_{i,-1})
            \sum_{\hat{\mathbf{g}}(t)}P_t(\hat{\mathbf{g}}(t))\Le\{
            f(\cdots,\underset{i\text{-th}}{\underline{\hat{g}_{i-1}(t)}},\underset{(i+1)\text{-th}}{\underline{\hat{g}_{i+1}(t)+\hat{g}_{i}(t)}},\cdots)
            -
            f(\hat{\mathbf{g}}(t))\Ri\}.\\
        }{}
    The contribution for Case 3 is given by 
        \eqsp{
        \left\langle \mr{d}f(\hat{\mbf{g}}(t))\right\rangle^{{\rm S};a}
        :=&\lambda\Delta\mr{d}t
        \sum_{i\geq0}\sum_{\hat{\mathbf{g}}(t)}\sum^{\hat{g}_i(t)-1}_{x=1}
        P_t(\hat{\mathbf{g}}(t))
        \left\{f(\cdots,\underset{i\text{-th}}{\underline{\hat{g}_{i}(t)-x}},\underset{(i+1)\text{-th}}{\underline{x}},\underset{(i+2)\text{-th}}{\underline{\hat{g}_{i+1}(t)}},\cdots)-f(\hat{\mathbf{g}}(t))\right\}.\\
        }{}
    The contribution for Case 4 is evaluated as 
        \eqsp{
            \left\langle \mr{d}f(\hat{\mbf{g}}(t))\right\rangle^{{\rm S};b}
            :=&\lambda\Delta\mr{d}t
            \sum_{i\leq 0}\sum_{\hat{\mathbf{g}}(t)}\sum^{\hat{g}_i(t)-1}_{x=1}P_t(\hat{\mathbf{g}}(t))
            \left\{f(\cdots,\underset{(i-2)\text{-th}}{\underline{\hat{g}_{i-1}(t)}},\underset{(i-1)\text{-th}}{\underline{x}},\underset{i\text{-th}}{\underline{\hat{g}_{i}(t)-x}},\cdots)-f(\hat{\mathbf{g}}(t))\right\}.\\
        }{}

    Since this identity holds for any $f(\hat{\mbf{g}}(t))$, let us focus on the specific case $f(\hat{\mbf{g}}(t))=\delta_{\hat{\mbf{g}}(t),\mbf{g}}$. The contributions for Case 1 and Case 2 are given by 
        \eqsp{
            \langle\mr{d}f(\hat{\mbf{g}}(t))\rangle^{\text{CM};a}
            =&\mr{d}t\sum_{i\geq 1}(v+\mu\delta_{i,1})
            \Le\{\sum_{\hat{g}_i(t)}P_t(\cdots,\underset{(i-1)\text{-th}}{\underline{g_{i-1}-\hat{g}_{i}(t)}},\underset{i\text{-th}}{\underline{\hat{g}_{i}(t)}},\underset{(i+1)\text{-th}}{\underline{g_{i}}},\cdots)
            -
            P_t(\mathbf{g})\Ri\}\\
            =&\mr{d}t\sum_{i\geq 1}(v+\mu\delta_{i,1})
            \Le\{\sum^{g_{i-1}-1}_{y=1}P_t(\cdots,\underset{(i-1)\text{-th}}{\underline{g_{i-1}-y}},\underset{i\text{-th}}{\underline{y}},\underset{(i+1)\text{-th}}{\underline{g_{i}}},\cdots)
            -
            P_t(\mathbf{g})\Ri\}
        }{}
    and
        \eqsp{
            \left\langle \mr{d}f(\hat{\mbf{g}}(t))\right\rangle^{{\rm CM};b}
            =&\mr{d}t\sum_{i\leq -1}(v+\mu\delta_{i,-1})
            \Le\{\sum_{\hat{g}_i(t)}
            P_t(\cdots,\underset{(i-1)\text{-th}}{\underline{g_{i}}},\underset{i\text{-th}}{\underline{\hat{g}_{i}(t)}},\underset{(i+1)\text{-th}}{\underline{g_{i+1}-\hat{g}_{i}(t)}},\cdots)
            -
            P_t(\mathbf{g})\Ri\}\\
            =&\mr{d}t\sum_{i\leq -1}(v+\mu\delta_{i,-1})
            \Le\{\sum^{g_{i+1}-1}_{y=1}
            P_t(\cdots,\underset{(i-1)\text{-th}}{\underline{g_{i}}},\underset{i\text{-th}}{\underline{y}},\underset{(i+1)\text{-th}}{\underline{g_{i+1}-y}},\cdots)
            -
            P_t(\mathbf{g})\Ri\},
        }{}
    where we replaced the dummy variable $\hat{g}_i(t)$ with $y$, and simplified the summation domain since the probability of the gap being 0 or less is zero.

    The average contributions for Case 3 and Case 4 are given by 
        \eqsp{
            \left\langle \mr{d}f(\hat{\mbf{g}}(t))\right\rangle^{{\rm S};a}
            =&\lambda\Delta\mr{d}t
            \sum_{i\geq0}\sum_{\hat{\mathbf{g}}(t)}\sum^\infty_{y=1}\delta_{y,\hat{g}_i(t)}\sum^{y-1}_{x=1}
            P_t(\hat{\mathbf{g}}(t))
            \left\{f(\cdots,\underset{i\text{-th}}{\underline{\hat{g}_{i}(t)-x}},\underset{(i+1)\text{-th}}{\underline{x}},\underset{(i+2)\text{-th}}{\underline{\hat{g}_{i+1}(t)}},\cdots)-f(\hat{\mathbf{g}}(t))\right\}\\
            =&\lambda\Delta\mr{d}t
            \sum_{i\geq0}
            \left\{\sum^\infty_{y=1}\sum^{y-1}_{x=1}\delta_{y,g_i+x}
            P_t(\cdots,\underset{i\text{-th}}{\underline{g_{i}+x}},\underset{(i+1)\text{-th}}{\underline{g_{i+2}}},\cdots)\delta_{x,g_{i+1}}-
            \sum^\infty_{y=1}\delta_{y,g_i}\sum^{y-1}_{x=1}
            P_t(\mathbf{g})\right\}\\
            =&\lambda\Delta\mr{d}t
            \sum_{i\geq0}
            \left\{\sum^\infty_{y=1}\delta_{y,g_i+g_{i+1}}
            P_t(\cdots,\underset{i\text{-th}}{\underline{g_{i}+g_{i+1}}},\underset{(i+1)\text{-th}}{\underline{g_{i+2}}},\cdots)\mbb{I}_{[g_{i+1}+1,\infty)}(y)-
            \sum^\infty_{y=1}\delta_{y,g_i}(y-1)
            P_t(\mathbf{g})\right\}\\
            =&\lambda\Delta\mr{d}t
            \sum_{i\geq0}
            \left\{
            P_t(\cdots,\underset{i\text{-th}}{\underline{g_{i}+g_{i+1}}},\underset{(i+1)\text{-th}}{\underline{g_{i+2}}},\cdots)-
            (g_i-1)
            P_t(\mathbf{g})\right\}
        }{}
    and
        \eqsp{
            \left\langle \mr{d}f(\hat{\mbf{g}}(t))\right\rangle^{{\rm S};b}
            =&\lambda\Delta\mr{d}t
            \sum_{i\leq 0}\sum_{\hat{\mathbf{g}}(t)}\sum^\infty_{y=1}\delta_{y,\hat{g}_i(t)}\sum^{y-1}_{x=1}P_t(\hat{\mathbf{g}}(t))
            \left\{f(\cdots,\underset{(i-2)\text{-th}}{\underline{\hat{g}_{i-1}(t)}},\underset{(i-1)\text{-th}}{\underline{x}},\underset{i\text{-th}}{\underline{\hat{g}_{i}(t)-x}},\cdots)-f(\hat{\mathbf{g}}(t))\right\}\\
            =&\lambda\Delta\mr{d}t
            \sum_{i\leq 0}
            \left\{\sum^\infty_{y=1}\sum^{y-1}_{x=1}\delta_{y,g_i+x}P_t(\cdots,\underset{(i-1)\text{-th}}{\underline{g_{i-2}}},\underset{i\text{-th}}{\underline{g_i+x}},\cdots)\delta_{x,g_{i-1}}-\sum^\infty_{y=1}\delta_{y,g_i}\sum^{y-1}_{x=1}P_t(\mbf{g})\right\}\\
            =&\lambda\Delta\mr{d}t
            \sum_{i\leq 0}
            \left\{\sum^\infty_{y=1}\delta_{y,g_i+g_{i-1}}P_t(\cdots,\underset{(i-1)\text{-th}}{\underline{g_{i-2}}},\underset{i\text{-th}}{\underline{g_i+g_{i-1}}},\cdots)\mbf{I}_{[g_{i-1}+1,\infty)}(y)-\sum^\infty_{y=1}\delta_{y,g_i}(y-1)P_t(\mbf{g})\right\}\\
            =&\lambda\Delta\mr{d}t
            \sum_{i\leq 0}
            \left\{P_t(\cdots,\underset{(i-1)\text{-th}}{\underline{g_{i-2}}},\underset{i\text{-th}}{\underline{g_i+g_{i-1}}},\cdots)-(g_i-1)P_t(\mbf{g})\right\}.\\
        }{}
        
    Finally, the left-hand side of \eqref{eq:mean_time_ev_gap} can be rewritten as 
    \eqsp{
    \left\langle \mr{d}f(\hat{\mathbf{g}}(t))\right\rangle
    =
    \sum_{\hat{\mathbf{g}}(t+dt)}P_{t+\mr{d}t}(\hat{\mathbf{g}}(t+dt))f(\hat{\mathbf{g}}(t+dt))
    -
    \sum_{\hat{\mathbf{g}}(t)}P_{t}(\hat{\mathbf{g}}(t))f(\hat{\mathbf{g}}(t))
    =
    P_{t+\mr{d}t}(\mathbf{g})-P_t(\mathbf{g}).
    }{}
    By taking the $\dd t\to 0$ limit, we obtain the exact master equation on the gap space

\subsection{BBGKY hierarchical equation on the gap space}
    By applying marginalisation of PDFs, we derive the BBGKY hierarchical equation---the time-evolution equation for the one-body PDF. Note that we define the marginalisation $\sum_{\mathbf{g}^{c;k}}[\dots]$ except for the $i$-th variable and the marginalisation $\sum_{\mathbf{g}^{c;k,l}}[\dots]$ except for the $i$-th and $l$-th variables, such that 
        \eq{
            \sum_{\mathbf{g}^{c;k}}[\dots]:= \left(\prod_{i\neq k}\sum^\infty_{g_i=1}\right) [\dots],\quad 
            \sum_{\mathbf{g}^{c;k,l}}[\dots]:=\left(\prod_{i\neq k, i\neq l}\sum^\infty_{g_i=1}\right) [\dots].
        }{}
        Using this operations, one-body and two-body PDFs are defined by 
        \eq{
            P^k_t(a):= \mbox{Prob}[\hat{g}_k=a] = \sum_{\mathbf{g}^{c;k}}P_t(\mathbf{g}), \quad 
            P^{k,l}_t(a,b):= \mbox{Prob}[\hat{g}_k=a,\hat{g}_{l}=b]:= \sum_{\mathbf{g}^{c;k,l}}P_t(\mathbf{g}).
        }{}

    \subsubsection{BBGKY hierarchical equation for the gap of the spread PDF}
        Let us focus on the gap of the spread PDF $P^{0}_t(g_0)$. From the exact master equation~\eqref{eq:app:ME-gap} on the gap space, we obtain the exact BBGKY hierarchical equation for the spread PDF:
           \eqsp{
                \pdiff{P^{0}_t(g_0)}{t}
                &=
                (v+\mu)
                \left\{\sum^{g_0-1}_{y=1}
                P_t^{0,1}(g_{0}-y,y)
                -P^{0}_t(g_{0})\right\}
                +
                (v+\mu)
                \left\{\sum^{g_0-1}_{y=1}
                P^{-1,0}_t(y,g_{0}-y)
                -P^0_t(g_{0})\right\}
                \\
                &+
                \lambda\Delta
                \left\{
                \sum_{g_1}P^0_t(g_0+g_{1})
                -
                (g_0-1)P^0_t(g_0)\right\}
                +
                \lambda\Delta
                \left\{\sum_{g_{-1}}P^0_t(g_0+g_{-1})
                -
                (g_0-1)P^0_t(g_{0})\right\}
            }{eq:app:BBGKY-spread-exact}

        \paragraph{Derivation.}
        Let us apply the marginalisation $\sum_{\mathbf{g}^{c;0}}[\dots]$ to the both hand sides of the master equation~\eqref{eq:app:ME-gap}. First, we have 
            \eqsp{
                &\sum_{\mathbf{g}^{c;0}}\sum_{i\geq 1}(v+\mu\delta_{i,1})
                \Le\{\sum^{g_{i-1}-1}_{y=1}P_t(\cdots,\underset{(i-1)\text{-th}}{\underline{g_{i-1}-y}},\underset{i\text{-th}}{\underline{y}},\underset{(i+1)\text{-th}}{\underline{g_{i}}},\cdots) - P_t(\mathbf{g})\Ri\} \\
                =& (v+\mu)\sum_{\mathbf{g}^{c;0}}
                \Le\{\sum^{g_0-1}_{y=1}P_t(\cdots,\underset{0\text{th}}{\underline{g_{0}-y}},\underset{1\text{st}}{\underline{y}},\underset{2\text{nd}}{\underline{g_{1}}},\cdots) - P_t(\mathbf{g})\Ri\}
                + 
                v\sum_{\mathbf{g}^{c;0}}\sum_{i\geq 2}
                \Le\{\sum^{g_{i-1}-1}_{y=1}P_t(\cdots,\underset{(i-1)\text{-th}}{\underline{g_{i-1}-y}},\underset{i\text{-th}}{\underline{y}},\underset{(i+1)\text{-th}}{\underline{g_{i}}},\cdots) - P_t(\mathbf{g})\Ri\} \\
                =& 
                (v+\mu)
                \left\{\sum^{g_0-1}_{y=1}
                P_t^{0,1}(g_{0}-y,y)
                -P^{0}_t(g_{0})\right\}.
            }{}
            In a parallel calculation, we obtain 
            \eqsp{
                \sum_{\mathbf{g}^{c;0}}\sum_{i\leq -1}(v+\mu\delta_{i,-1})
                \Le\{\sum^{g_{i+1}-1}_{y=1}
                P_t(\cdots,\underset{(i-1)\text{-th}}{\underline{g_{i}}},\underset{i\text{-th}}{\underline{y}},\underset{(i+1)\text{-th}}{\underline{g_{i+1}-y}},\cdots)
                -
                P_t(\mathbf{g})\Ri\}
                = (v+\mu) \left\{\sum^{g_0-1}_{y=1}
                P^{-1,0}_t(y,g_{0}-y)
                -P^0_t(g_{0})\right\}.
            }{}
        We also obtain 
            \eqsp{
                &\lambda\Delta
                \sum_{\mathbf{g}^{c;0}}\sum_{i\geq0}
                \left\{
                P_t(\cdots,\underset{i\text{-th}}{\underline{g_{i}+g_{i+1}}},\underset{(i+1)\text{-th}}{\underline{g_{i+2}}},\cdots)-
                (g_i-1)
                P_t(\mathbf{g})\right\} \\ 
                =&
                \lambda\Delta
                \left\{
                \sum_{g_1}P^0_t(g_0+g_{1})
                -
                (g_0-1)P^0_t(g_0)\right\}
                +
                \lambda\Delta
                \sum_{i\geq1}
                \left\{\sum_{g_i}\sum_{g_{i+1}}P^{0,i}_t(g_0,g_i+g_{i+1})
                -
                \sum_{g_i}(g_i-1)P^{0,i}_t(g_0,g_i)\right\} \\
                =& \lambda\Delta
                \left\{
                \sum_{g_1}P^0_t(g_0+g_{1})
                -
                (g_0-1)P^0_t(g_0)\right\}, 
            }{}
        where we use 
            \eqsp{
                &\sum_{g_i}\sum_{g_{i+1}}P^{0,i}_t(g_0,g_i+g_{i+1})
                -
                \sum_{g_i}(g_i-1)P^{0,i}_t(g_0,g_i)\\
                =&\sum_{g_i}\sum_{g_{i+1}}\sum_{y}P^{0,i}_t(g_0,y)\delta_{y,g_i+g_{i+1}}
                -
                \sum_{g_i}(g_i-1)P^{0,i}_t(g_0,g_i)\\
                =&\sum_{g_i}\sum_{y}P^{0,i}_t(g_0,y)\mbb{I}_{[1+g_i,\infty)}(y)
                -
                \sum_{g_i}(g_i-1)P^{0,i}_t(g_0,g_i)\\
                =&\sum_{y}\sum^{y-1}_{g_i=1}P^{0,i}_t(g_0,y)
                -
                \sum_{g_i}(g_i-1)P^{0,i}_t(g_0,g_i)=0.
            }{}
        Likewise, we obtain 
            \eqsp{
                \lambda\Delta \sum_{\mathbf{g}^{c;0}}\sum_{i\leq 0}
                \left\{P_t(\cdots,\underset{(i-1)\text{-th}}{\underline{g_{i-2}}},\underset{i\text{-th}}{\underline{g_i+g_{i-1}}},\cdots)-(g_i-1)P_t(\mbf{g})\right\} 
                = 
                \lambda\Delta
                \left\{\sum_{g_{-1}}P^0_t(g_0+g_{-1})
                -
                (g_0-1)P^0_t(g_{0})\right\}.
            }{}        
        In summary, we obtain Eq.~\eqref{eq:app:BBGKY-spread-exact}.        

    \subsubsection{BBGKY hierarchical equation for the gap PDF}
        By applying the marginalisation $\sum_{\mathbf{g}^{c;k}}[\dots]$ to the both hand sides of the master equation~\eqref{eq:app:ME-gap} with $k\geq 1$, we obtain the BBGKY hierarchical equation for the gap PDF. Finally, in the sum of Eq.~\eqref{eq:app:ME-gap}, the cancellation order and market order contribute only for $1\leq i \leq k+1$, and the submission contributes only for $1\leq i \leq k$; other contributions are cancelled out:
            \begin{align}
                &\sum_{\mathbf{g}^{c;k}}\sum_{i\geq k+2}(v+\mu\delta_{i,1})
                \Le\{\sum^{g_{i-1}-1}_{y=1}P_t(\cdots,\underset{(i-1)\text{-th}}{\underline{g_{i-1}-y}},\underset{i\text{-th}}{\underline{y}},\underset{(i+1)\text{-th}}{\underline{g_{i}}},\cdots)
                -
                P_t(\mathbf{g})\Ri\} = 0,\\
                &\sum_{\mathbf{g}^{c;k}}\sum_{i\leq -1}(v+\mu\delta_{i,-1})
                \Le\{\sum^{g_{i+1}+1}_{y=1}
                P_t(\cdots,\underset{(i-1)\text{-th}}{\underline{g_{i}}},\underset{i\text{-th}}{\underline{y}},\underset{(i+1)\text{-th}}{\underline{g_{i+1}-y}},\cdots)
                -
                P_t(\mathbf{g})\Ri\} = 0,\\
                &\lambda\Delta
                \sum_{\mathbf{g}^{c;k}}\sum_{i\geq k+1}
                \left\{
                P_t(\cdots,\underset{i\text{-th}}{\underline{g_{i}+g_{i+1}}},\underset{(i+1)\text{-th}}{\underline{g_{i+2}}},\cdots)-
                (g_i-1)
                P_t(\mathbf{g})\right\}=0,\\
                &\lambda\Delta
                \sum_{\mathbf{g}^{c;k}}\sum_{i\leq 0}
                \left\{P_t(\cdots,\underset{(i-1)\text{-th}}{\underline{g_{i-2}}},\underset{i\text{-th}}{\underline{g_i+g_{i-1}}},\cdots)-(g_i-1)P_t(\mbf{g})\right\}=0.
            \end{align}        
        We then obtain 
            \eqsp{
                &\pdiff{P^k_t(g_k)}{t}
                =
                \sum^{k}_{i=1}(v+\mu\delta_{i,1})
                \left\{
                P^{k+1}_t(g_k)
                -P^k_t(g_k)\right\}
                +
                \underset{i=k+1}{\underline{v
                \left\{\sum^{g_k-1}_{y=1}
                P^{k,k+1}_t(g_{k}-y,y)
                -P^k_t(g_{k})\right\}}}\\
                &+
                \lambda\Delta
                \sum^{k-2}_{i=0}
                \left\{\sum_{g_i,g_{i+1}}P^{i,k-1}_t(g_i+g_{i+1},g_k)
                -
                \sum_{g_i}(g_i-1)P^{i,k}_t(g_i,g_k)\right\}\\
                &+
                \underset{i=k-1}{\underline{\lambda\Delta
                \left\{\sum_{g_{k-1}}P^{k-1}_t(g_{k-1}+g_{k})
                -
                \sum_{g_{k-1}}(g_{k-1}-1)P^{k-1,k}_t(g_{k-1},g_{k})\right\}}}
                +
                \underset{i=k}{\underline{\lambda\Delta
                \left\{\sum_{g_{k+1}}P^k_t(g_k+g_{k+1})
                -
                (g_k-1)P^k_t(g_k)\right\}}}
                \\
                &=
                (vk+\mu)
                \left\{
                P^{k+1}_t(g_k)
                -P^k_t(g_k)\right\}
                +
                v
                \left\{\sum^{g_k-1}_{y=1}
                P^{k,k+1}_t(g_{k}-y,y)
                -P^k_t(g_{k})\right\}
                \\
                &+
                \lambda\Delta
                \sum^{k-2}_{i=0}
                \sum_{g_i}(g_i-1)
                \left\{P^{i,k-1}_t(g_i,g_k)
                -P^{i,k}_t(g_i,g_k)\right\}\\
                &
                +
                \lambda\Delta
                \left\{\sum_{g_{k-1}}P^{k-1}_t(g_{k-1}+g_{k})
                -
                \sum_{g_{k-1}}(g_{k-1}-1)P^{k-1,k}_t(g_{k-1},g_{k})\right\}
                +
                \lambda\Delta
                \left\{\sum_{g_{k+1}}P^k_t(g_k+g_{k+1})
                -
                (g_k-1)P^k_t(g_k)\right\}.\\
            }{}
        Here, we use the equation 
            \eqsp{
                \sum_{g_i,g_{i+1}}P^{i,k-1}_t(g_i+g_{i+1},g_k)
                =\sum_{g_i,g_{i+1},y}P^{i,k-1}_t(y,g_k)\delta_{y,g_i+g_{i+1}}
                =\sum_{y}\sum^{y-1}_{g_i=1}P^{i,k-1}_t(y,g_k)
                =\sum_{y}(y-1)P^{i,k-1}_t(y,g_k),
            }{}
        where the dummy variable $y$ can be replaced by $g_{i}$.

\subsection{Boltzmann-equation for the spread and gap-size PDFs}
    Let us apply the mean-field approximation, such that $\left\{\hat{g}_i(t)\right\}^{\infty}_{i=-\infty}$ are all independent of each other. Also, let us assume the symmetry of the LOB between bid and ask sides. In other words, we assume the following relations: 
        \eq{
            P_t(\left\{g_i\right\}^{\infty}_{i=-\infty}) \simeq\prod^{\infty}_{j=-\infty}P^{j}_t(g_j), \quad 
            P^{-j}_t(g_j)=P^{j}_t(g_j),
        }{}
    where $P^j_t(g_j)$ is the $j$-th gap PDF. We than obtain the Boltzmann equations for the spread PDF $P^0_t(g_o)$,
    \begin{itembox}[l]{Boltzmann equation for the spread PDF}
        \vspace{-5mm}
        \eqsp{
            \pdiff{P^{0}_t(g_0)}{t}
            =
            2(v+\mu)\left[\sum^{\infty}_{y=1}P^{0}_t(g_0-y)P^{1}_t(y)-P^0_t(g_0)\right]
            +2\lambda\Delta\left[\sum^{\infty}_{y=1}P^0_t(g_0+y)-(g_0-1)P^0_t(g_0)\right]
        }{eq:app:BoltzmannSpread}
    \end{itembox}
    and the ask-side gap PDF $P^k_t(g_k)$ with $k\geq1$, 
    \begin{itembox}[l]{Boltzmann equation for the ask-side gap PDF}
        \vspace{-5mm}
        \eqsp{
            \pdiff{P^k_t(g_k)}{t}
            =&
            (vk+\mu)\left[P^{k+1}_t(g_k)-P^k_t(g_k)\right]
            +v\left[\sum^{g_k-1}_{y=1}P^{k}_t(g_k-y)P^{k+1}_t(y)-P^k_t(g_k)\right]\\
            &
            +\lambda\Delta\sum^{k-2}_{i=0}(\left\langle g_i\right\rangle -1)
            \left[P^{k-1}_t(g_k)-P^{k}_t(g_k)\right]
            +\lambda\Delta\left[\sum^{\infty}_{y=1}P^{k-1}_t(g_k+y)
            -(\langle g_{k-1}\rangle -1)P^{k}_t(g_k)\right]
            \\
            &
            +\lambda\Delta\left[\sum^{\infty}_{y=1}P^k_t(g_k+y)-(g_k-1)P^k_t(g_k)\right].
        }{eq:app:BoltzmannGap}
    \end{itembox}

\subsection{Closed mean-field equation for the average spread and average gap}
    From the mean-field master equation for the spread PDF~\eqref{eq:app:BoltzmannSpread}, the time-evolution equation for the average spread $\left\langle g_0\right\rangle$ is given by 
    \eq{
        \diff{\left\langle \hat{g}_0\right\rangle}{t}
        =
        2\left(\mu+v\right)\left\langle \hat{g}_1\right\rangle
        -\lambda\Delta\left(\left\langle \hat{g}^2_0\right\rangle
        -\left\langle \hat{g}_0\right\rangle\right).
    }{}
    Also, from the mean-field master equation for the gap PDF~\eqref{eq:app:BoltzmannGap}, the average gap $\langle g_k\rangle$ obeys 
    \eqsp{
        \pdiff{\langle g_k\rangle}{t}
        =&
        (vk+\mu)\left[\langle g_{k+1}\rangle-\langle g_k\rangle\right]
        +v\langle g_{k+1}\rangle
        +\lambda\Delta\left[\frac{1}{2}(\langle g^2_{k-1}\rangle-\left\langle g_{k-1}\right\rangle)-(\left\langle g_{k-1}\right\rangle -1)\left\langle g_k\right\rangle\right]\\
        &+\lambda\Delta\sum^{k-2}_{i=0}(\langle g_i\rangle-1)
        \left[\langle g_{k-1}\rangle-\langle g_k\rangle\right]
        -\lambda\Delta
        \frac{1}{2}(\langle g^2_k\rangle-\left\langle g_k\right\rangle).
    }{}
    In the steady state, the average spread and average gap size follows the following nonlinear equation: 
    \begin{itembox}[l]{Average spread and gap-size equation}
        \vspace{-3mm}
        \begin{equation}\label{eq:steady_gap}
            \begin{cases}
                0= &
                2\left(\mu+v\right) \left(\left\langle \hat{g}_1\right\rangle\right)
                -\lambda\Delta\left(\left\langle \hat{g}^2_0\right\rangle
                -\left\langle \hat{g}_0\right\rangle\right) \\
                0= & \{\mu+(k+1)v\}\langle g_{k+1}\rangle
                -\frac{\lambda\Delta}{2}(\langle g^2_{k}\rangle-\langle g_{k}\rangle)
                -\lambda\Delta \langle g_{k}\rangle
                \sum^{k-1}_{i=0}(\langle g_{i}\rangle-1).
            \end{cases}
        \vspace{-2mm}
        \end{equation}
    \end{itembox}
    This equation corresponds to the previous results of E.~Smith {\it et al.} 2003 (i.e., Eqs.~{(50)} and {(53)} in Ref.~\cite{ZIOB-QFin2003}). Thus, we have provided the mathematical foundation for their mean-field gap-size equation via our kinetic theory. 

\subsection{Scaling analysis}
    In E.~Smith {\it et al.} 2003, they applied several approximations to the mean-field equation~\eqref{eq:steady_gap} to obtain the following recursive formula:
    \begin{equation}\label{eq:Smith_gap_comp}
        \begin{cases}
            0= &
            \left(\mu+v\right) \left(\left\langle \hat{g}_1\right\rangle\right)
            -\lambda\Delta\left(\left\langle \hat{g}_0\right\rangle^2
            -\left\langle \hat{g}_0\right\rangle\right) \\
            0= & \{\mu+(k+1)v\}\langle g_{k+1}\rangle
            -\lambda\Delta \langle g_{k}\rangle
            \sum^{k}_{i=0}(\langle g_{i}\rangle-1)
        \end{cases}.
    \end{equation}
    By assuming the boundary condition
        \eqsp{
            \lim_{k\to\infty}\langle g_k \rangle=1+\frac{v}{\lambda\Delta},
        }{}
    they numerically evaluated $\la g_0\ra$ by solving Eq.~\eqref{eq:Smith_gap_comp}. They argue a scaling relation for the spread
        \eq{
            \left\langle s\right\rangle=\la g_0\ra\Delta\sim O\left(\frac{\mu}{\lambda}\right)
        }{}
    via their numerical simulations, while there has been no mathematical derivation yet.

    In this Appendix, we mathematically prove their argument. In other words, we prove the following proposition: 
    \begin{itembox}[l]{Scaling law for the average spread}
        \vspace{-3mm}
        \begin{equation}
            \lim_{k\to\infty}\langle g_k \rangle=1+\frac{v}{\lambda\Delta}
            \quad \Longrightarrow \quad 
            \left\langle g_0\right\rangle\leq1+\frac{\mu+v}{\lambda\Delta}.
            \vspace{-2mm}
        \end{equation}
    \end{itembox}
    This result readily implies that $\la s\ra=\Delta \la g_0\ra = O(\mu/\lambda)$ and thus supports the numerical argument in Ref.~\cite{ZIOB-QFin2003}.

    \begin{proof}
        We prove the above statement by proving its contraposition: 
            \begin{equation}
                \left\langle g_0\right\rangle > 1+\frac{\mu+v}{\lambda\Delta} 
                \quad \Longrightarrow \quad 
                \lim_{k\to\infty}\langle g_k \rangle\neq 1+\frac{v}{\lambda\Delta}.
                \label{prop:app:contraposition}
            \end{equation}            
        By assuming an initial condition $\left\langle g_0\right\rangle > 1+(\mu+v)/(\lambda\Delta)$, we obtain 
            \eqsp{
                \langle g_{1}\rangle > \langle g_{0}\rangle.
            }{}
        from the recurrence relation~\eqref{eq:Smith_gap_comp}. Here we prove
            \begin{equation}
                \label{eq:app:induction}
                \langle g_{k}\rangle > \langle g_{k-1}\rangle, \quad
                \langle g_{k}\rangle>\left\langle g_0\right\rangle
                \quad \Longrightarrow \quad 
                \langle g_{k+1}\rangle > \langle g_{k}\rangle, \quad
                \langle g_{k+1}\rangle>\left\langle g_0\right\rangle.
            \end{equation}
        as mathematical induction. Indeed, from Eq.~\eqref{eq:Smith_gap_comp}, we obtain
            \eqsp{
                \frac{(\mu+kv)\langle g_{k}\rangle}{\lambda\Delta \langle g_{k-1}\rangle}
                =\sum^{k-1}_{i=0}(\langle g_{i}\rangle-1).
            }{}
        We then obtain 
            \eqsp{
                \frac{\{\mu+(k+1)v\}\langle g_{k+1}\rangle}{\lambda\Delta \langle g_{k}\rangle}
                 -\frac{(\mu+kv)\langle g_{k}\rangle}{\lambda\Delta \langle g_{k-1}\rangle}
                =\langle g_{k}\rangle-1,
            }{}
        or equivalently,  
            \eqsp{
                 \frac{\langle g_{k+1}\rangle}{\langle g_{k}\rangle}
                =\frac{(\mu+kv)\langle g_{k}\rangle+\lambda\Delta \langle g_{k-1}\rangle(\langle g_{k}\rangle-1)}{\{\mu+(k+1)v\}\langle g_{k}\rangle}.
            }{eq:gap_three_zenka}
        Assuming $\langle g_{k}\rangle > \langle g_{k-1}\rangle$ and $\langle g_{k}\rangle>\left\langle g_0\right\rangle > 1+(\mu+v)/(\lambda\Delta)$,
        we obtain
            \eqsp{
                 \frac{\langle g_{k+1}\rangle}{\langle g_{k}\rangle}
                >\frac{\mu+kv+\lambda\Delta (\langle g_{k}\rangle-1)}{\{\mu+(k+1)v\}}
                >1.
            }{}
        We therefore show the proposition~\eqref{eq:app:induction} via mathematical induction. This fact implies that $\{g_k\}_k$ is monotonically increasing and satisifies the following relation: 
            \eqsp{
                \lim_{k\to\infty}\langle g_{k}\rangle>\left\langle g_0\right\rangle > 1+\frac{\mu+v}{\lambda\Delta}.
            }{}
        Thus, we show the contrapositive proposition~\eqref{prop:app:contraposition}. 
    \end{proof}

\end{document}